\documentclass[twocolumn,twocolappendix]{aastex63}
\usepackage{url}
\usepackage[colorlinks=true]{hyperref}
\usepackage{graphicx}
\usepackage{overpic}

\newcommand{\G}{\mathcal{G}}

\newcommand{\appropto}{\mathrel{\vcenter{\offinterlineskip\halign{\hfil$##$\cr\propto\cr\noalign{\kern2pt}\sim\cr\noalign{\kern-2pt}}}}}
\newcommand{\bjdtdb}{\ensuremath{\rm {BJD_{TDB}}}}
\newcommand{\feh}{\ensuremath{\left[{\rm Fe}/{\rm H}\right]}}
\newcommand{\teff}{\ensuremath{T_{\rm eff}}}
\newcommand{\logg}{\ensuremath{\,{\rm log}\,{g}}}

\newcommand{\msun}{\ensuremath{\,{\rm M_\Sun}}}
\newcommand{\rsun}{\ensuremath{\,{\rm R_\Sun}}}
\newcommand{\lsun}{\ensuremath{\,{\rm L_\Sun}}}
\newcommand{\mearth}{\ensuremath{\,{\rm M_{\oplus}}}}

\newcommand{\mstar}{\ensuremath{\,M_{\rm *}}}
\newcommand{\rstar}{\ensuremath{\,R_{\rm *}}}
\newcommand{\mj}{\ensuremath{\,{\rm M_{\rm J}}}}
\newcommand{\rj}{\ensuremath{\,{\rm R_{\rm J}}}}

\newcommand{\mplanet}{\ensuremath{\,M_{\rm P}}}
\newcommand{\rp}{\ensuremath{\,R_{\rm P}}}
\newcommand{\fave}{\langle F \rangle}
\newcommand{\fluxcgs}{10$^9$ erg s$^{-1}$ cm$^{-2}$}

\submitjournal{ApJS}
\shortauthors{Wang et al.}

\shorttitle{New Photometry, New Parameters}
\begin{document}

\title{Transiting Exoplanet Monitoring Project (TEMP). VI. The Homogeneous Refinement of System Parameters for 39 Transiting Hot Jupiters with \textbf{127} New Light Curves}

% \affiliation{School of Physics and Astronomy, Sun Yat-Sen University, Zhuhai, 519082, China; \href{mailto:wangyhao5@mail.sysu.edu.cn}{wangyhao5@mail.sysu.edu.cn}}
% \author[0000-0002-0376-6365]{Xian-Yu Wang}
% \affiliation{National Astronomical Observatories, Chinese Academy of Sciences, Beijing 100012, China; \href{mailto:xianyu\_wang@nao.cas.cn}{xianyu\_wang@nao.cas.cn}}
% \affiliation{University of Chinese Academy of Sciences, Beijing, 100049, China}

\correspondingauthor{Xian-Yu Wang}
\email{xianyu\_wang@nao.cas.cn}
\correspondingauthor{Yong-Hao Wang}
\email{wangyhao5@mail.sysu.edu.cn}

\affiliation{School of Physics and Astronomy, Sun Yat-Sen University, Zhuhai, 519082, China}
\author[0000-0002-0376-6365]{Xian-Yu Wang}
\affiliation{National Astronomical Observatories, Chinese Academy of Sciences, Beijing 100012, China}
\affiliation{University of Chinese Academy of Sciences, Beijing, 100049, China}

\author[0000-0003-0261-6362]{Yong-Hao Wang}
\affiliation{School of Physics and Astronomy, Sun Yat-Sen University, Zhuhai, 519082, China}

\author[0000-0002-7846-6981]{Songhu Wang}
\affiliation{Department of Astronomy, Indiana University, Bloomington, IN 47405, USA}

\author[0000-0001-8037-1984]{Zhen-Yu Wu}
\affiliation{National Astronomical Observatories, Chinese Academy of Sciences, Beijing 100012, China}
\affiliation{University of Chinese Academy of Sciences, Beijing, 100049, China}

\author[0000-0002-7670-670X]{Malena Rice}
\affiliation{Department of Astronomy, Yale University, New Haven, CT 06511, USA}

\author{Xu Zhou}
\affiliation{National Astronomical Observatories, Chinese Academy of Sciences, Beijing 100012, China}

\author[0000-0001-8870-3146]{Tobias C. Hinse}
\affiliation{Institute of Astronomy, Faculty of Physics, Astronomy and Informatics, Nicolaus Copernicus University, Grudziadzka 5, 87-100 Torun, Poland}
\affiliation{Chungnam National University, Department of Astronomy, Space Science $\&$ Geology, 34134 Daejeon, Republic of Korea}

\author[0000-0001-5162-1753]{Hui-Gen Liu}
\affiliation{School of Astronomy and Space Science and Key Laboratory of Modern Astronomy and Astrophysics in Ministry of Education, Nanjing University, Nanjing 210093, China}

\author[0000-0002-0378-2023]{Bo Ma}
\affiliation{School of Physics and Astronomy, Sun Yat-Sen University, Zhuhai, 519082, China}

\author{Xiyan Peng}
\affiliation{National Astronomical Observatories, Chinese Academy of Sciences, Beijing 100012, China}

\author[0000-0003-3491-6394]{Hui Zhang}
\affiliation{School of Astronomy and Space Science and Key Laboratory of Modern Astronomy and Astrophysics in Ministry of Education, Nanjing University, Nanjing 210093, China}
\affiliation{Shanghai Astronomical Observatory, Chinese Academy of Sciences, Shanghai 200030, China}

\author[0000-0003-0454-7890]{Cong Yu}
\affiliation{School of Physics and Astronomy, Sun Yat-Sen University, Zhuhai, 519082, China}

\author[0000-0003-1680-2940]{Ji-Lin Zhou}
\affiliation{School of Astronomy and Space Science and Key Laboratory of Modern Astronomy and Astrophysics in Ministry of Education, Nanjing University, Nanjing 210093, China}

\author[0000-0002-3253-2621]{Gregory Laughlin}
\affiliation{Department of Astronomy, Yale University, New Haven, CT 06511, USA}

\begin{abstract}
We present 127 new transit light curves for 39 hot Jupiter systems, obtained over the span of five years by two ground-based telescopes. A homogeneous analysis of these newly collected light curves together with archived spectroscopic, photometric, and Doppler velocimetric data using EXOFASTv2 leads to a significant improvement in the physical and orbital parameters of each system. All of our stellar radii are constrained to accuracies of better than 3\%.  The planetary radii for 37 of our 39 targets are determined to accuracies of better than $5\%$. Compared to our results, the literature eccentricities are preferentially overestimated due to the Lucy-Sweeney bias. Our new photometric observations therefore allow for significant improvement in the orbital ephemerides of each system. Our correction of the future transit window amounts to a change exceeding $10\,{\rm min}$ for ten targets at the time of JWST's launch, including a $72\,{\rm min}$ change for WASP-56. The measured transit mid-times for both literature light curves and our new photometry show no significant deviations from the updated linear ephemerides, ruling out in each system the presence of companion planets with masses greater than $0.39 - 5.0\, \mearth$, $1.23 - 14.36\, \mearth$, $1.65 - 21.18\, \mearth$, and $0.69 - 6.75\, \mearth$ near the 1:2, 2:3, 3:2, and 2:1 resonances with the hot Jupiters , respectively, at a confidence level of $\pm 1\,\sigma$. The absence of resonant companion planets in the hot Jupiter systems is inconsistent with the conventional expectation from disk migration.
\end{abstract}
% \keywords{planets and satellites: fundamental parameters --- stars: fundamental parameters --- \\techniques: photometric \vspace{1.5cm} }


\section{Introduction}
Thousands of exoplanets have been detected to date; among them, transiting hot Jupiters are the most observationally accessible class of exoplanets and hence provide the highest signal-to-noise data (see, e.g., new discoveries from the Transiting Exoplanet Survey Satellite (TESS): \citealt{Wangs2019, Canas2019, Jones2019, Addison2020, Brahm2020, Davis2020, Jordan2020}). These transiting ``hot Jupiters" provide us with unique laboratories to test theories of exoplanet formation and evolution with relatively high precision over short time scales.

% Thousands of exoplanets have been detected to date, and perhaps one of the most valuable findings has been that of short-period giant planets that transit in front of their bright host stars once per orbit as seen from Earth (see, e.g., new discoveries from the Transiting Exoplanet Survey Satellite (TESS): \citealt{Wangs2019, Canas2019, Jones2019, Addison2020, Brahm2020, Davis2020, Jordan2020}). These transiting ``hot Jupiters" provide us with unique laboratories to test theories of exoplanet formation and evolution with relatively high precision over short time scales.

The frequency of a planet's transits encodes the orbital period, and the depth of these transits encodes the planetary radius, assuming we know the stellar radius. A combination of photometric radius measurements and Doppler mass determinations enables us to measure the bulk density of exoplanets, helping us to understand what they are made of and how they formed. Some hot Jupiters have unusually low densities by comparison with the expected values from structural models \citep{Laughlin2011}. The dominant mechanism that slows down their convective cooling is still poorly known \citep{Thorngren2018}. 

Moreover, transiting hot Jupiters around bright stars enable the study of their stellar obliquities via the Rossiter-Mclaughlin effect (\citealt{Holt1893, Schlesinger1910, Rossiter1924, McLaughlin1924, Queloz2000, Hebrard2008, Albrecht2012, Zhou2016, Wang2018a, Addison2018, Winn2015} and references within), planetary atmospheres using transmission spectroscopy (\citealt{Charbonneau2002, Kreidberg2014, Deming2017} and references within), and thermal emission through secondary eclipses \citep{Charbonneau2005,Deming2005} and phase variations \citep{Borucki2009, Knutson2012}. Transiting hot Jupiters around bright stars also present the opportunity to measure planetary oblateness \citep{Carter2010}, rotation rates \citep{Seager2002}, orbital variations (e.g. transit timing variations (TTVs), \citealt{Holman2006, Millholland2016, Wu2018}; orbital decay, \citealt{Schlaufman2013, Penev2018}, orbital procession, \citealt{Bouma2019}), and planetary system architectures \citep{Becker2015, Canas2019, Huang2020}. These observational properties afford a unique opportunity for us to understand the underlying physics of tidal and perturbation theory, as well as planet formation mechanisms.

Although the number and variety of exoplanets have been dramatically expanded by space missions (\textit{CoRoT}, \citealt{Auvergne2009}; \textit{Kepler}, \citealt{Borucki2010}; \textit{K2}, \citealt{Howell2014}; \textit{TESS}, \citealt{Ricker2015}), most of the currently known hot Jupiters were discovered by ground-based transit surveys, such as SuperWASP \citep{Pollacco2006}, HATNet \citep{Bakos2004}, HATSouth \citep{Bakos2013}, OGLE \citep{Udalski2002}, TRES \citep{Alonso2004}, QES \citep{Alsubai2013}, CSTAR \citep{Wang2014}, KELT \citep{Pepper2007}, XO \citep{McCullough2005}, MASCARA \citep{Talens2017}, MEarth \citep{Irwin2009}, and NGTS \citep{Wheatley2018}.

Some hot Jupiters that are discovered by ground-based transit surveys receive only a handful of photometric follow-up observations. We therefore initiated a geographically dispersed Transiting Exoplanet Monitoring Project (TEMP, \citealt{Wang2018b}) by leveraging abundant 1m-class telescopes, to which we have access, to gather long-term, high-quality photometry for currently known hot Jupiters. Considering the target visibility, telescope size, and photometric precision that we usually achieve, we set the following three criteria for target selection: 1). Targets are above a declination of $-20^{\circ}$; 2). The magnitudes of targets are between $V=7.5$ and $V=14.5$; and 3). The planetary transit has depth $>0.005\,$mag. TEMP offers a powerful tool for us to achieve several goals. First, we can refine the orbital and physical parameters of the known transiting hot Jupiters discovered with ground-based photometric surveys. Second, we can identify statistically significant TTVs, which can be caused not only by planet-planet interactions \citep{Wangs2017}, but also by tidal dissipation, as well as apsidal precession caused by the stellar quadrupole moment, general relativity, and long-period planetary/stellar companions \citep{Pal2008}.

In this paper, we present 127 new photometric light curves for 39 transiting hot Jupiters. By jointly analyzing our light curves together with archived photometric and Doppler velocimetric data, as well as stellar information, we report refined system parameters for these targets.
 
This paper is organized as follows. In Section 2, we describe the new photometric observations and their reduction. Section 3 details the technique we used to estimate the system parameters. Section 4 discusses our results and their implications. 

\section{Observations and Data Reduction}
\begin{figure}
    % \hskip -0.5cm\
    \includegraphics[width=0.5\textwidth,trim=40 55 40 100, clip]{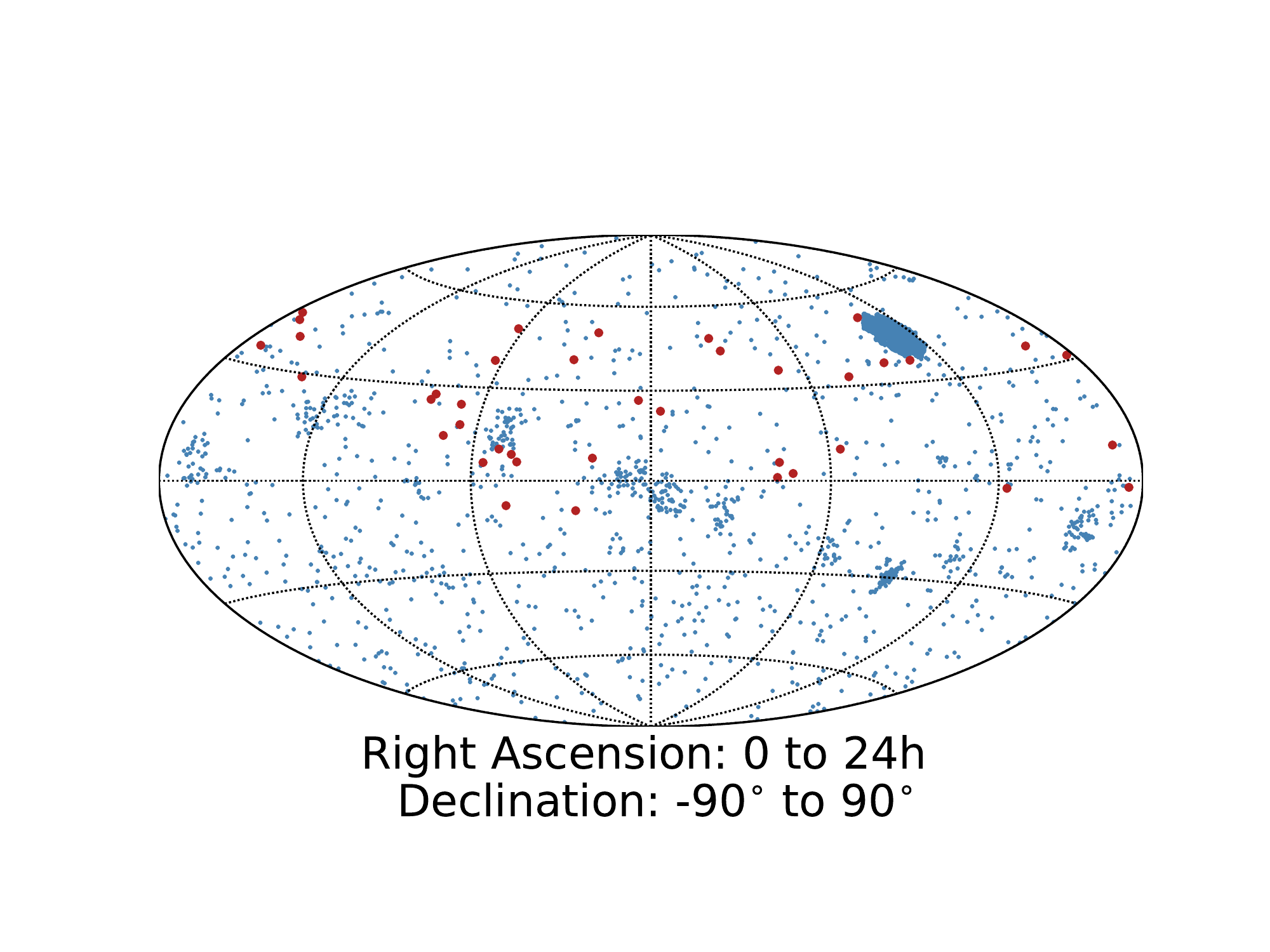}
    \caption{The distribution of observed TEMP targets (red dots). We collected 127 transit light curves for 39 hot Jupiter systems between Nov 11, 2013 and May 08, 2018. The brightest target in our sample is WASP-38 ($V = 9.48$)  and the faintest one is HAT-P-53 ($V = 13.73$). All other currently known exoplanets are displayed in blue.}
    \label{fig:temp_source}
\end{figure}

\begin{figure}[t]
    \centering
    \includegraphics[width=0.5\textwidth]{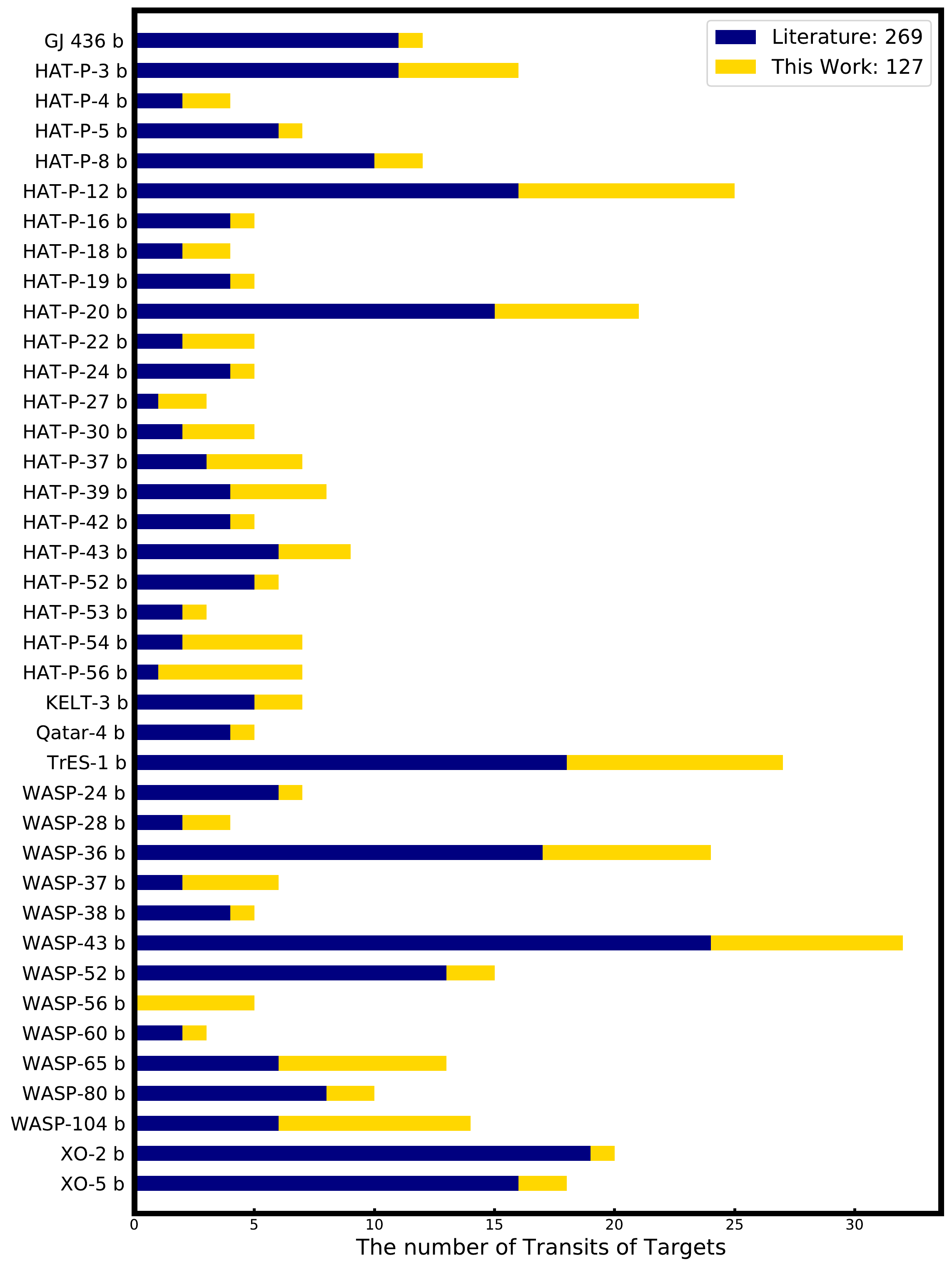}
    \caption{The number of transit light curves we used in this study (396 in total). The blue bar represents the number of archived photometric transits for each target (269 in total). The yellow bar represents the number of transits that we collected (127 in total).}
    \label{fig:temp_number}
\end{figure}

We collected 127 transits for 39 hot Jupiters with the $60/90\,{\rm cm}$ Schmidt telescope and the $60\,{\rm cm}$ telescope at the Xinglong Station of the National Astronomical Observatory of China (NAOC), which lies about $75\, {\rm miles}$ ($120\,{\rm km}$) to the northeast of Beijing. A total of 23,519 exposures ($\sim$ 454 hours, spread across 102 nights) were collected between Nov 11, 2013 and May 8, 2018. Figure~\ref{fig:temp_source} shows the distribution of TEMP targets across the sky. The allocation of transit observations reported in this work is displayed for each target in Figure~\ref{fig:temp_number}, described in detail in the following sections.

\subsection{Xinglong $60/90\,$cm Schmidt Telescope}\label{sss:xl60_90}

We observed 49 transits of 24 planets in $R$-band using the Xinglong $60/90\,{\rm cm}$ Schmidt telescope. The telescope is equipped with a 4K$\times$4K CCD, which gives a pixel scale of $1.38''\,{\rm pixel^{-1}}$ and a field of view of $94'\times94'$. A $512 \times 512$ pixel ($\sim 12'\times12'$) subframe was used to significantly reduce the readout time from $1.5\,{\rm minutes}$ to $4$ seconds. 

More details about this telescope and instrument are given by  \citet{Zhou1999, Zhou2001}.

\subsection{Xinglong $60\,$cm Telescope}\label{sss:xl60}

We acquired 78 light curves for 33 planets in $R$ band with the Xinglong $60\,$cm Telescope. Because of frequent equipment updates and tests, the Xinglong $60\,{\rm cm}$ telescope has a complicated history of CCD use. Observations taken between Jan 17, 2014 and Oct 21 2014 were conducted with a $512\times512$ CCD, which covers a field of view of $17'\times 17'$ with a plate scale of $1.95''\,\rm{pixel}^{-1}$. Observations collected during May 05 to Oct 22, 2015 or during Jun 18, 2016 to Feb 14, 2018 were performed with a $\rm {1K\times1K}$ CCD, which covers a field of view of $17'\times 17'$ with a plate scale of $0.99''\,\rm{pixel}^{-1}$. Observations made between Oct 23, 2015 and May 16, 2016 were obtained with a $\rm {2K\times2K}$ CCD, which covers a field of view of $36'\times 36'$ with a plate scale of $1.06''\,\rm{pixel}^{-1}$.

\bigskip
The time of each observation taken by both telescopes was automatically set to the precise current time using the GPS function. The beginning time of each exposure was recorded in the frame header using the UTC time standard, which was then converted to ${\rm BJD_{\rm TDB}}$ as described in \citet{Eastman2010}. To avoid non-linear effects of the CCD, the defocusing technique, which increases the duty cycle of observations to reduce the Poisson and the scintillation noise \citep{Southworth2009}, was used for bright stars. The data were reduced following a standard photometric procedure described in \citet{WangX2018}, \citet{Wang2018a, Wang2018b}, \citet{WangY2017}, and \citet{WangY2019}. We first applied standard bias and flat-field corrections to all frames, then performed aperture photometry using SExtractor \citep{Bertin1996}. We identified the best aperture for both the target and reference stars as the one that minimized the root mean square (RMS) of the final differential light curves, which were obtained by comparing our target with several reference stars in the field. Highly discrepant points and/or linear trends presented in these light curves were removed. Furthermore, we quantified the quality of each light curve by its photometric noise rate \citep[$pnr$;][]{Fulton2011}, which is defined as 
\begin{equation}
pnr = \frac{\rm rms}{\sqrt{\Gamma}}
\end{equation}
where the rms is the root mean square of the fitting residuals and $\rm \Gamma$ is the median number of exposures per minute. Then, using the K-means clustering method \citep{MacQueen1967}, the light curves were divided into three groups: golden, mediocre, and bad. The results of the classification show that there are four bad light curves, including HAT-P-37 on May 31 2017; WASP-36 on Mar 02 2016; and WASP-37 on Mar 26 2017 and on Apr 20 2018. We present these bad light curves in Table 1 but did not include them in the light curve fit. A summary of the observations is provided in Table~\ref{tab:obs}. The final light curves are presented in Table~\ref{tab:lcs} and shown in Figure~\ref{fig:lcs}.

\section{Light curve analysis}
\subsection{Planetary System Parameters from Global Analysis}
\label{fit:glo}

To refine the system parameters for our observed hot Jupiters, we used EXOFASTv2 \citep{Eastman2013, Eastman2017, Eastman2019} to simultaneously fit all published Doppler velocimetric and photometric data together with our newly collected light curves for each system. EXOFASTv2 enables simultaneous modeling of multi-band transits and multi-set radial velocities by combining the AMOEBA symplectic solver with a differential evolution Markov chain Monte Carlo (DE-MCMC; \citealt{TerBraak2006}) algorithm.

For each system, we adopted Gaussian priors from the Exoplanet Orbit Database\footnote{http://exoplanets.org/} for $\teff$, $\feh$, transit mid-times ($T_0$), and orbital period ($P$). We imposed priors on the quadratic limb-darkening coefficients from \citet{Claret2018} based on $\teff$, $\feh$, and filter wavelength. We applied the Gaussian prior on stellar parallax to the corrected Gaia DR2 parallax \citep{GaiaCollaboration2018, Stassun2018}.
We constrained the $V$-band extinction ($A_{V}$) for each system by using the galactic dust maps presented in \citet{Schlafly2011}. We also used stellar isochrones from the MESA Isochrones \& Stellar Tracks (MIST) catalog and constructed spectral energy distributions (SEDs) from a series of photometric catalogs\footnote{
Galex \citep{Bianchi2011}, 
Tycho-2 \citep{Hog2000}, 
UCAC4 \citep{Zacharias2013}, 
APASS \citep{Henden2016}, 
2MASS \citep{Cutri2003}, 
WISE \citep{Cutri2013}, 
Gaia \citep{GaiaCollaboration2016}, 
the \textit{Kepler} INT Survey \citep{Greiss2012}, 
the UBV Photoelectric Catalog \citep{Mermilliod1994}, and the Stroemgren-Crawford uvby photometry catalog \citep{Paunzen2015}.} to estimate the host stellar parameters.

\begin{figure*}
    \centering
    \includegraphics[width=1\textwidth]{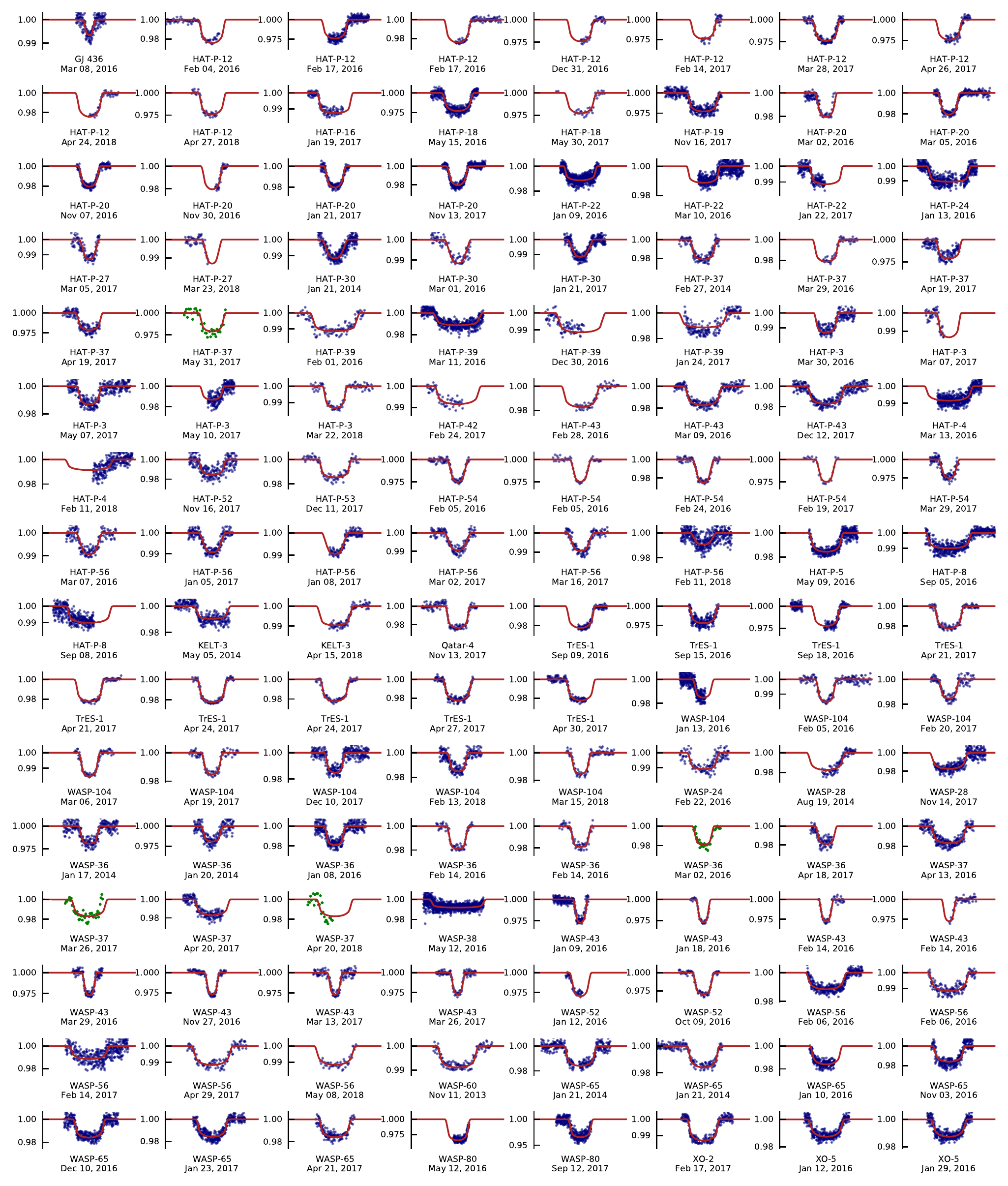}
    \caption{127 follow-up light curves that we collected for this study. 
    These light curves were fitted simultaneously with the archived photometric, Doppler velocimetric, and spectroscopic data to estimate the system parameters. \textbf{Four bad light curves are plotted in green.} The solid red lines show the best-fitting models.
    }
    \label{fig:lcs}
\end{figure*}

\begin{figure*}
    \centering
    \includegraphics[width=0.9\textwidth]{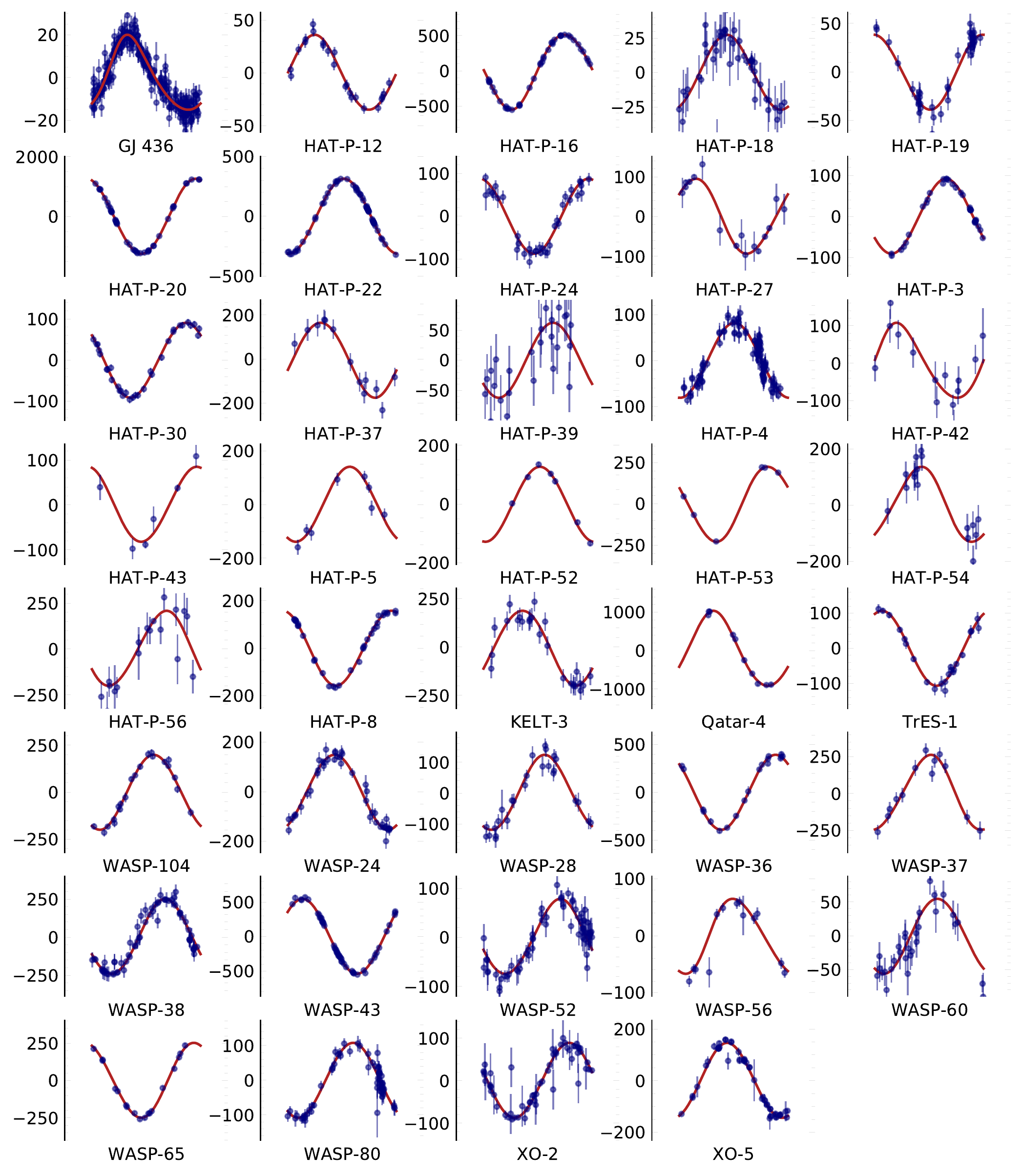}
    \caption{Archived Doppler velocimetric measurements (m/s) of our TEMP targets phase-folded with the updated orbital periods determined in this study. The best-fitting Keplerian orbital solutions from the joint radial-velocity, light-curve, and spectroscopic modeling are overplotted as solid red lines. 
    }
    \label{fig:rvs}
\end{figure*}

All the fittings were converged by the Gelman-Rubin diagnostic ($<1.01$, \citealt{Gelman1992}). All of our new light curves and phased-folded Doppler velocities are compared in Figure~\ref{fig:lcs} and Figure~\ref{fig:rvs} to the respective best-fitting models.

\subsection{Transit Mid-Times from Individual Fitting}\label{fit:ind}

Following the routine described in \cite{WangX2018}, we used the JKTEBOP tool \citep{Southworth2004a, Southworth2004b} to fit each observed transit and to subsequently derive the associated transit mid-times. For each target, we fixed all global parameters to the results derived from the global fitting in Section~\ref{fit:glo}. The transit mid-time ($T_0$) and baseline flux ($F_{0}$) are the only two free parameters that we fit.

The best-fitting transit mid-times for each target were obtained by using the Levenberg-Marquardt non-linear least-squares fitting algorithm \citep{Press1992}. To obtain reliable uncertainties for the transit mid-times, the bootstrapping method, Monte Carlo simulations, and the residual-shift method were each employed independently. We chose the largest uncertainty derived from the above three methods for a conservative estimate. 

\begin{figure*}
    \centering
    \includegraphics[width=0.9\textwidth]{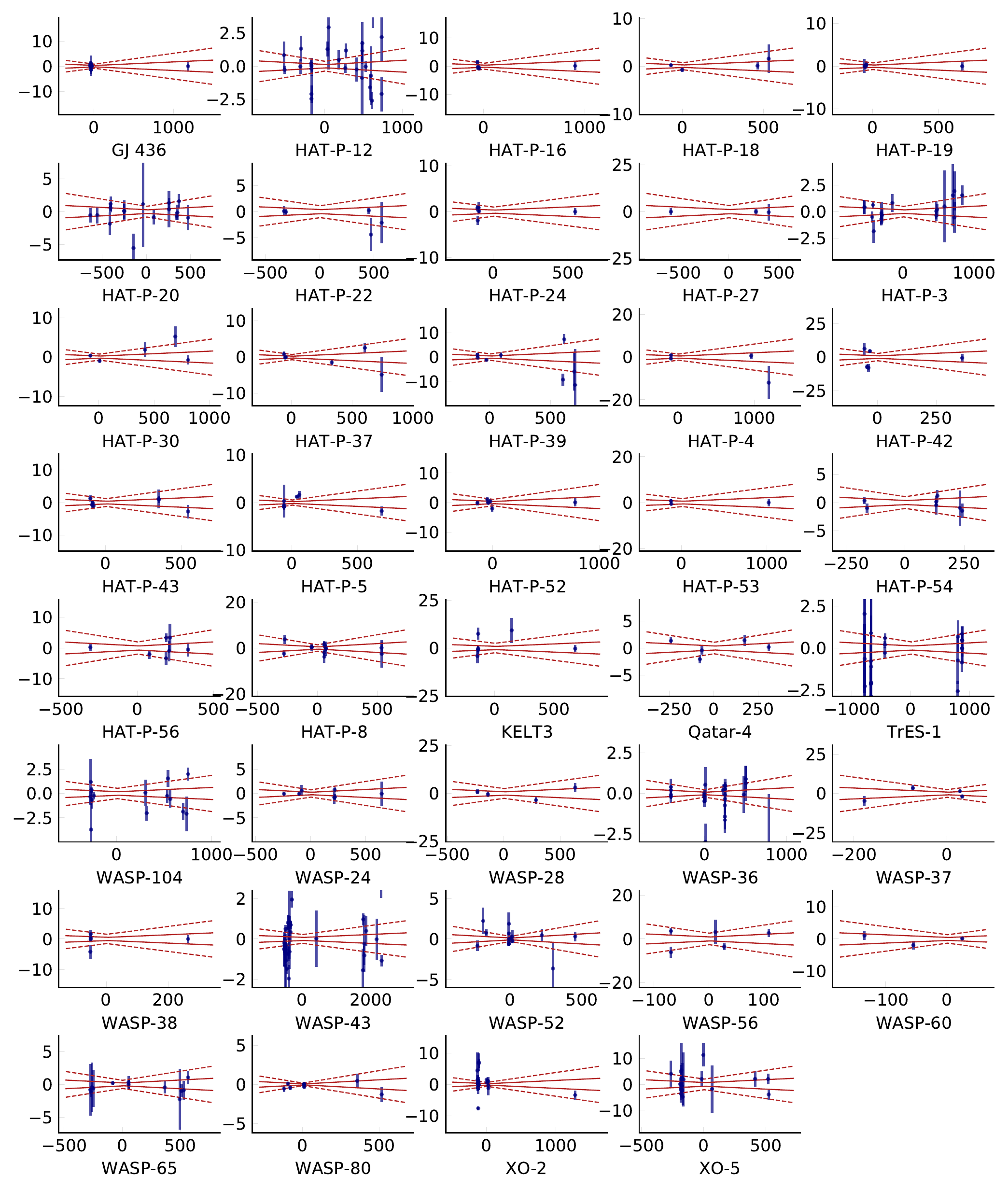}
    \caption{Transit timing variations (mins) for our targets, as compared to the best-fitting ephemerides reported in this study, shown as a function of epoch. The solid (dashed) lines indicate the propagation of $\pm 1\,\sigma$ ($\pm 3\,\sigma$) errors in the updated linear ephemerides. No statistically significant TTVs are detected in our sample at levels of $\pm 3\,\sigma$.       
    }
    \label{fig:ttvs}
\end{figure*}

The resulting transit mid-times for each target are reported in Table~\ref{tab:midtimes}. Based on the transit mid-times, we derived the updated linear orbital ephemerides for each target using the weighted least-squares fitting algorithm. We chose the epoch in our observations that corresponds to the minimum covariance with the orbital period as the reference epoch \citep{Mallonn2019}. The deviations of transit mid-times for each target from the updated linear orbital ephemerides determined in this work are plotted in Figure~\ref{fig:ttvs}.

\begin{figure*}[htbp]
    \centering
    \includegraphics[width=0.32\textwidth,trim=0 10 50 60,clip]{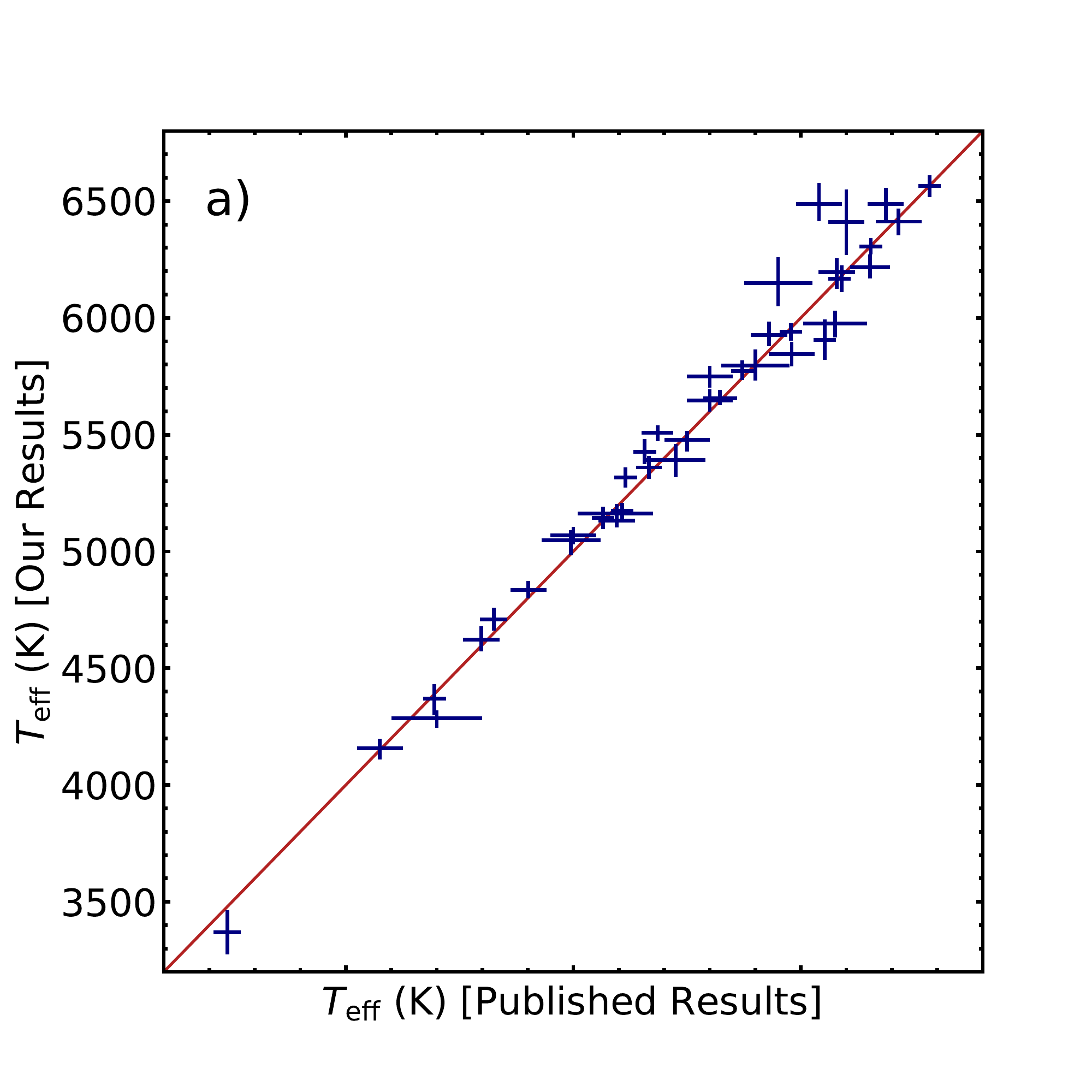}
    \includegraphics[width=0.32\textwidth,trim=0 10 50 60,clip]{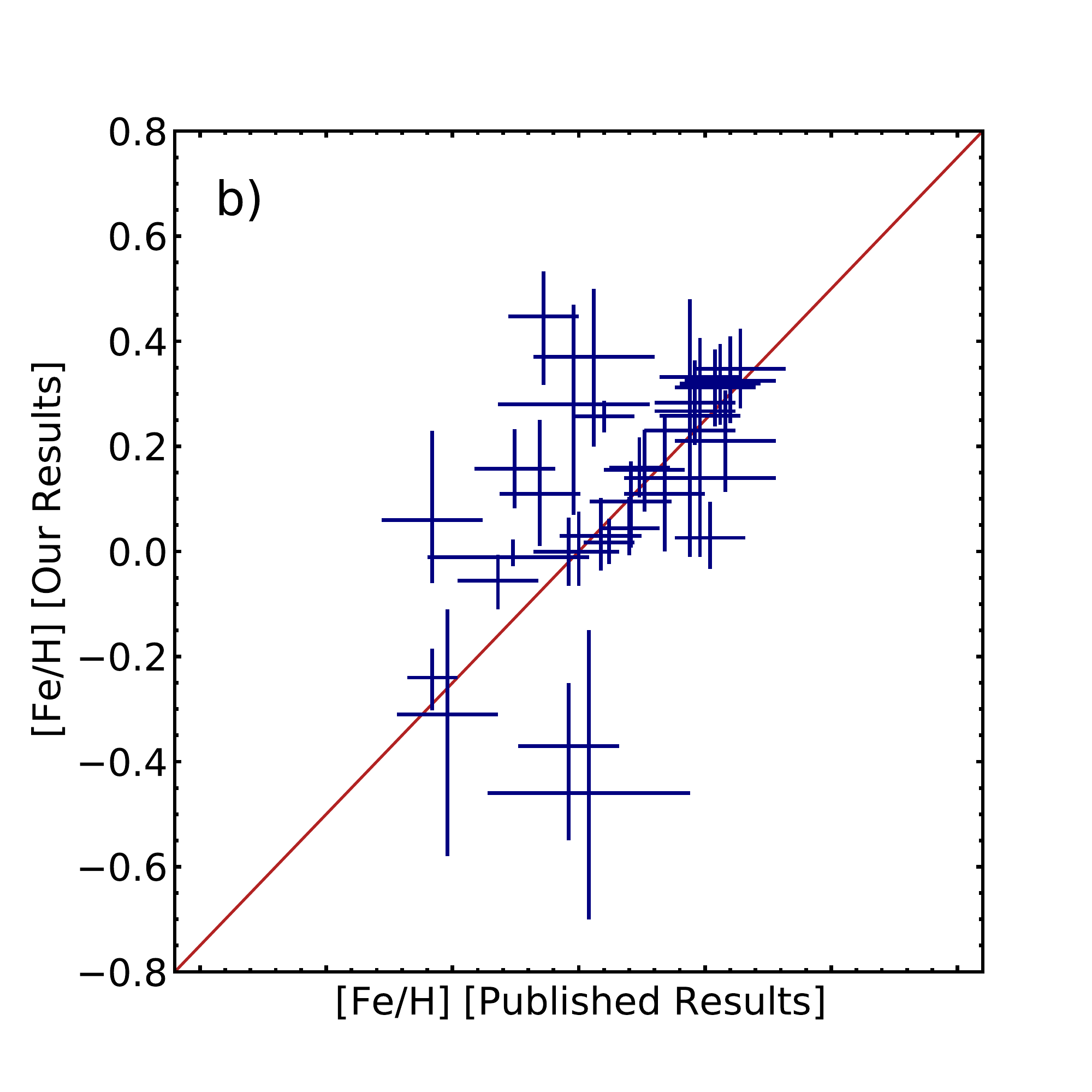}
    \includegraphics[width=0.32\textwidth,trim=0 10 50 60,clip]{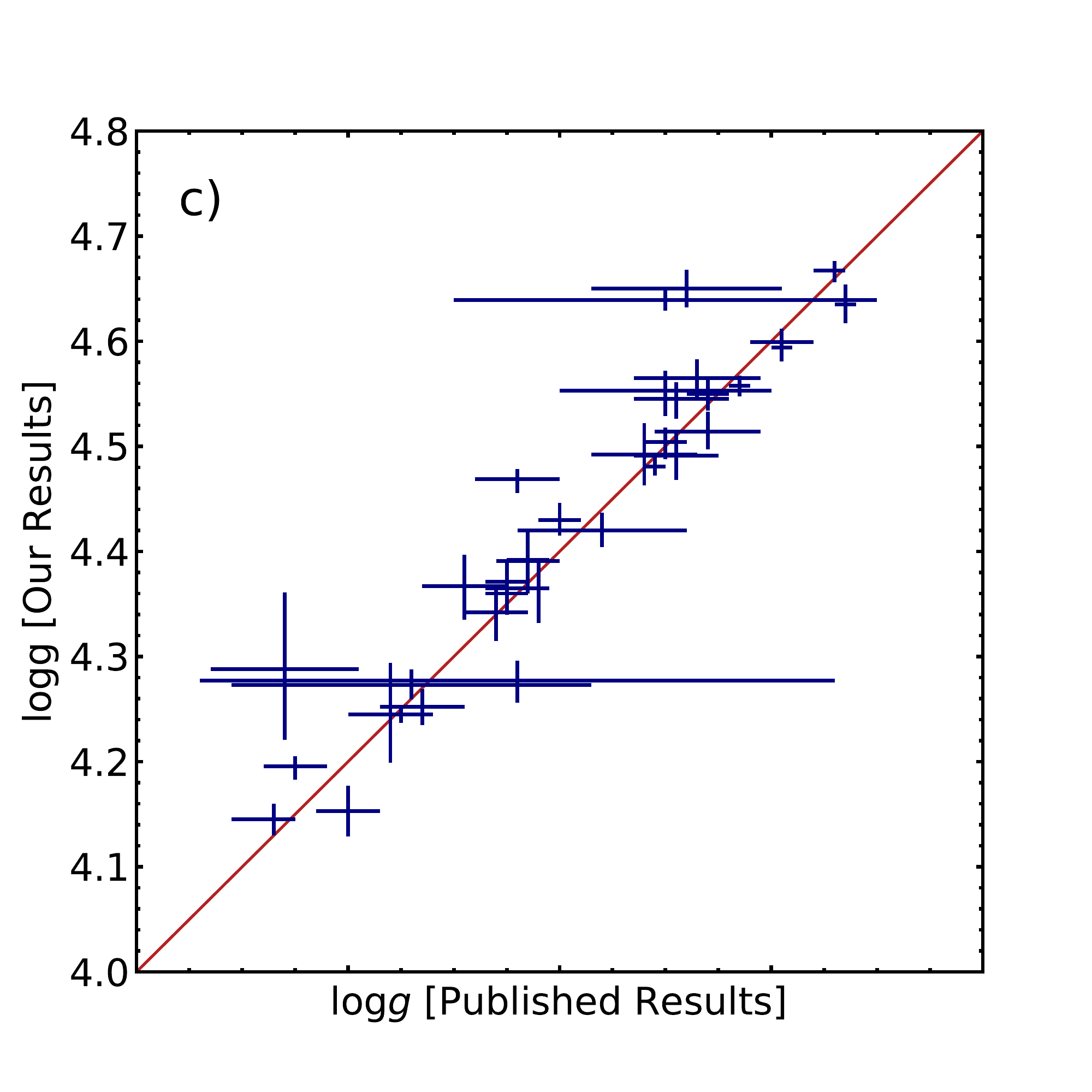}
    \includegraphics[width=0.32\textwidth,trim=0 10 50 60,clip]{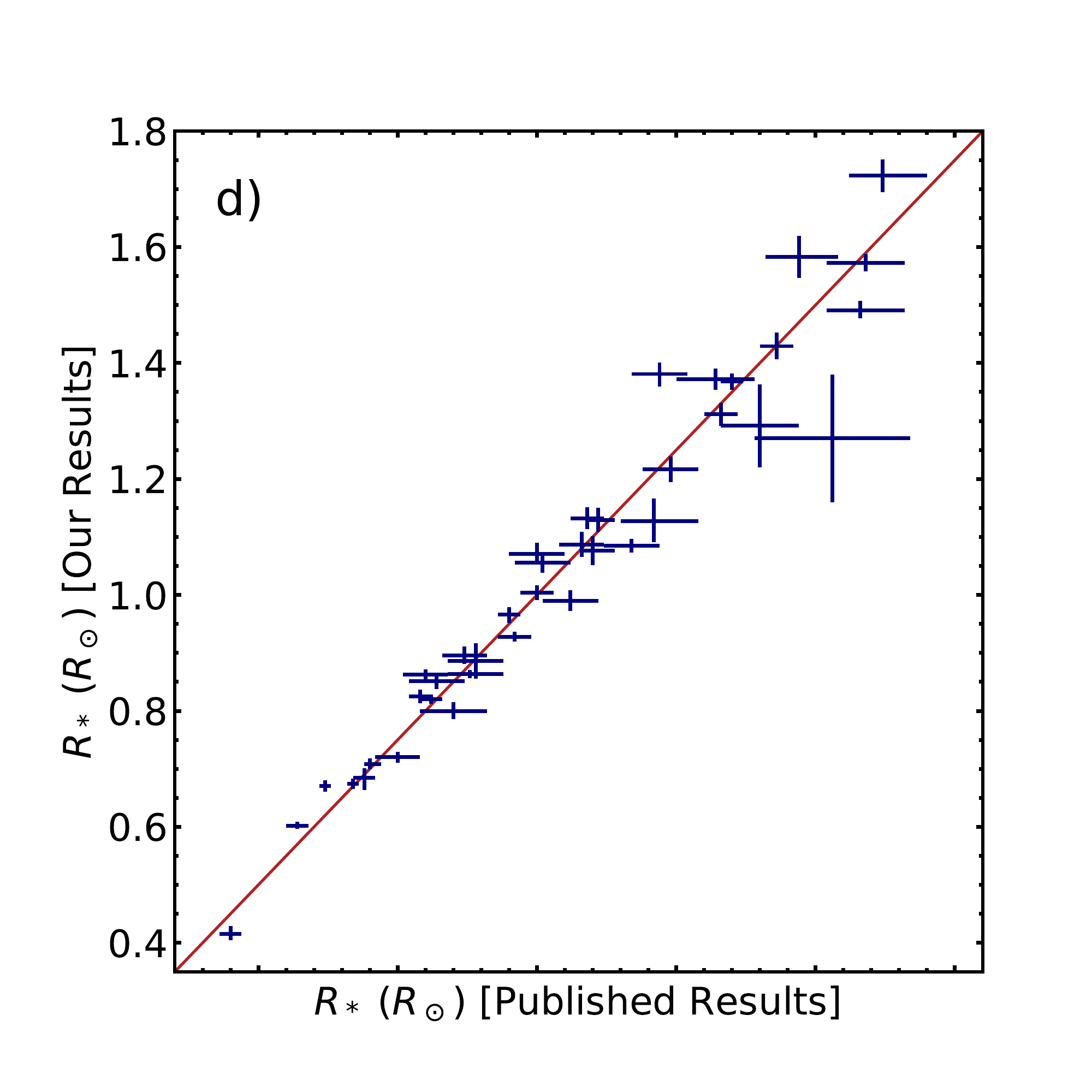}
    \includegraphics[width=0.32\textwidth,trim=0 10 50 60,clip]{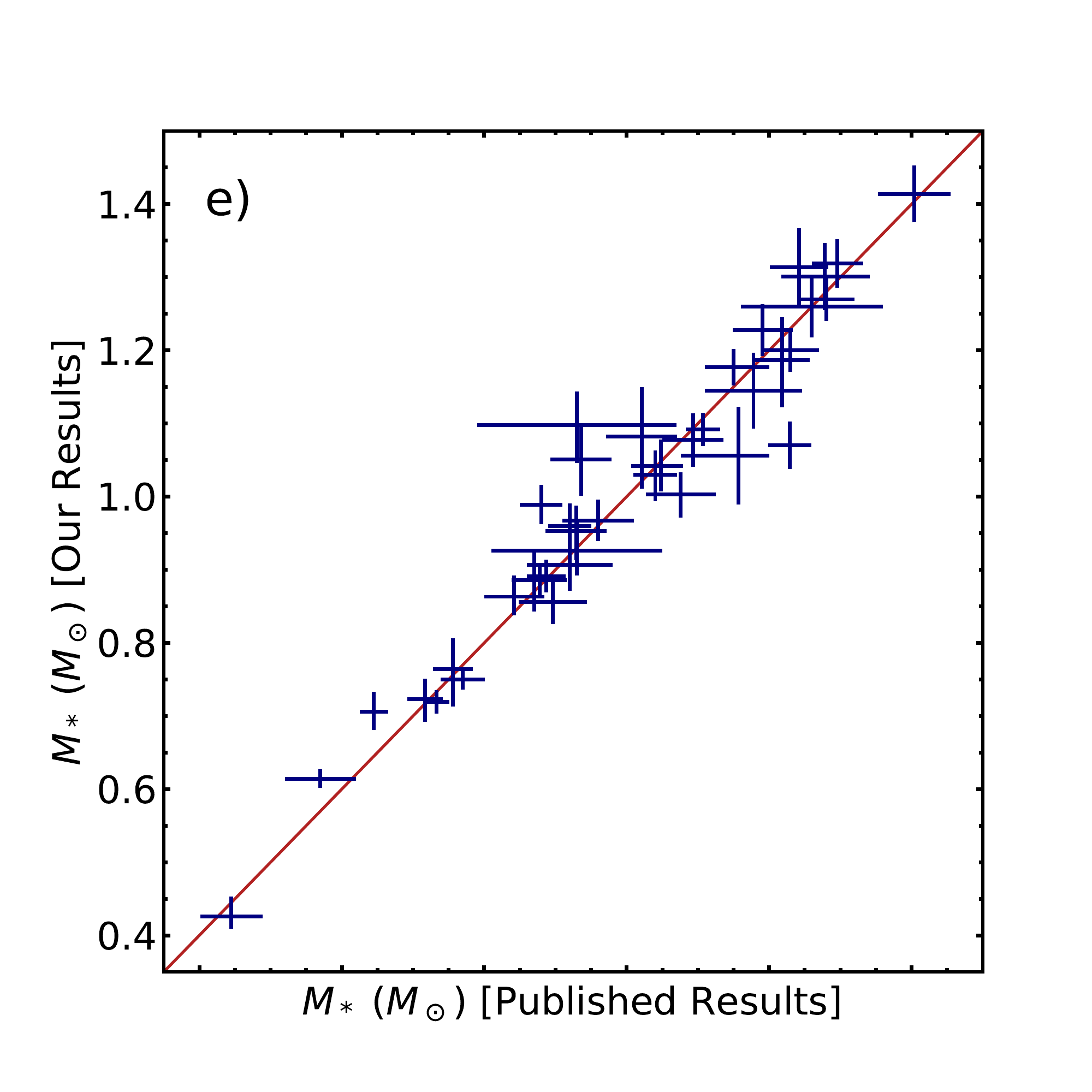}
    \includegraphics[width=0.32\textwidth,trim=0 10 50 60,clip]{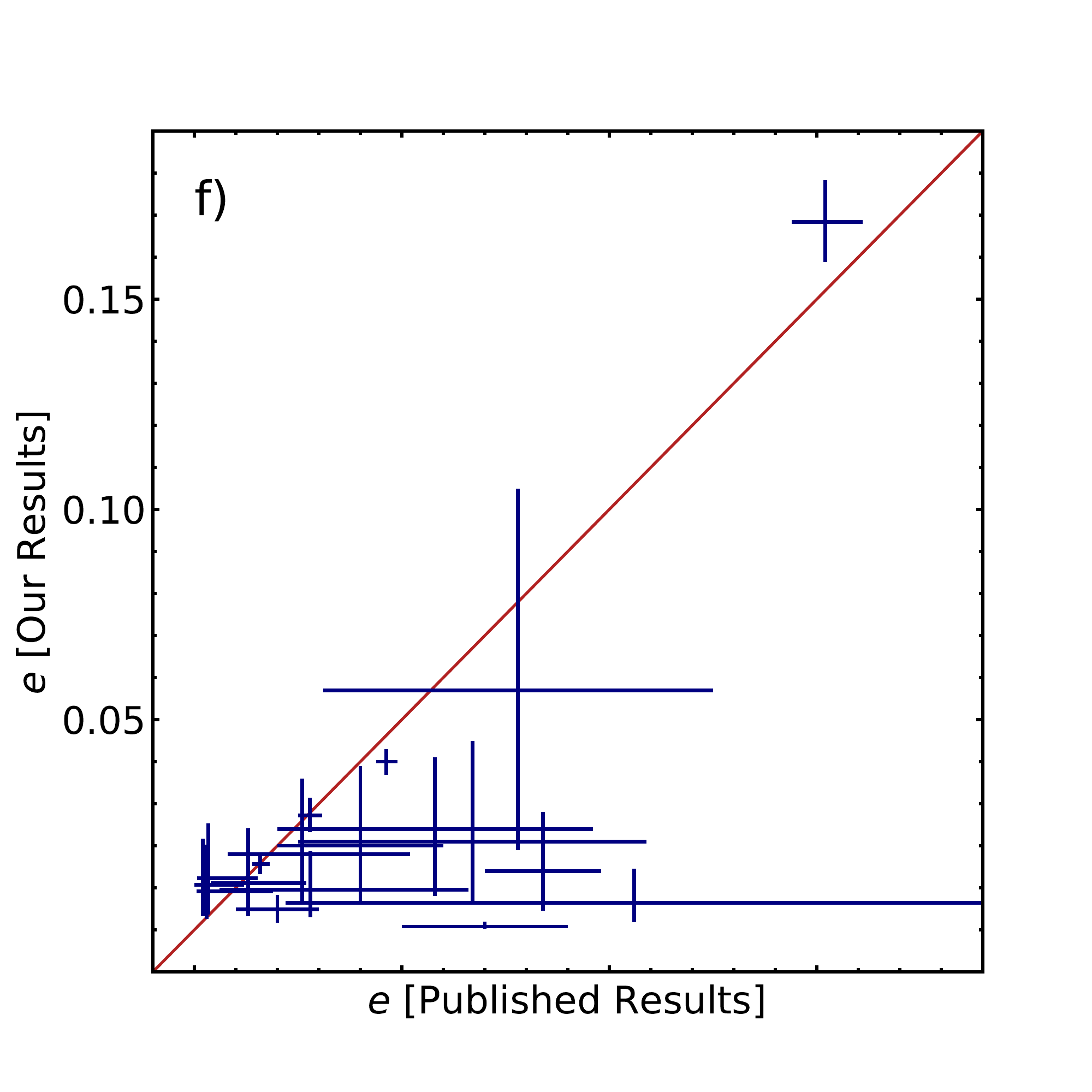}
    \includegraphics[width=0.32\textwidth,trim=0 10 50 60,clip]{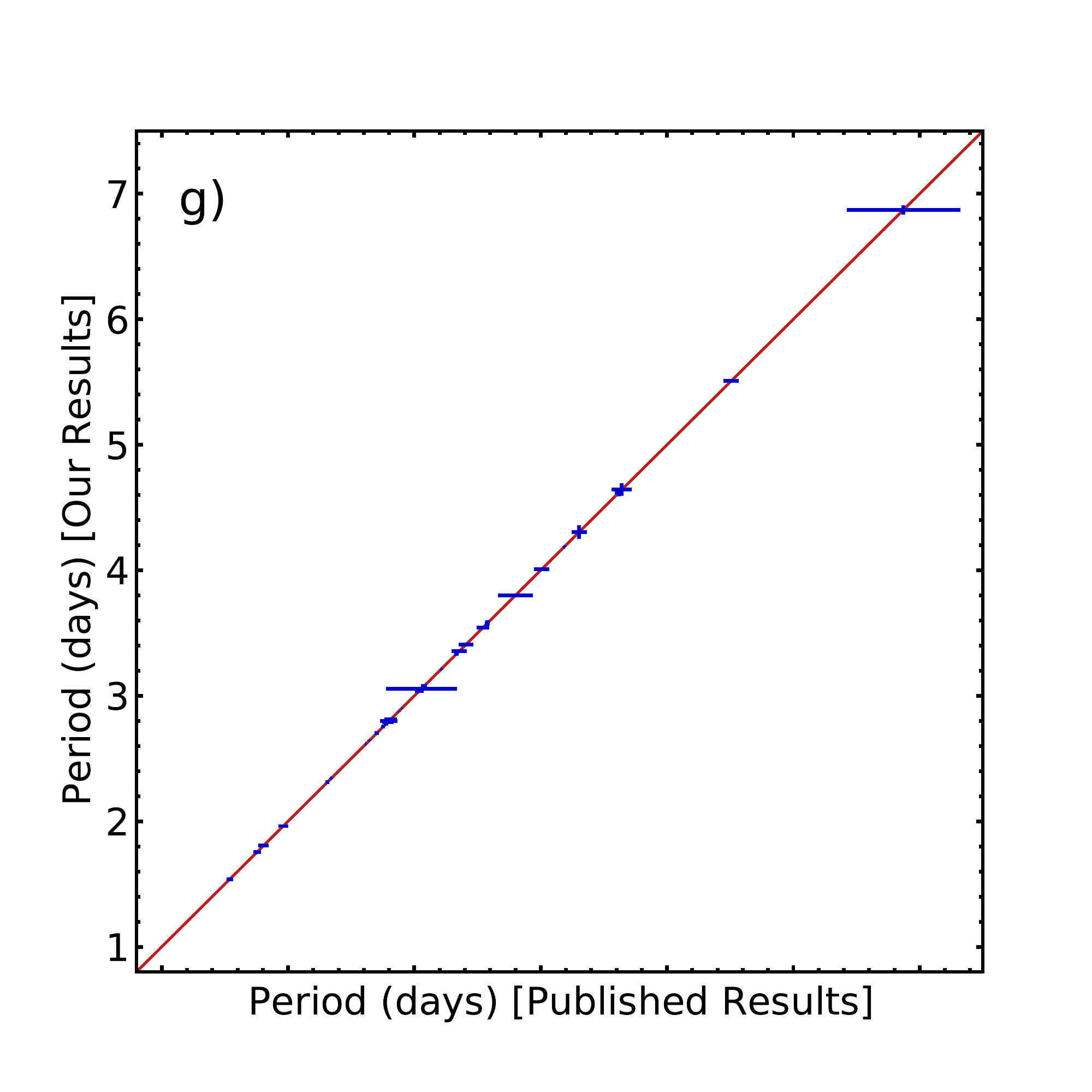}
    \includegraphics[width=0.32\textwidth,trim=0 10 50 60,clip]{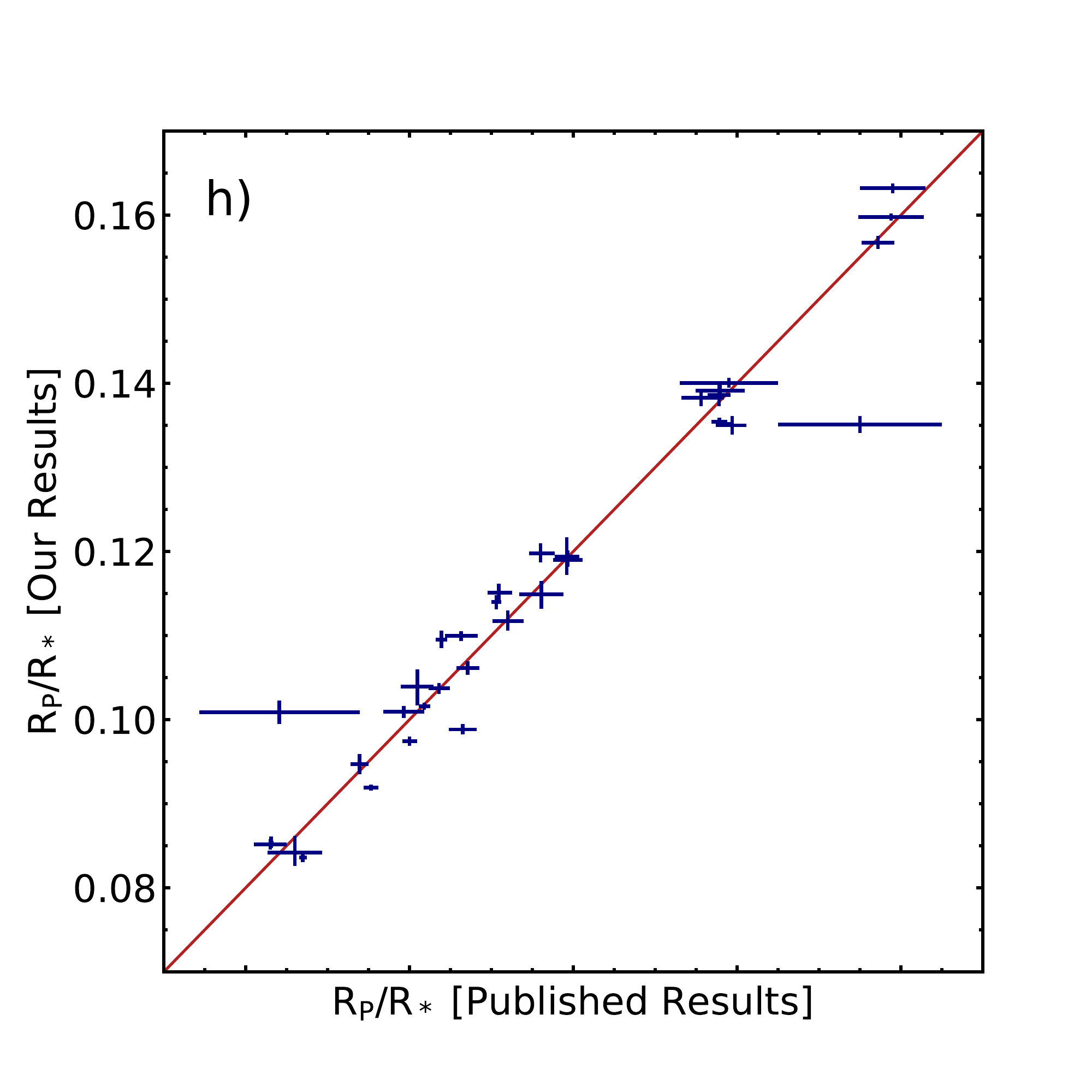}
    \includegraphics[width=0.32\textwidth,trim=0 10 50 60,clip]{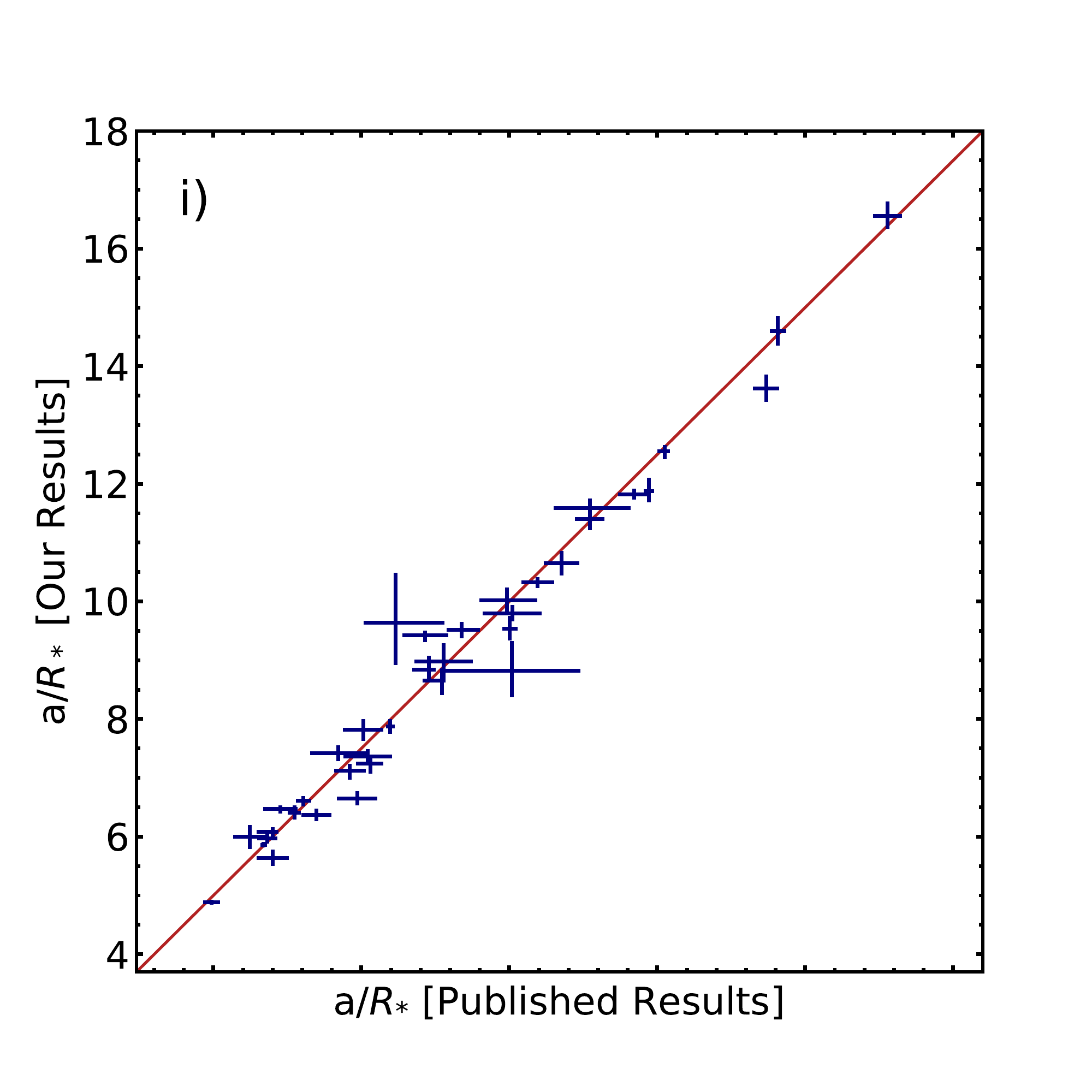}
    \includegraphics[width=0.32\textwidth,trim=0 10 50 60,clip]{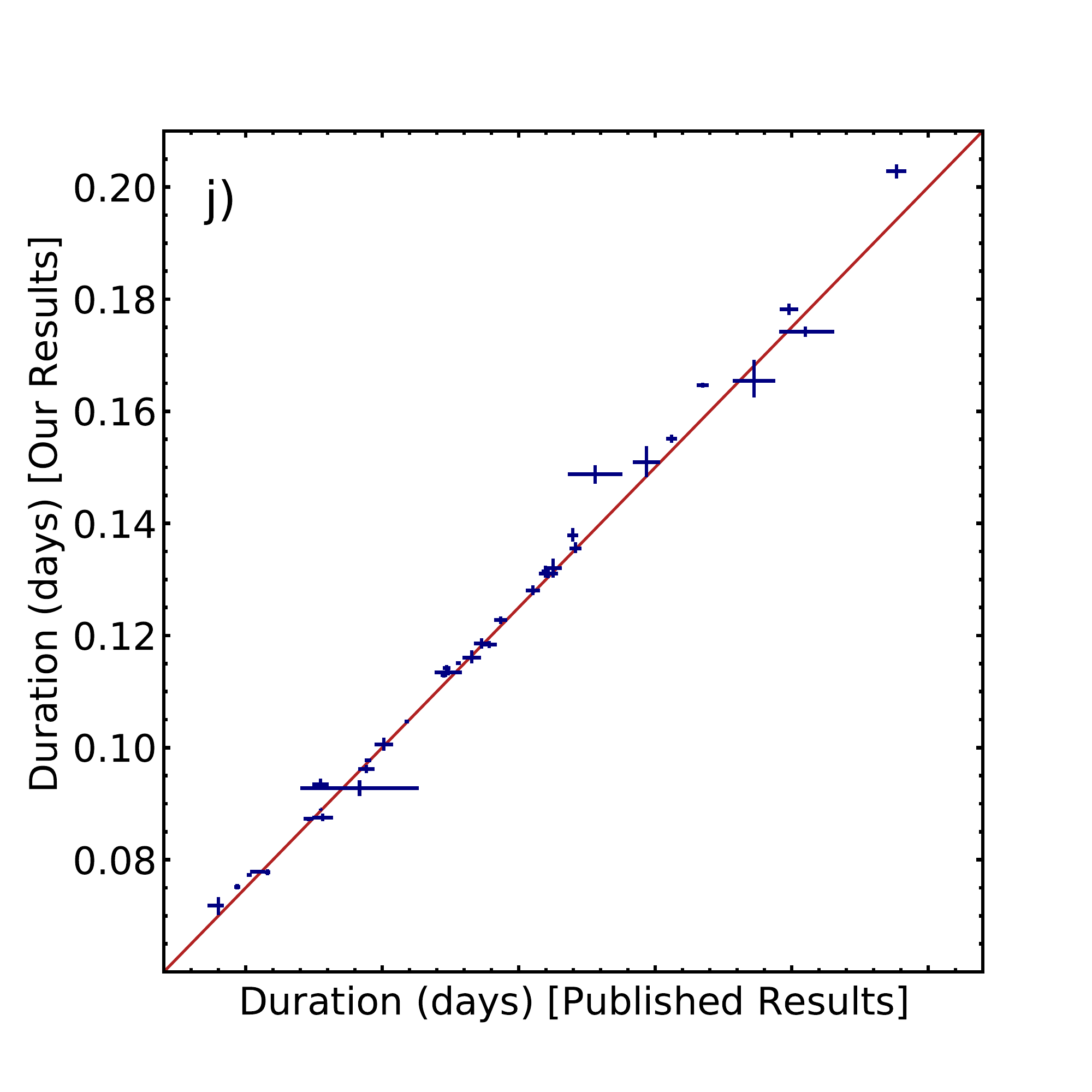}
    \includegraphics[width=0.32\textwidth,trim=0 10 50 60,clip]{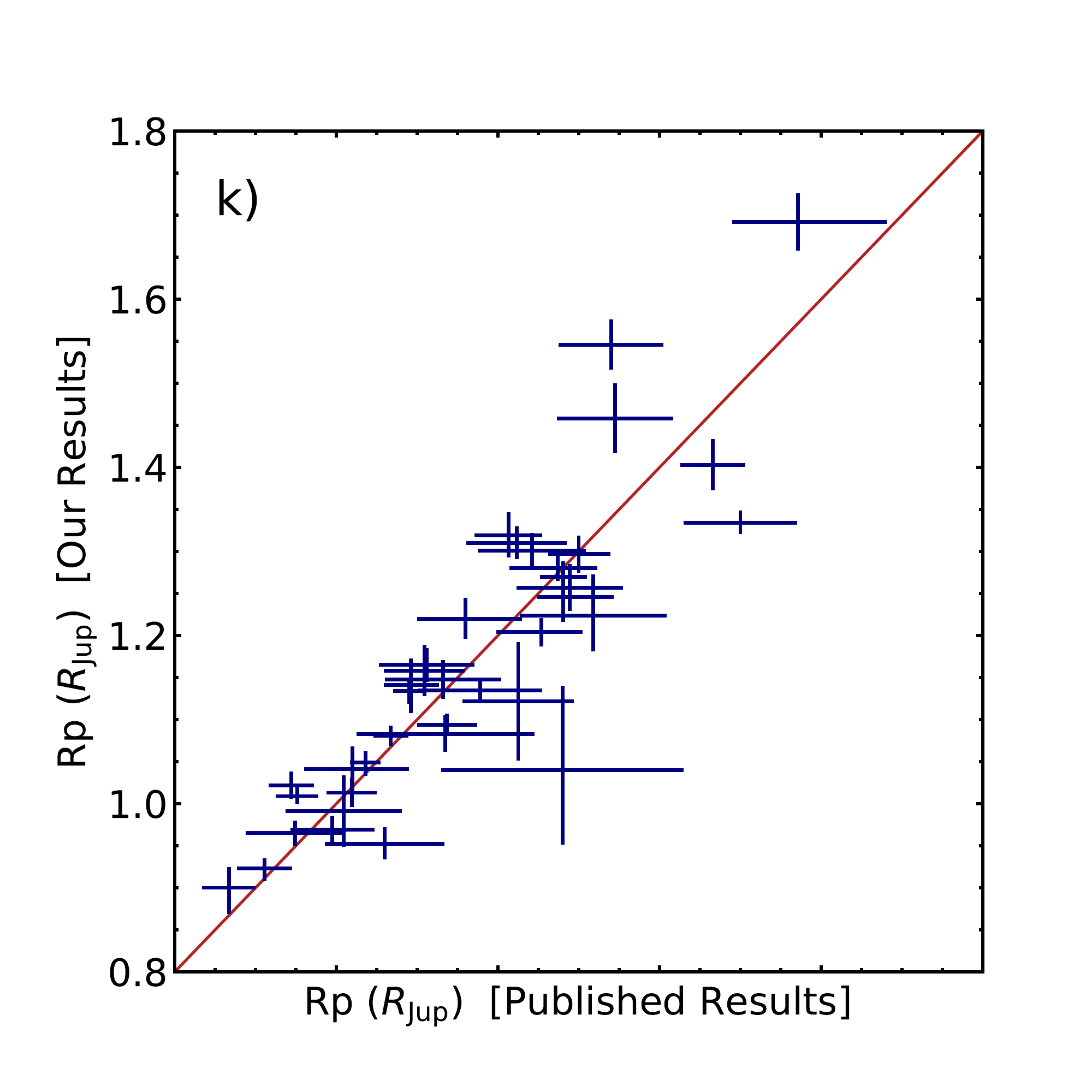}
    \includegraphics[width=0.32\textwidth,trim=0 10 50 60,clip]{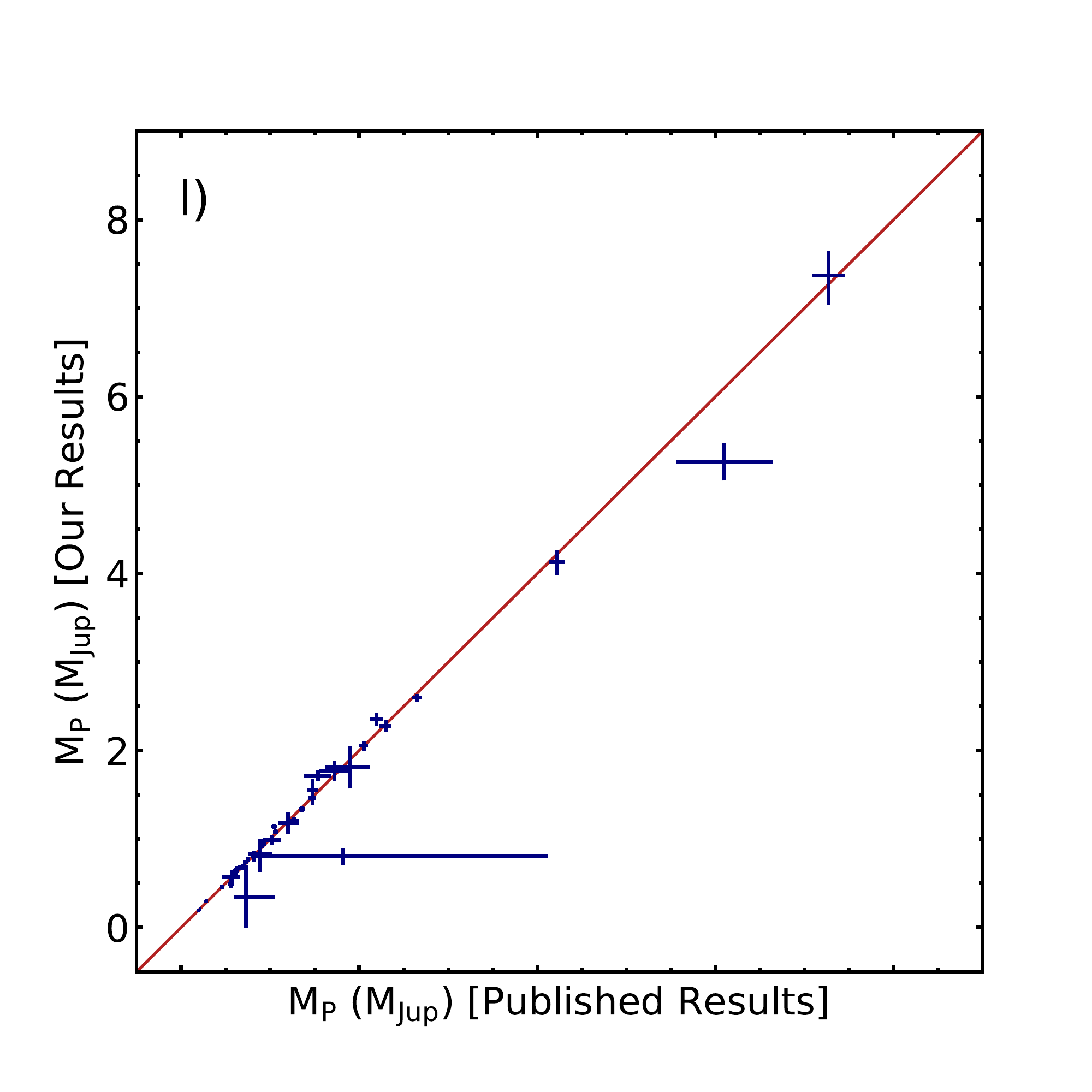}
    \caption{Comparison of the 
    effective temperature ($\teff$, Panel a),
    metallicity ($\feh$, Panel b),
    surface gravity ($\logg$, Panel c), 
    stellar radius ($\rstar$, Panel d), 
    stellar mass ($\mstar$, Panel e), 
    orbital eccentricity ($e$, Panel f), 
    orbital period ($P$, Panel g),
    planet-to-star radius ratio ($\rp / \rstar$, Panel h), 
    scaled semi-major axis ($a / \rstar$, Panel i),
    transit duration ($T_{14}$, Panel j),
    planetary radius ($\rp$, Panel k), and
    planetary mass ($\mplanet$, Panel l) 
    derived in this work and in the literature for our 39 target hot Jupiter systems. The solid lines represent exact agreement.}
    \label{fig: agreement}
\end{figure*}

\begin{figure*}[htbp]
    \centering
    % stellar 
    \includegraphics[width=0.32\textwidth,trim=20 15 35 60,clip]{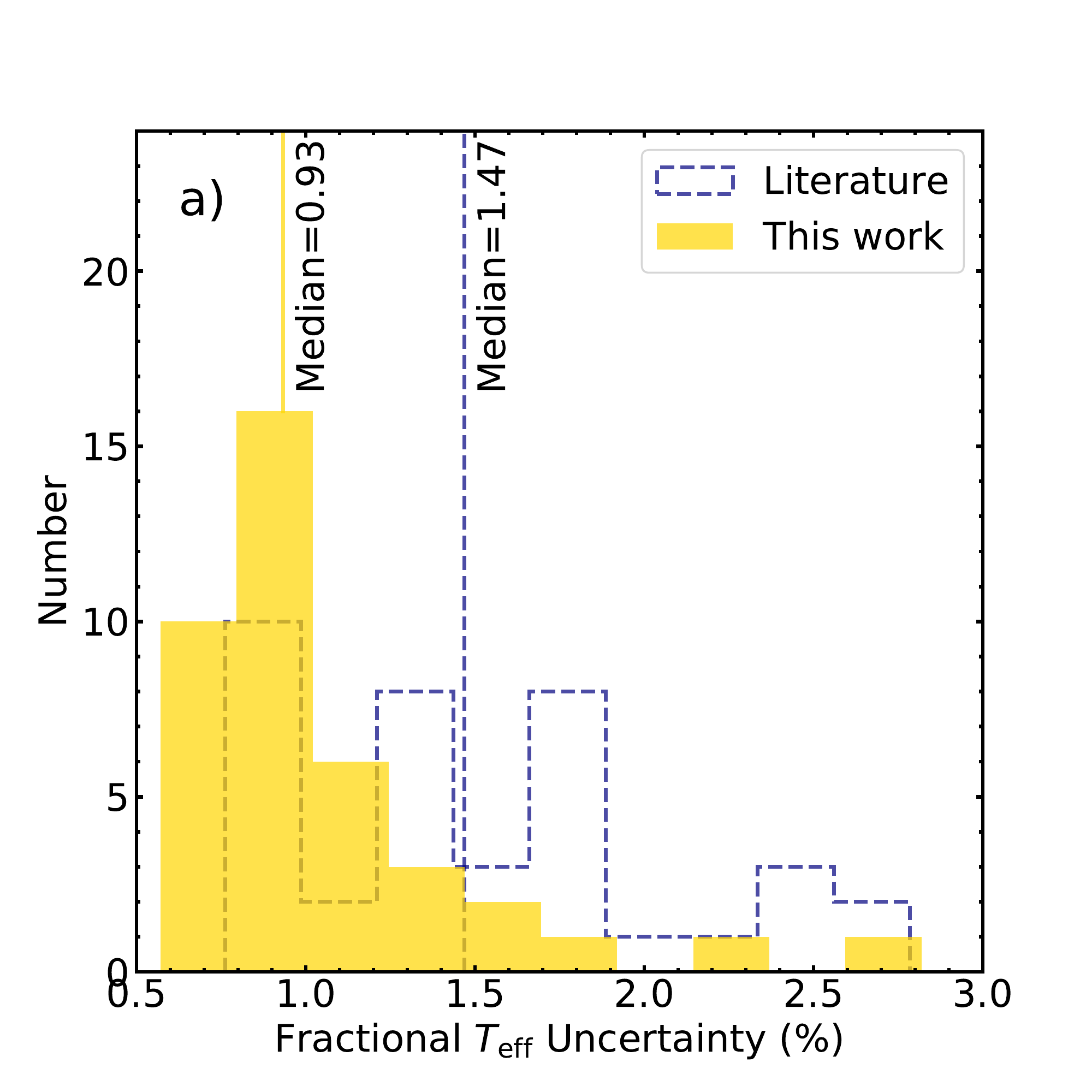}
    \includegraphics[width=0.32\textwidth,trim=20 15 35 60,clip]{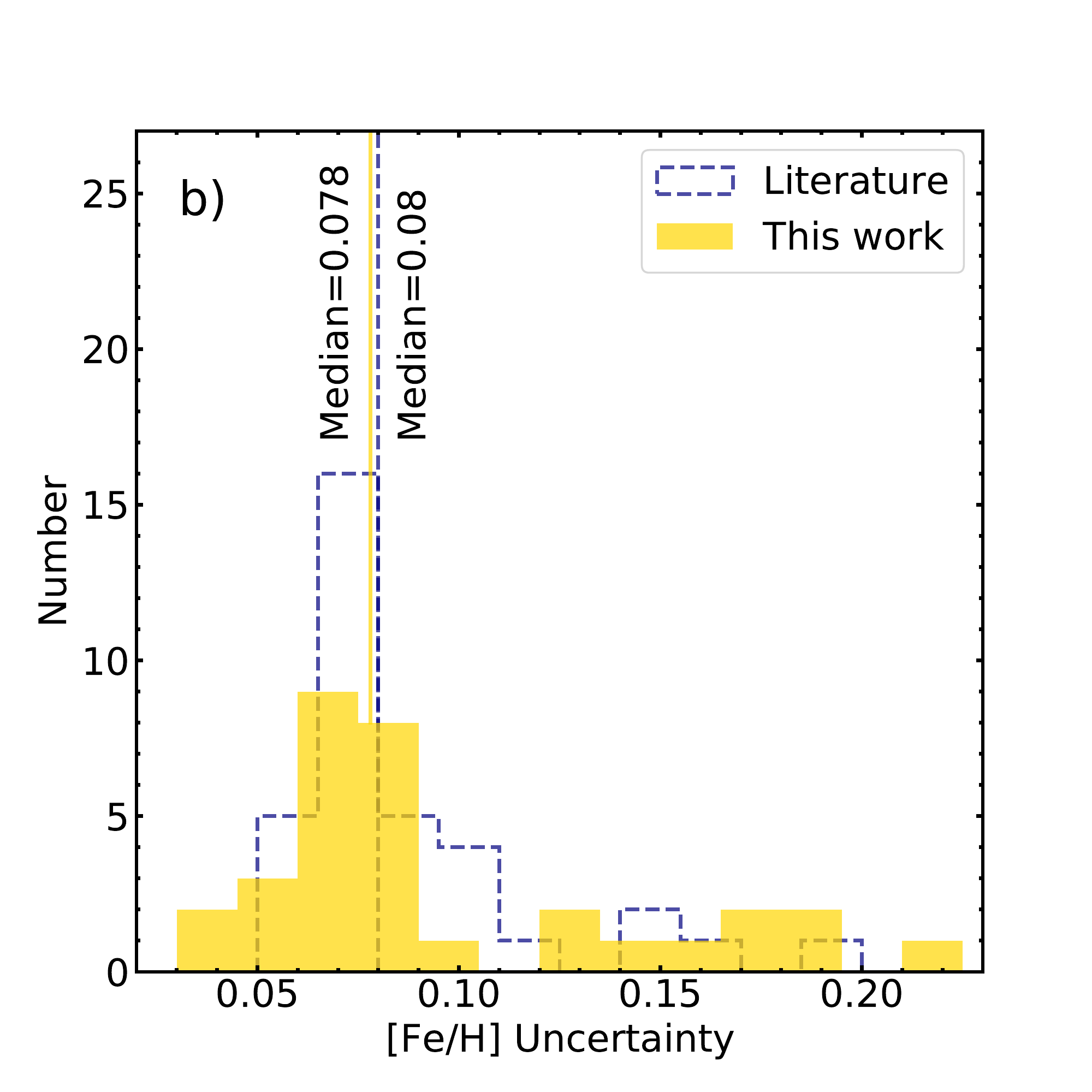}
    \includegraphics[width=0.32\textwidth,trim=20 15 35 60,clip]{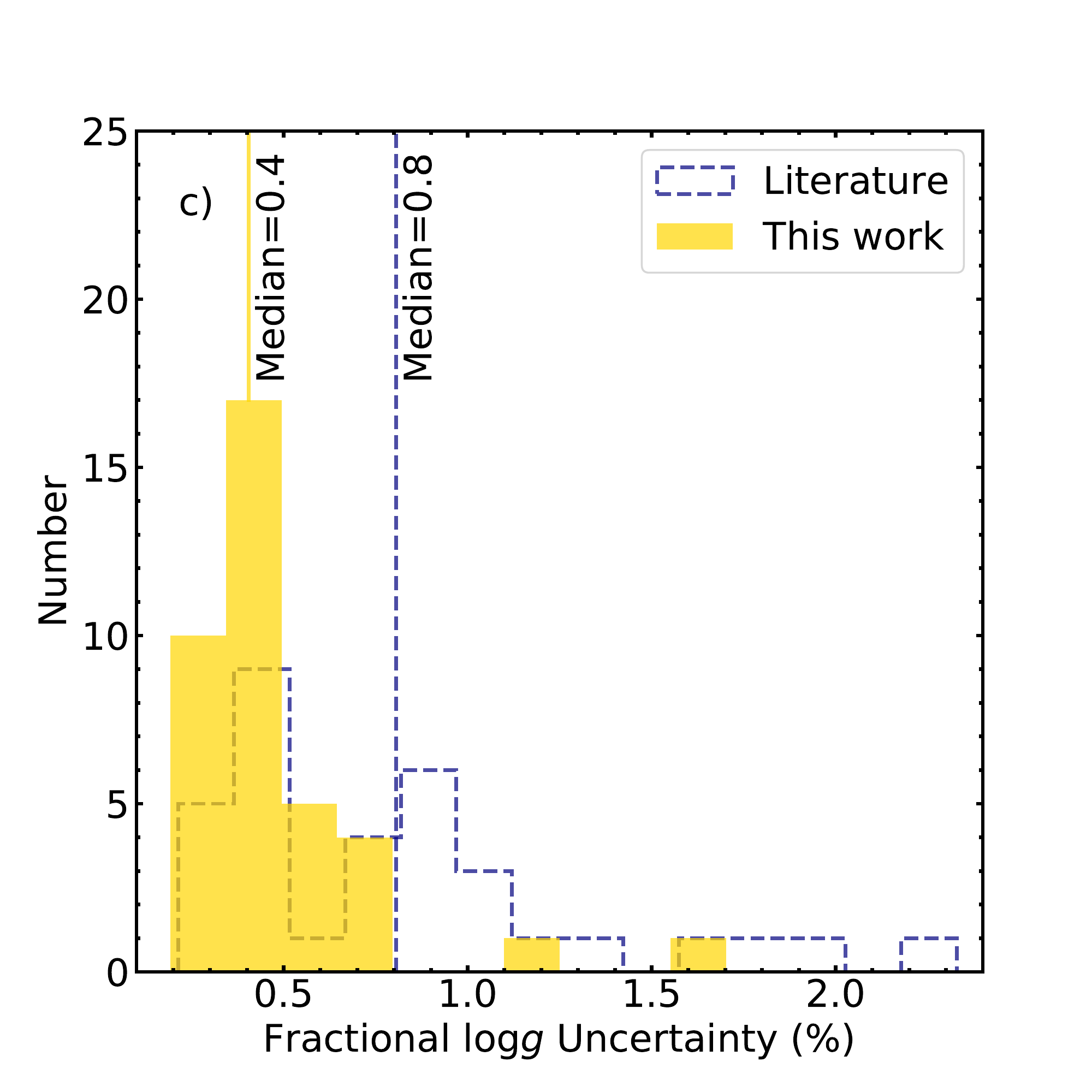}
    \includegraphics[width=0.32\textwidth,trim=20 15 35 60,clip]{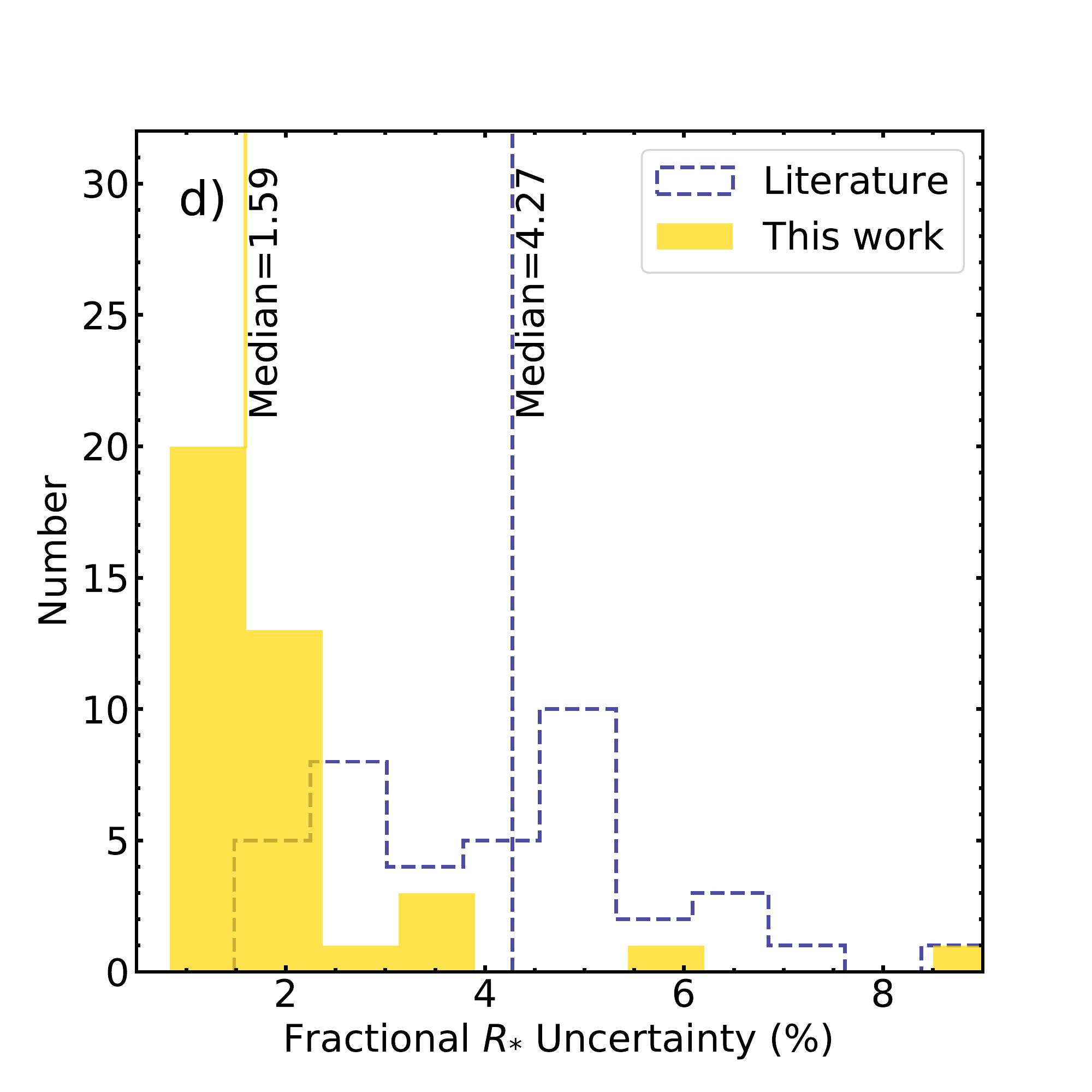}
    \includegraphics[width=0.32\textwidth,trim=20 15 35 60,clip]{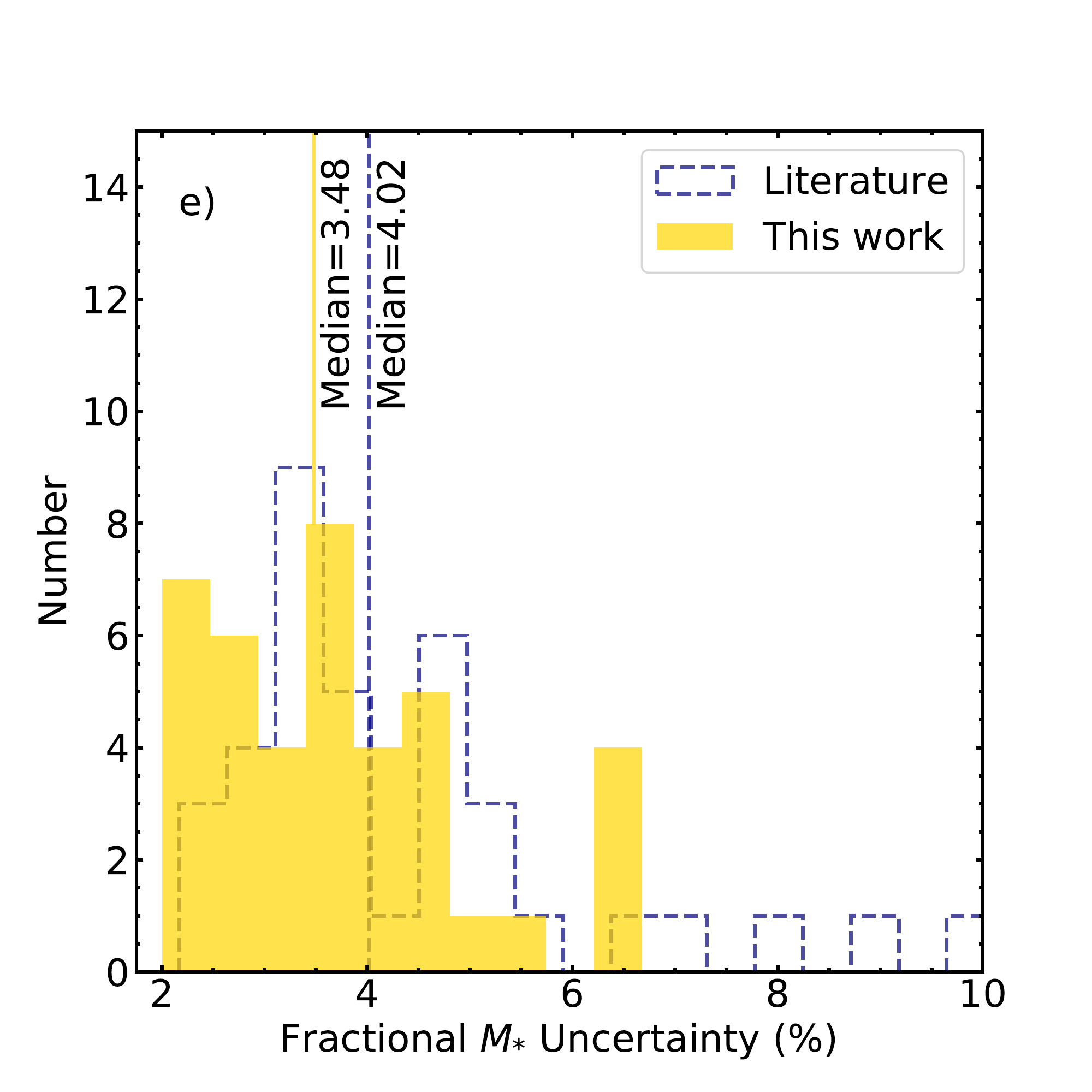}
     % planet
    \includegraphics[width=0.32\textwidth,trim=20 15 35 60,clip]{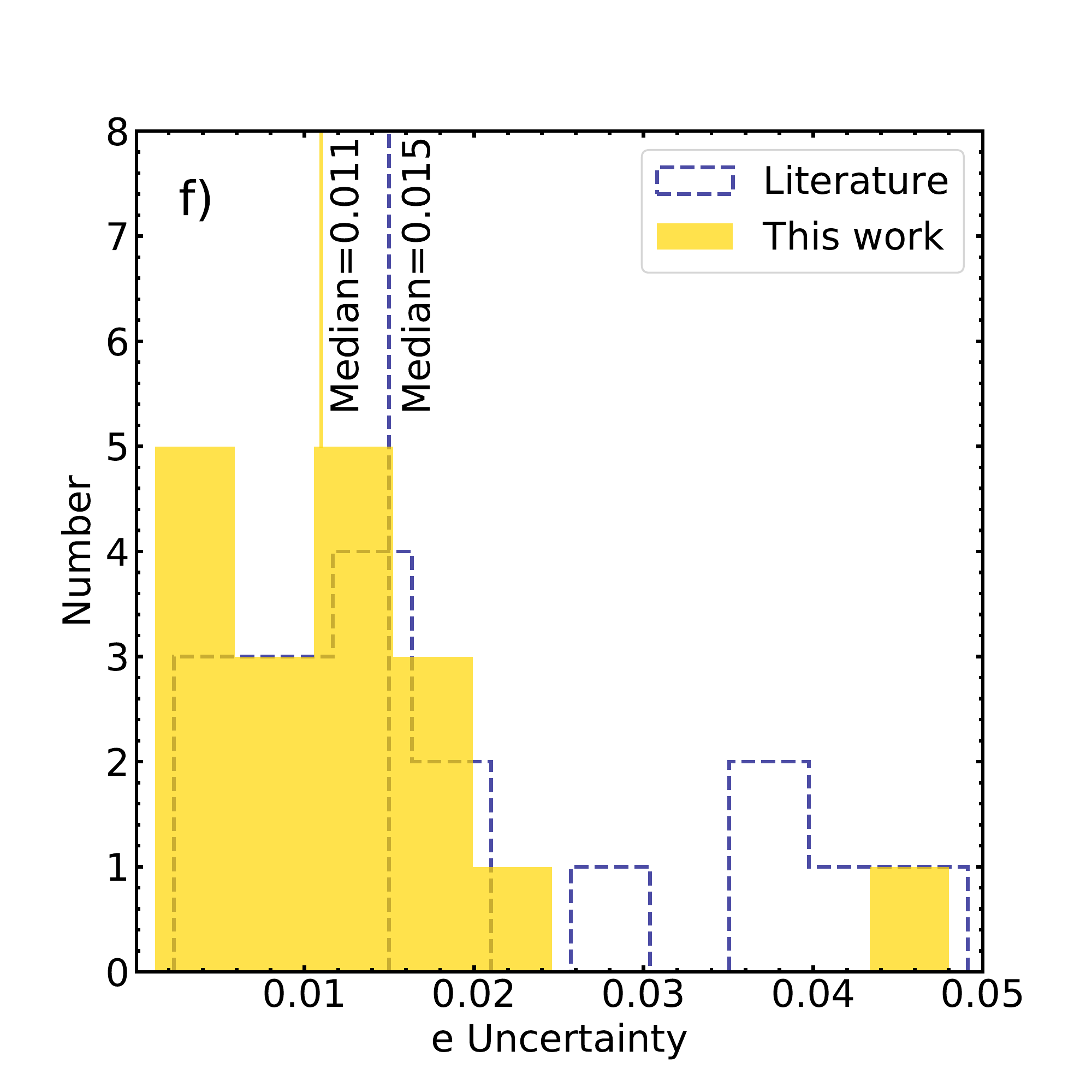}
    \includegraphics[width=0.32\textwidth,trim=20 15 35 60,clip]{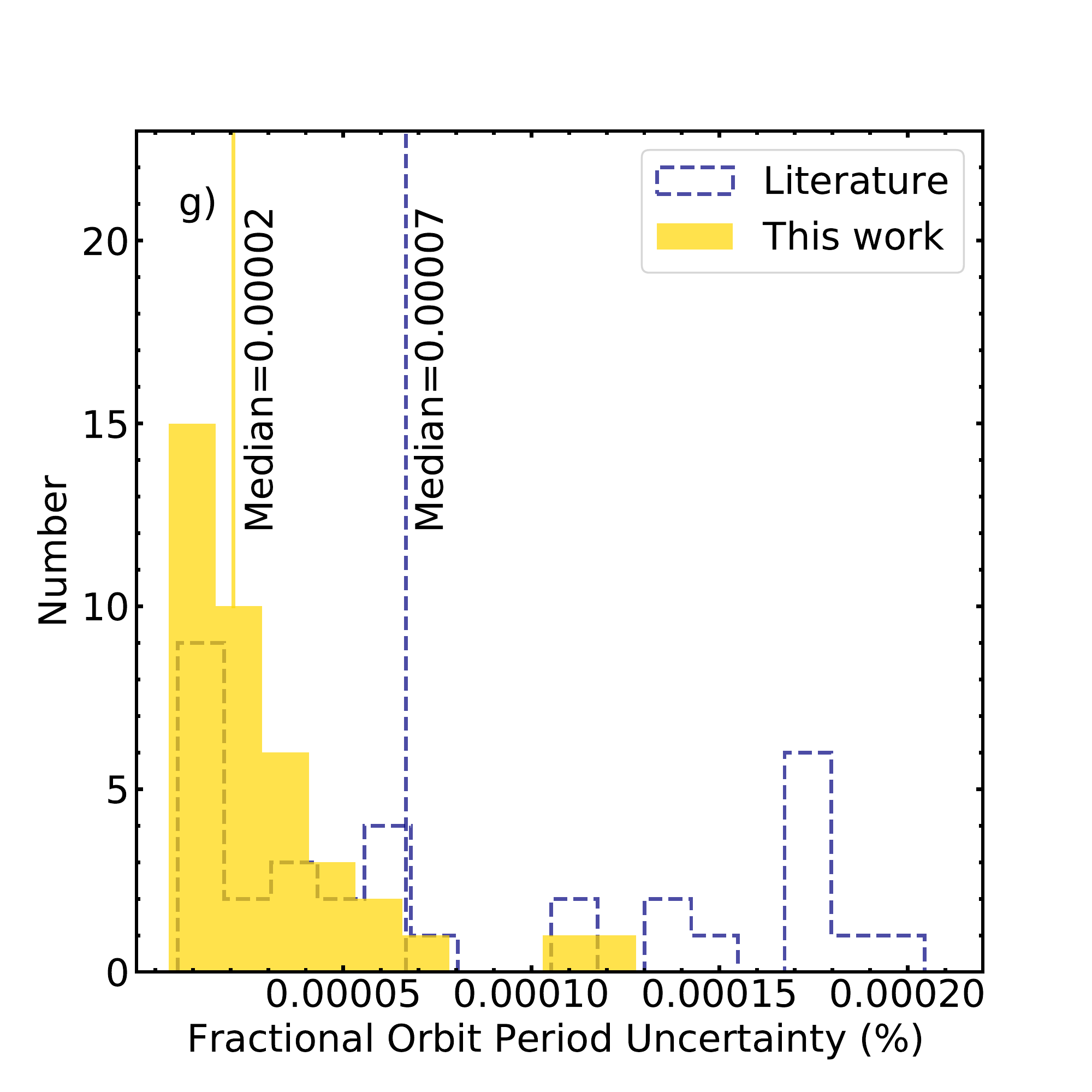}
    \includegraphics[width=0.32\textwidth,trim=20 15 35 60,clip]{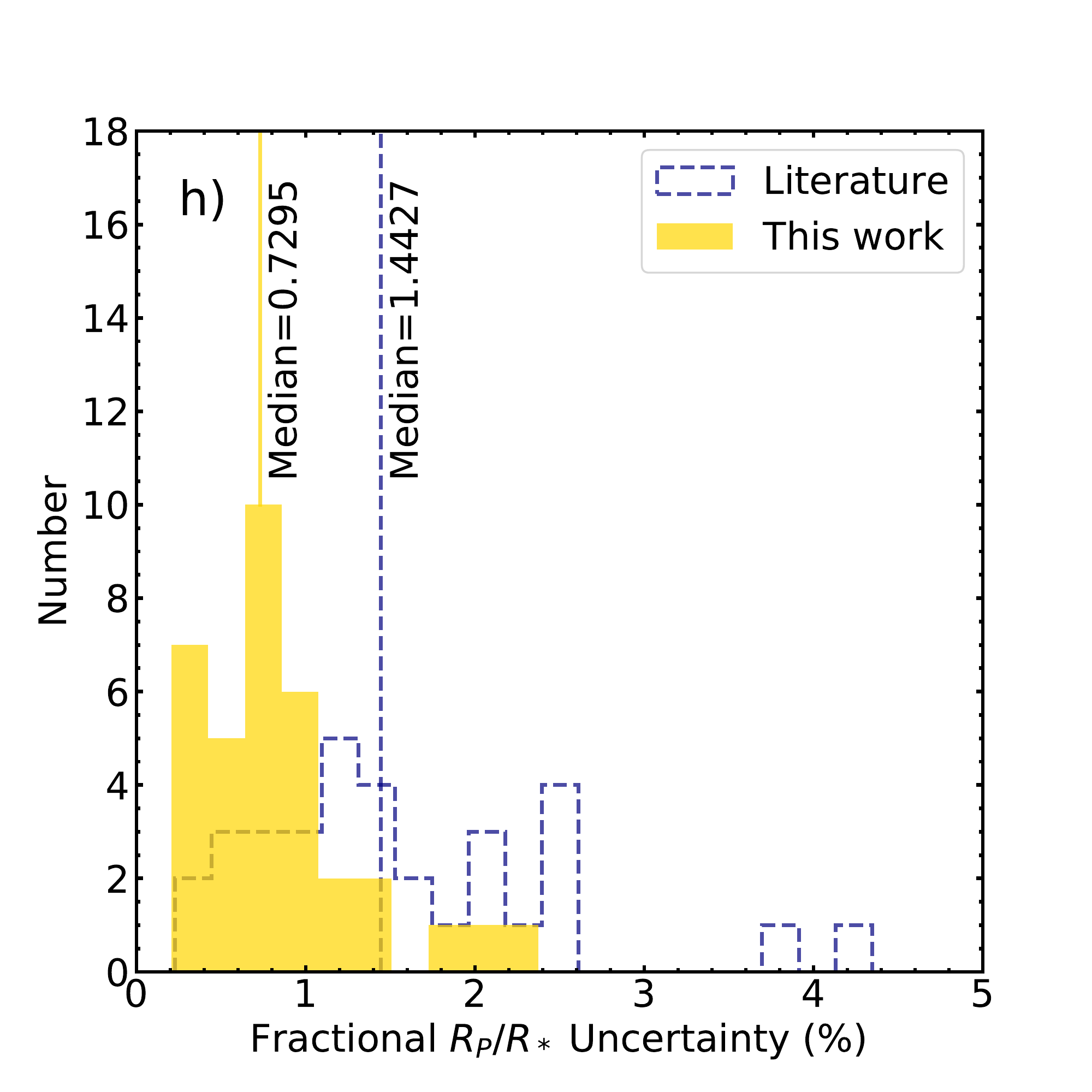}
    \includegraphics[width=0.32\textwidth,trim=20 15 35 60,clip]{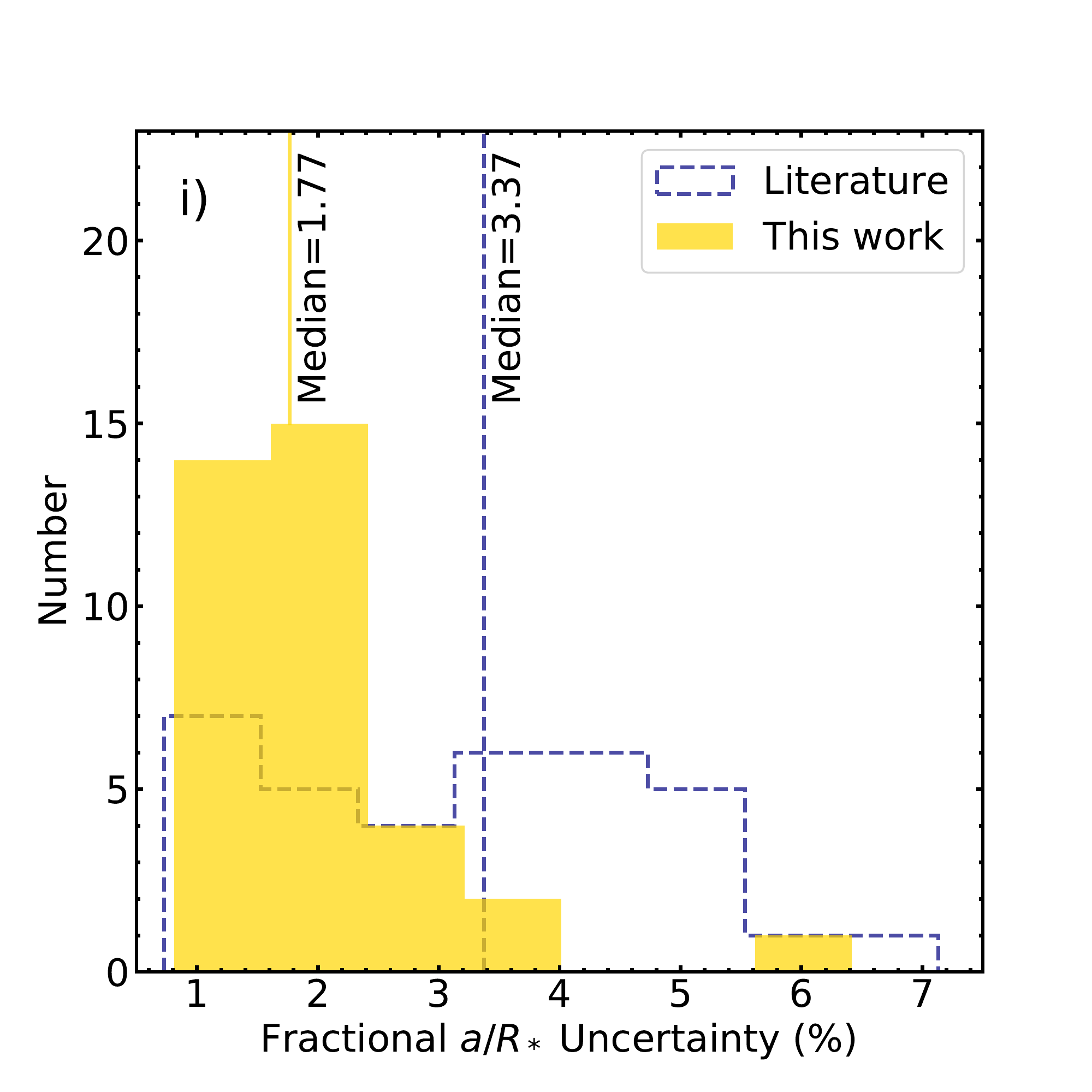}
    \includegraphics[width=0.32\textwidth,trim=20 15 35 60,clip]{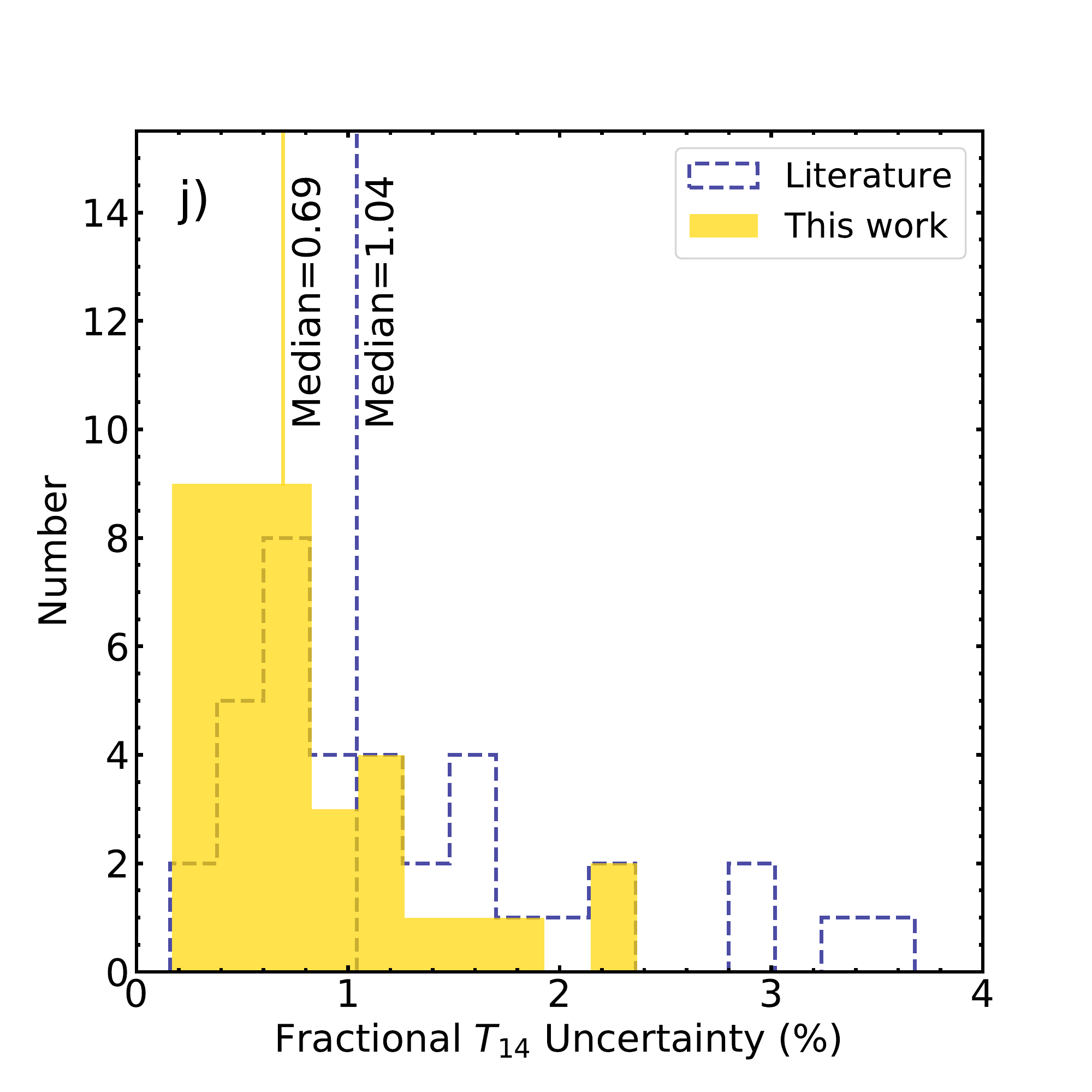}
    \includegraphics[width=0.32\textwidth,trim=20 15 35 60,clip]{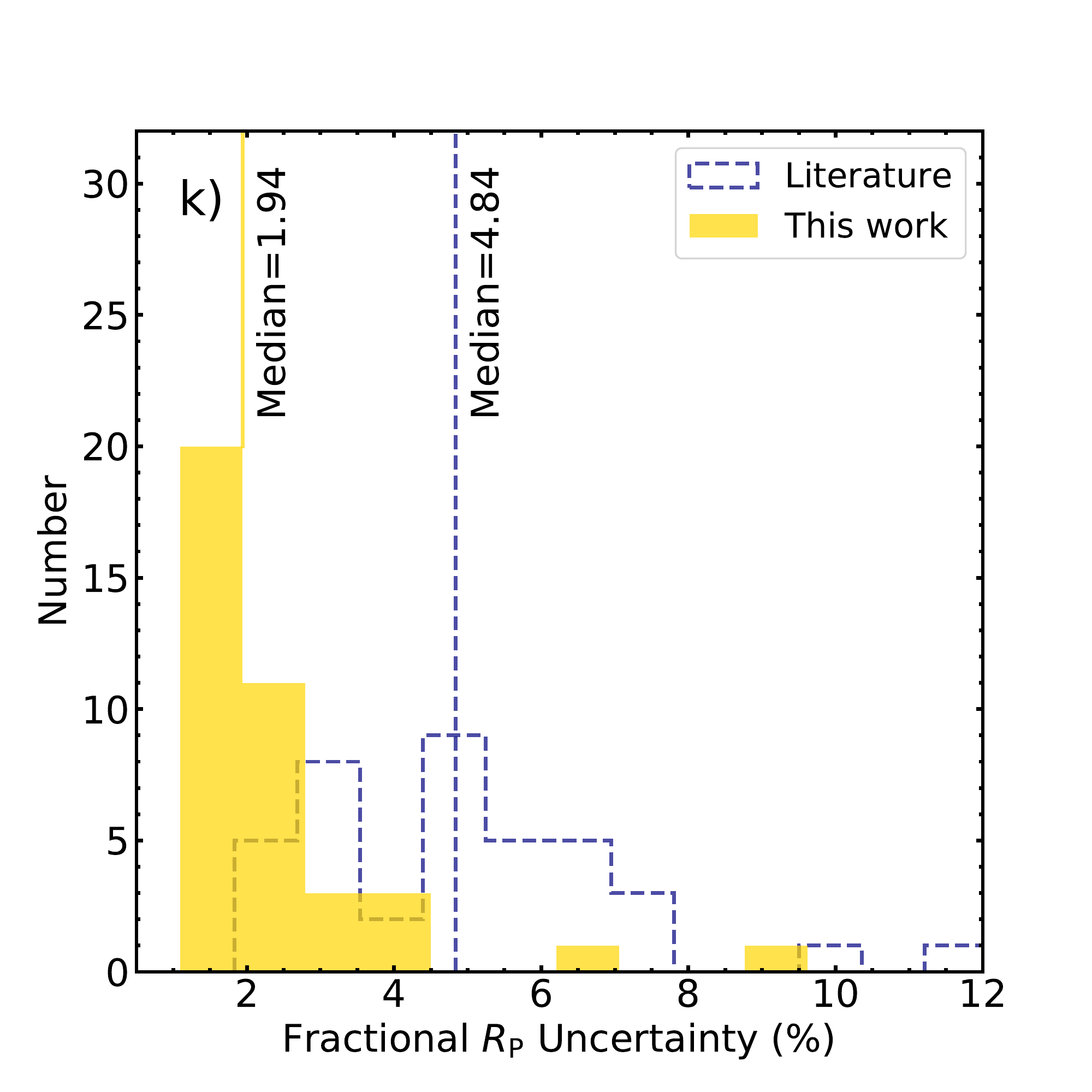}
    \includegraphics[width=0.32\textwidth,trim=20 15 35 60,clip]{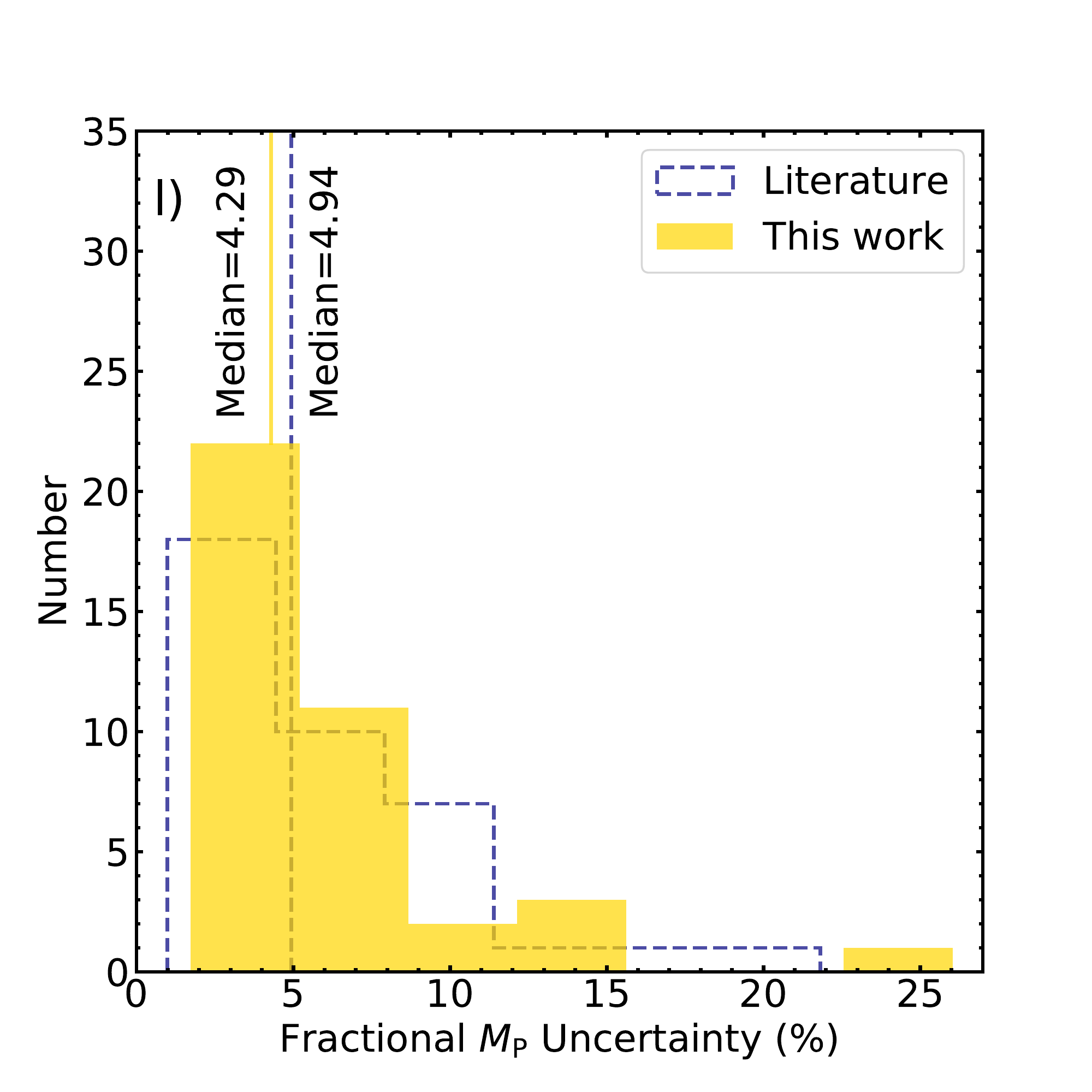}
    % Transit
    \caption{Distribution of uncertainties in
    fractional effective temperature ($\sigma_{\teff}/\teff$, Panel a), 
    stellar metallicity ($\sigma_{\feh}$, Panels b), 
    fractional surface gravity ($\sigma_{\logg}/\logg$, Panel c), 
    fractional stellar radius ($\sigma_{\rstar}/\rstar$, Panel d), 
    fractional stellar mass ($\sigma_{\mstar}/\mstar$, Panel e), 
    eccentricity ($\sigma_{e}$, Panel f), 
    fractional orbital period ($\sigma_{P}/P$, Panel g), 
    fractional planet-to-star radius ratio ($\sigma_{\rp/\rstar}/\rp/\rstar$, Panel h), 
    fractional scaled semi-major axis ($\sigma_{a/\rstar}/a/\rstar$, Panel i), 
    % fractional impact parameter ($\sigma_{b}/b$, Panel k), 
    fractional transit duration  ($\sigma_{T_{14}}/{T_{14}}$, Panel g),
    fractional planetary radius ($\sigma_{\rp}/\rp$, Panel k), and
    fractional planetary mass ($\sigma_{\mplanet}/\mplanet$, Panel l)
    from this work (yellow) compared to the literature values (blue dashed line).}
    \label{fig: Fractional parameters Uncertainty}
\end{figure*}

\begin{figure}
    \centering
    \includegraphics[width=0.49\textwidth]{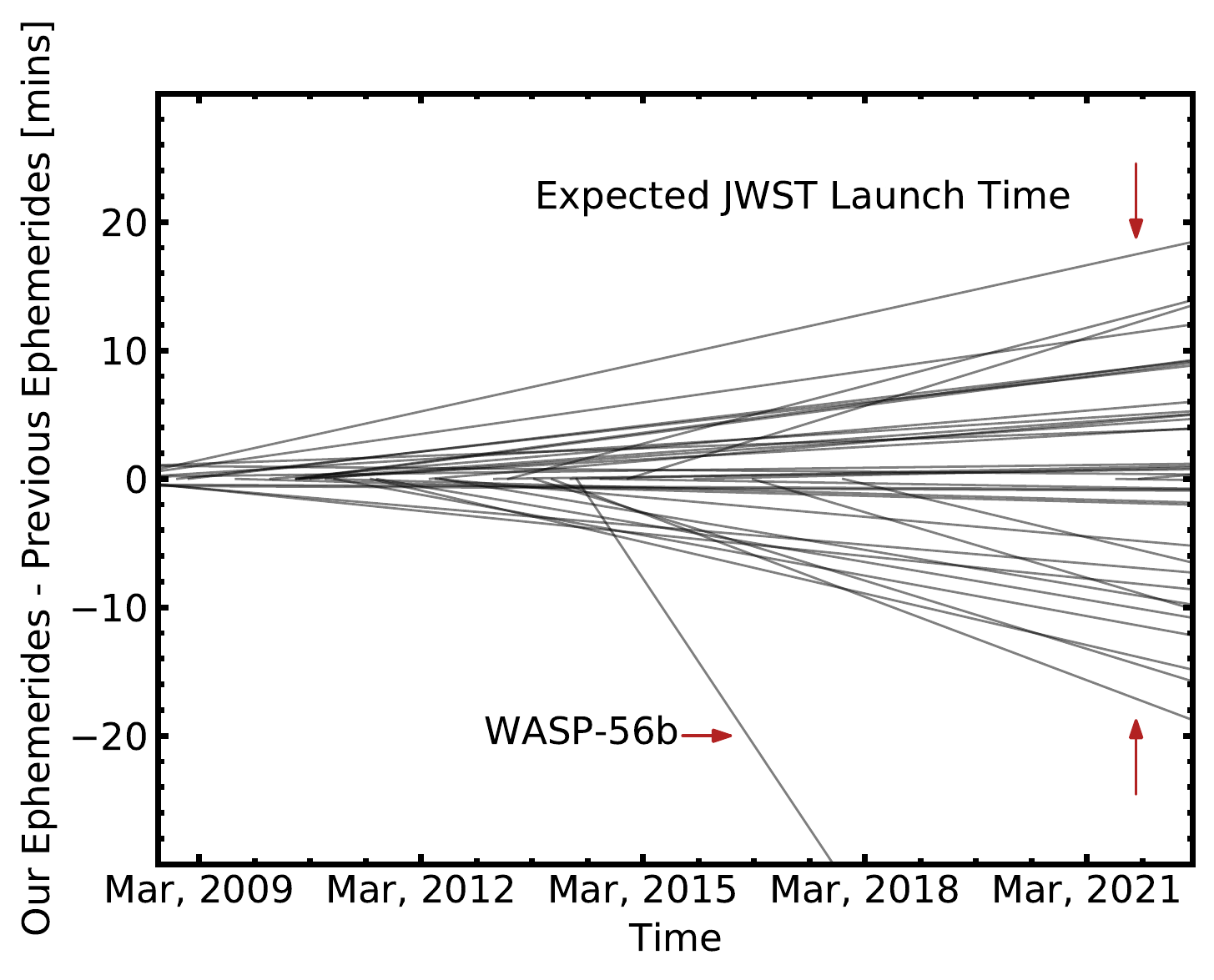}
    \caption{The difference between calculated (predicted) transit windows determined using our updated ephemerides and literature ephemerides. The difference for 29 out of 39 targets are within $10\,{\rm min}$ at the time of launch of JWST. For the other 10 targets, our predicted transit mid-times differ by more than $10\,{\rm min}$ from the literature ephemerides; for WASP-56b, the deviation amounts to about $72\,{\rm min}$.
    }
    \label{fig: Ephemerises}
\end{figure}

\section{Results and Discussion}

The headline result of this work is the refinement of system parameters for 39 transiting hot Jupiters.
These are collected and compared to the literature results in Table~\ref{tab:WASP-36.}.
The stellar parameters and the most relevant planetary and orbital parameters are aggregated in Table~\ref{tab:stellar_parameters} and in Table~\ref{tab:planetary_parameters}, respectively. 

We show a comparison between system parameters from this work and from previous studies in Figure~\ref{fig: agreement}. Our results are based on a larger number of transits than previous measurements, so we are confident that our results with increased precision (as shown in Figure~\ref{fig: Fractional parameters Uncertainty}) are preferred over past literature values.

\textbf{Stellar Physical Parameters}: 

Our stellar parameters are in reasonable agreement ($2\,\sigma$) with published results (See Panels a, b, c, d, and e in Figure ~\ref{fig: agreement}), with the exception of a $2.6\,\sigma$ larger $\logg$ value derived for HAT-P-22. The derived system parameters showing $>2\,\sigma$ discrepancies are marked in bold in Table~\ref{tab:WASP-36.}.

Although $\teff$ and $\feh$ can be independently constrained from the SED and transit limb darkening, the improvement in precision of these two parameters from our analysis is not significant (See Panels a and b in Figure~\ref{fig: Fractional parameters Uncertainty}). These parameters are typically best constrained by spectroscopic priors, while our work focuses on extending and refining the photometric baseline for the observed systems.

Compared to literature solutions, we obtained more precise \logg\ values (See Panel c in Figure~\ref{fig: Fractional parameters Uncertainty}), since the transit alone provides an additional constraint on $\logg$ through limb darkening. We also significantly improved the precision of the host stellar radii, since a combination of the SED fits to the broad-band photometry, the upper limit on the $V$-band extinction from galactic dust maps, and a parallax constraint from Gaia DR2 precisely determines the bolometric luminosity and therefore stellar radius. All of our stellar radii are constrained to accuracies of better than 3\% (See Panel d in Figure~\ref{fig: Fractional parameters Uncertainty}). By comparison, previous studies attained the same level of accuracy for only 28 of the 39 systems.

The transit alone provides an independent and additional constraint on the stellar density through its duration \citep{Seager2003}. The stellar mass is, therefore, more precisely constrained in our study (See Panel e in Figure~\ref{fig: Fractional parameters Uncertainty}) due to our improved stellar radius and stellar density measurements, together with the SED fits to the broadband photometry and MIST fits to $\teff$, $\feh$, and $\logg$.

\textbf{Radial Velocity Parameters}: 

Since we did not obtain any new radial velocity measurements in this work, it is not surprising that the radial velocity parameters derived from our analysis typically agree with previous results in both errors and values.

Although the eccentricities that we found agree with published values within $2\,\sigma$, literature studies are systemically biased towards larger eccentricities (See Panel f in Figure~\ref{fig: agreement}) for these nearly circular hot Jupiter systems due to the eccentricity boundary at zero \citep{Wittenmyer2013, Wittenmyer2019}. In this work, instead of sampling directly in $e$, we sample uniformly in $e\,\mathrm{cos}\,\omega_*$ and $e\,\mathrm{sin}\,\omega_*$, which helps us to reduce the Lucy-Sweeney bias \citep{Lucy1971}.

\textbf{Transit Parameters}:

The majority of the final photometric parameters for our systems --- the orbital period ($P$, Panel g in Figure~\ref{fig: agreement} and Figure~\ref{fig: Fractional parameters Uncertainty}), the planet-to-star radius ratio ($R_{\rm P}/R_*$, Panel h in Figure~\ref{fig: agreement} and Figure~\ref{fig: Fractional parameters Uncertainty}),  the scaled semi-major axis ($a/R_*$, Panel i in Figure~\ref{fig: agreement} and Figure~\ref{fig: Fractional parameters Uncertainty}), and the transit duration ($T_{14}$, Panel j in Figure~\ref{fig: agreement} and Figure~\ref{fig: Fractional parameters Uncertainty}) --- are in agreement with but more precise than the published values. 

For several systems that show $>2\,\sigma$ discrepancies in the transit parameters between our analysis and previous literature values (marked in bold in Table~\ref{tab:WASP-36.}), we are confident that our results are more reliable since our results are based on more extensive transit data than previous studies.

\textbf{Planetary Physical Parameters}:

Although we found a larger planetary radius for GJ 436 and a smaller planetary radius for HAT-P-54 compared to previous work, the planetary physical parameters ($R_{\rm P}$ and $M_{\rm P}$) obtained in this study are generally in reasonable agreement with those from the previous literature (See Panel k and l in Figure~\ref{fig: agreement}), but with higher precision (see Panel k and l in Figure~\ref{fig: Fractional parameters Uncertainty}). This is especially true for planetary radius, as our results are based on more extensive photometric data. The planetary radii for almost all of our targets (37 out of 39) are determined to accuracies of better than $5\%$. By comparison, previous studies attained the same level of accuracy for only 21 of the 39 systems.

\textbf{Transit Timing Variations}:

Our new photometric observations allow for significant improvements in the orbital ephemerides ($T_0$ and $P$) of each system, which are useful to accurately predict future transit events. By 31 Oct, 2021, the currently scheduled James Webb Space Telescope (JWST) \citep{Gardner2006} launch date, if the literature ephemerides were left unimproved, the accumulated errors of predicted transit mid-times would exceed $1\,\rm{hour}$ for systems like WASP-56 (see Figure~\ref{fig: Ephemerises}), although 29 of our 39 targets are within $10\,{\rm min}$ of previous predictions.

The refined orbital ephemerides also enable a search for TTVs. The measured transit mid-times of our targets are consistent with our updated linear ephemerides within $3\,\sigma$, except for several outliers (See Figure~\ref{fig:ttvs}). Although the origin of these outliers could be astrophysical, it is most likely that 1). The uncertainties of transit mid-times for the light curves taken under poor weather conditions are underestimated because of the presence of correlated noise or 2). The transit mid-times of very incomplete transits are imprecisely determined. Additional photometry is needed to uncover the nature of these discrepancies.

Our null results agree with previous TTV searches for hot Jupiter companion planets, conducted using both the \textit{Kepler} dataset \citep{Steffen2012} and ground-based photometric follow-up programs (for example: TLC, \citealt{Holman2006}; YETI, \citealt{Neuhauser2011}; Trappsit, \citealt{Gillon2012}; Taste, \citealt{Nascimbeni2011}; TraMoS, \citealt{Cortes2020}; TEMP, \citealt{Wang2018b}; and \citealt{Mallonn2019}).

\textbf{Constraints on Additional Planets}:

The absence of statistically significant TTVs provides constraints on the upper mass limits for any putative companion planets in the observed hot Jupiter systems.

Using dynamical simulations with REBOUND \citep{Rein2012, Rein2015}, we searched for the masses of putative companion planets that would produce TTVs whose RMS values are the same as those measured in Section~\ref{fit:ind}. This delineates the mass-period parameter space where additional close-in companion planets would have been detected if they existed, and, conversely, where these planets would have been remained undetected.

Our simulations assume that the putative companion planets are coplanar with the hot Jupiters, and both planets are on circular orbits. Compared with non-coplanar or eccentric orbits, the coplanar and circular configuration induces smaller TTVs and therefore provides a more conservative estimate of the putative companion planets' upper mass limits, as discussed by \citet{Agol2016}.   

We stepped through the period ratio of putative companion planets and known hot Jupiters from 1:5 to 5:1 in 100 steps. The sample resolution was tripled when in proximity to resonances where the largest planetary TTVs are likely to arise \citep{Agol2005, Holman2005}. For each step, the approximated upper mass limit of the putative planet was obtained iteratively by linear interpolation with an initial mass guess of $1\mearth$ and with a convergence tolerance of $|O_{\rm TTV\,RMS}-C_{\rm TTV\,RMS}|< $1\,${\rm s}$.

\begin{figure*}
    \centering
    \includegraphics[width=1\textwidth]{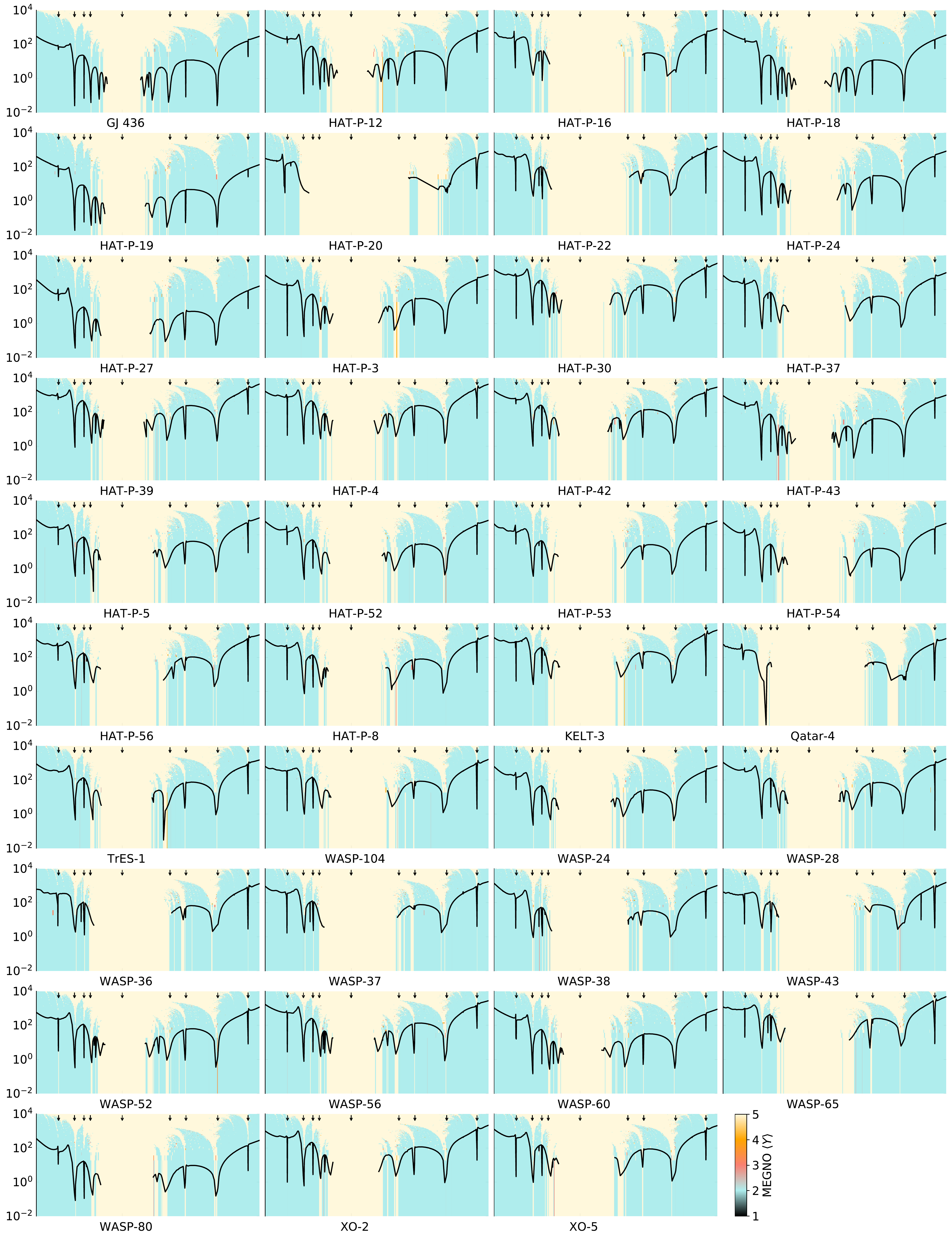}
    \caption{Upper mass limits ($\mearth$) as a function of the orbital period ratio of the putative companion planets and the transiting hot Jupiters. The black solid curves represent the mass-period region where putative companion planets would produce TTVs with RMS values matching those that we measured for our targets. The mass constraints are most restrictive near the low-order mean-motion resonances (marked by arrows). We use MEGNO Factor ($\left \langle Y \right \rangle$) to identify quasi-periodic regions ($\left \langle Y \right \rangle$=2, color coded blue), as well as chaotic and possibly unstable regions ($\left \langle Y \right \rangle \neq$ 2, color coded black, red, and yellow). 
    }
    \label{fig:MEGNO}
\end{figure*}

We did not explore the TTV behavior induced by unstable putative companion planets. The regions of orbital instability were identified using the Mean Exponential Growth factor of Nearby Orbits (MEGNO $\left \langle Y \right \rangle$ ; \citealt{Cincotta1999, Cincotta2000, Cincotta2003, Hinse2010}), which can efficiently distinguish quasi-periodic ($\left \langle Y \right \rangle \to 2$ as t$\to \infty $) or chaotic motion ($\left \langle Y \right \rangle$ $>$ 2 as t$\to \infty$). The MEGNO maps (see Figure~\ref{fig:MEGNO}) are gridded in period ratio-mass space, with 500 evenly spaced values on each axis spanning $0.1<P_{\rm putative \ planet}/P_{\rm hot \ Jupiter}<3.5$ and $10^{-2} \mearth < M_{\rm putative \ planet} < 10^{4} \mearth$ for the period ratio of putative companion planet and known hot Jupiter (x-axis) and mass of putative companion planet (y-axis), respectively. At each grid point, the putative companion planet was integrated together with the known hot Jupiter for $1,000\,$ years, at which point we calculated the MEGNO factor $\left \langle Y \right \rangle$.

MEGNO also identifies the locations of mean motion resonances. As illustrated in Figure~\ref{fig:MEGNO}, the mass constraints from TTVs are more restrictive at mean-motion resonances (especially the low-order mean-motion resonances) by comparison with general orbital configurations. We rule out the presence of putative companion planets with masses greater than $0.39 - 5.00\,\mearth$, $1.23 - 14.36\,\mearth$, $1.65 - 21.18\,\mearth$, and $0.69 - 6.75\,\mearth$ near the 1:2, 2:3, 3:2, and 2:1 resonances, respectively. For a given location in a given hot Jupiter system, the upper mass limit of the putative companion planet ($m_{\rm putative \ planet}$) is a function of the orbital period of the hot Jupiter ($P$), the TTV amplitude ($\Delta$), and the host stellar mass ($\mstar$), roughly scaling as $m_{\rm putative \ planet} = P \cdot \Delta/M_{*}$. Since the systems in our sample have a similar range of stellar mass ($\mstar$ ranging from $0.426$ to $1.41\,\msun$) and orbital period ($P$ ranging from $0.81$ to $6.87\,$ days), the upper limits of the putative companion planet masses are dominantly set by the TTV amplitudes ($\Delta$ ranging from $11$ to $585\,$s) of our hot Jupiters.

The absence of nearby resonant companion planets in hot Jupiter systems is inconsistent with the conventional expectation from disk migration. Although disk migration does not provide a precise quantitative prediction of the typical occurrence rate of resonant pairs, the absence of resonant companion planets across the hot Jupiter sample, as suggested by this work and previous publications \citep{Steffen2012, Montalto2012}, strongly disfavors disk migration. This absence is compatible with violent high-eccentricity tidal migration, which ejects any original close-in companion planets. We cannot rule out \textit{in-situ} formation, since, although \textit{in-situ} formation tends to form hot Jupiters accompanied by nearby planets \citep{Boley2016, Batygin2016}, those planets are not necessarily in or near mean motion resonance.

A dearth of additional transit signal detections in the \textit{Kepler} hot Jupiter sample initially appeared to place a very strong constraint on the occurrence rate of non-resonant companion planets in those systems \citep{Steffen2012, Huang2017}. However, these findings, which are generally quoted as evidence for the high-eccentricity origin of hot Jupiters, could instead result from observational bias, since the companion planets can hide in exterior and/or inclined orbits \citep{Millholland2016}, or in the observational noise. Previously hidden close-in companion planets in existing hot Jupiter systems have begun to be discovered via the transit method as better photometric precision is achieved (WASP-47, \citealt{Becker2015}; Kepler-730, \citealt{Canas2019}; TOI-1130, \citealt{Huang2020}), as well as through the radial velocity method (WASP-148: \citealt{Hebrard2020}). These systems cannot be explained by high-eccentricity tidal migration, but they are consistent with \textit{in-situ} formation.

\clearpage

\startlongtable
\begin{deluxetable*}{lccccccccc}
\tablecaption{Log of Observations of the TEMP Targets Presented in This Work  \label{tab:obs}}
\tablehead{ 
 \colhead{Planet}  &\colhead{Date}   &\colhead{Time}    &\colhead{Telescope} &\colhead{Typical Exposure Time} &\colhead{$\rm N_{\rm data}$} &\colhead{PNR} &\colhead{Quality}\\
\colhead{ }  &\colhead{(UTC)}  &\colhead{(UTC)}  &\colhead{ }  &\colhead{(s)} &\colhead{ } &\colhead{ }&\colhead{ }}
\startdata 
GJ 436 b & Mar 08, 2016 & 13:17:15-15:50:10 & Xinglong 60/90 cm Schmidt & 40 & 174 & 0.12& Golden\\
\hline
HAT-P-12 b & Feb 04, 2016 & 17:46:07-22:19:55 & Xinglong 60/90 cm Schmidt & 200 & 76 & 0.32& Mediocre\\
      & Feb 17, 2016 & 17:37:22-21:05:46 & Xinglong 60 cm & 30 & 323 & 0.15& Golden\\
      & Feb 17, 2016 & 17:57:53-21:53:21 & Xinglong 60/90 cm Schmidt & 140 & 94 & 0.17& Golden\\
      & Dec 31, 2016 & 20:25:13-22:31:45 & Xinglong 60/90 cm Schmidt & 150 & 48 & 0.21& Mediocre\\
      & Feb 14, 2017 & 19:49:37-21:33:24 & Xinglong 60/90 cm Schmidt & 80 & 41 & 0.29& Mediocre\\
      & Mar 28, 2017 & 12:31:25-15:59:05 & Xinglong 60 cm & 60 & 134 & 0.20& Mediocre\\
      & Apr 26, 2017 & 12:11:48-14:18:28 & Xinglong 60/90 cm Schmidt & 80 & 73 & 0.27& Mediocre\\
      & Apr 24, 2018 & 14:15:57-16:48:05 & Xinglong 60 cm & 180 & 44 & 0.21& Mediocre\\
      & Apr 27, 2018 & 17:40:03-19:48:32 & Xinglong 60 cm & 180 & 36 & 0.25& Mediocre\\
\hline
HAT-P-16 b & Jan 19, 2017 & 10:16:44-13:17:18 & Xinglong 60/90 cm Schmidt & 80 & 106 & 0.13& Golden\\
\hline
HAT-P-18 b & May 15, 2016 & 12:30:12-16:33:04 & Xinglong 60 cm & 30 & 397 & 0.17& Golden\\
      & May 30, 2017 & 14:01:18-18:10:23 & Xinglong 60 cm & 30 & 34 & 0.41& Mediocre\\
\hline
HAT-P-19 b & Nov 16, 2017 & 10:25:29-15:18:45 & Xinglong 60 cm & 25 & 363 & 0.22& Mediocre\\
\hline
HAT-P-20 b & Mar 02, 2016 & 14:41:46-17:19:44 & Xinglong 60/90 cm Schmidt & 60 & 90 & 0.26& Mediocre\\
      & Mar 05, 2016 & 12:16:08-17:04:46 & Xinglong 60/90 cm Schmidt & 55 & 243 & 0.13& Golden\\
      & Nov 07, 2016 & 19:09:22-22:00:40 & Xinglong 60 cm & 30 & 195 & 0.11& Golden\\
      & Nov 30, 2016 & 20:33:31-21:37:17 & Xinglong 60/90 cm Schmidt & 90 & 35 & 0.16& Golden\\
      & Jan 21, 2017 & 13:09:29-15:38:39 & Xinglong 60 cm & 30 & 165 & 0.13& Golden\\
      & Nov 13, 2017 & 16:52:34-20:33:30 & Xinglong 60 cm & 18 & 321 & 0.13& Golden\\
\hline
HAT-P-22 b & Jan 09, 2016 & 15:27:06-18:51:35 & Xinglong 60 cm & 10 & 770 & 0.09& Golden\\
      & Mar 10, 2016 & 17:32:40-21:23:25 & Xinglong 60 cm & 20 & 539 & 0.13& Golden\\
      & Jan 22, 2017 & 15:49:49-18:07:29 & Xinglong 60 cm & 20 & 181 & 0.14& Golden\\
\hline
HAT-P-24 b & Jan 13, 2016 & 15:04:46-20:11:18 & Xinglong 60 cm & 20 & 570 & 0.11& Golden\\
\hline
HAT-P-27 b & Mar 05, 2017 & 17:32:14-19:57:17 & Xinglong 60/90 cm Schmidt & 70 & 94 & 0.21& Mediocre\\
      & Mar 23, 2018 & 16:28:44-18:28:21 & Xinglong 60 cm & 130 & 47 & 0.23& Mediocre\\
\hline
HAT-P-3 b & Mar 30, 2016 & 12:50:45-16:02:01 & Xinglong 60/90 cm Schmidt & 60 & 180 & 0.19& Golden\\
      & Mar 07, 2017 & 15:35:15-17:11:25 & Xinglong 60/90 cm Schmidt & 110 & 48 & 0.21& Mediocre\\
      & May 07, 2017 & 13:13:30-18:36:43 & Xinglong 60 cm & 40 & 310 & 0.21& Mediocre\\
      & May 10, 2017 & 12:18:24-14:34:42 & Xinglong 60 cm & 10 & 221 & 0.21& Mediocre\\
      & Mar 22, 2018 & 12:19:33-17:37:22 & Xinglong 60 cm & 170 & 91 & 0.15& Golden\\
\hline
HAT-P-30 b & Jan 21, 2014 & 13:52:02-15:38:27 & Xinglong 60 cm & 30 & 299 & 0.13& Golden\\
      & Mar 01, 2016 & 14:06:32-17:28:20 & Xinglong 60/90 cm Schmidt & 100 & 80 & 0.20& Golden\\
      & Jan 21, 2017 & 15:43:00-19:14:28 & Xinglong 60 cm & 18 & 301 & 0.12& Golden\\
\hline
HAT-P-37 b & Feb 27, 2014 & 18:35:54-22:01:09 & Xinglong 60 cm & 90 & 120 & 0.25& Mediocre\\
      & Mar 29, 2016 & 17:59:16-21:01:16 & Xinglong 60/90 cm Schmidt & 180 & 56 & 0.21& Mediocre\\
      & Apr 19, 2017 & 17:11:22-20:22:21 & Xinglong 60 cm & 50 & 146 & 0.27& Mediocre\\
    %   & Apr 19, 2017 & 17:11:22-20:22:21 & Xinglong 60 cm & 50 & 146 & 0.27& Mediocre\\
      & May 31, 2017 & 16:09:27-19:41:51 & Xinglong 60 cm & 40 & 39 & 0.58& Bad\\
\hline
HAT-P-39 b & Feb 01, 2016 & 11:54:55-17:38:54 & Xinglong 60/90 cm Schmidt & 160 & 109 & 0.28& Mediocre\\
      & Mar 11, 2016 & 11:34:31-16:53:11 & Xinglong 60 cm & 30 & 529 & 0.16& Golden\\
      & Dec 30, 2016 & 15:05:30-18:22:49 & Xinglong 60/90 cm Schmidt & 120 & 89 & 0.28& Mediocre\\
      & Jan 24, 2017 & 11:49:28-16:40:59 & Xinglong 60 cm & 60 & 170 & 0.32& Mediocre\\
\hline
HAT-P-4 b & Mar 13, 2016 & 17:42:52-21:29:26 & Xinglong 60 cm & 20 & 534 & 0.09& Golden\\
      & Feb 11, 2018 & 17:39:19-21:02:29 & Xinglong 60 cm & 20 & 254 & 0.24& Mediocre\\
\hline
HAT-P-42 b & Feb 24, 2017 & 15:37:30-18:40:52 & Xinglong 60/90 cm Schmidt & 200 & 55 & 0.19& Golden\\
\hline
HAT-P-43 b & Feb 28, 2016 & 13:40:38-17:22:45 & Xinglong 60/90 cm Schmidt & 200 & 61 & 0.23& Mediocre\\
      & Mar 09, 2016 & 11:32:30-17:33:03 & Xinglong 60 cm & 80 & 233 & 0.19& Golden\\
      & Dec 12, 2017 & 16:10:28-22:34:57 & Xinglong 60 cm & 50 & 307 & 0.24& Mediocre\\
\hline
HAT-P-5 b & May 09, 2016 & 16:08:34-20:18:52 & Xinglong 60 cm & 20 & 435 & 0.13& Golden\\
\hline
HAT-P-52 b & Nov 16, 2017 & 16:12:38-20:18:02 & Xinglong 60 cm & 40 & 222 & 0.34& Mediocre\\
\hline
HAT-P-53 b & Dec 11, 2017 & 10:22:06-15:16:34 & Xinglong 60 cm & 180 & 85 & 0.25& Mediocre\\
\hline
HAT-P-54 b & Feb 05, 2016 & 10:58:23-14:26:31 & Xinglong 60 cm & 120 & 93 & 0.19& Golden\\
      & Feb 05, 2016 & 11:06:12-15:13:04 & Xinglong 60/90 cm Schmidt & 290 & 48 & 0.27& Mediocre\\
      & Feb 24, 2016 & 11:11:08-14:51:16 & Xinglong 60/90 cm Schmidt & 160 & 78 & 0.23& Mediocre\\
      & Feb 19, 2017 & 11:12:32-14:09:23 & Xinglong 60 cm & 60 & 25 & 0.39& Mediocre\\
      & Mar 29, 2017 & 11:27:29-13:55:49 & Xinglong 60 cm & 50 & 110 & 0.30& Mediocre\\
\hline
HAT-P-56 b & Mar 07, 2016 & 11:18:29-15:34:21 & Xinglong 60/90 cm Schmidt & 60 & 165 & 0.12& Golden\\
      & Jan 05, 2017 & 16:02:17-19:51:19 & Xinglong 60/90 cm Schmidt & 60 & 191 & 0.11& Golden\\
      & Jan 08, 2017 & 12:37:08-15:25:31 & Xinglong 60/90 cm Schmidt & 70 & 121 & 0.09& Golden\\
      & Mar 02, 2017 & 11:22:09-15:12:53 & Xinglong 60/90 cm Schmidt & 80 & 141 & 0.13& Golden\\
      & Mar 16, 2017 & 11:16:49-14:54:48 & Xinglong 60/90 cm Schmidt & 80 & 121 & 0.14& Golden\\
      & Feb 11, 2018 & 13:17:10-17:37:21 & Xinglong 60 cm & 10 & 423 & 0.23& Mediocre\\
\hline
HAT-P-8 b & Sep 05, 2016 & 11:43:00-17:56:37 & Xinglong 60 cm & 20 & 505 & 0.13& Golden\\
      & Sep 08, 2016 & 12:28:56-16:00:13 & Xinglong 60 cm & 14 & 347 & 0.16& Golden\\
\hline
KELT-3 b & May 05, 2014 & 12:03:08-16:44:55 & Xinglong 60 cm & 60 & 265 & 0.23& Mediocre\\
      & Apr 15, 2018 & 12:33:32-16:00:34 & Xinglong 60 cm & 100 & 92 & 0.09& Golden\\
\hline
Qatar-4 b & Nov 13, 2017 & 10:24:09-15:07:17 & Xinglong 60 cm & 100 & 130 & 0.24& Mediocre\\
\hline
TrES-1 b & Sep 09, 2016 & 12:01:48-14:30:25 & Xinglong 60 cm & 60 & 119 & 0.12& Golden\\
      & Sep 15, 2016 & 12:26:32-14:58:10 & Xinglong 60 cm & 18 & 216 & 0.16& Golden\\
      & Sep 18, 2016 & 11:22:10-16:27:05 & Xinglong 60 cm & 22 & 218 & 0.16& Golden\\
      & Apr 21, 2017 & 15:58:58-20:11:03 & Xinglong 60 cm & 100 & 116 & 0.16& Golden\\
      & Apr 21, 2017 & 16:55:05-20:22:28 & Xinglong 60/90 cm Schmidt & 160 & 73 & 0.13& Golden\\
      & Apr 24, 2017 & 16:39:39-20:12:48 & Xinglong 60 cm & 100 & 93 & 0.12& Golden\\
      & Apr 24, 2017 & 16:43:20-20:17:45 & Xinglong 60/90 cm Schmidt & 160 & 76 & 0.16& Golden\\
      & Apr 27, 2017 & 17:06:20-20:26:14 & Xinglong 60/90 cm Schmidt & 80 & 132 & 0.16& Golden\\
      & Apr 30, 2017 & 16:57:51-20:18:08 & Xinglong 60/90 cm Schmidt & 70 & 145 & 0.14& Golden\\
\hline
WASP-104 b & Jan 13, 2016 & 20:14:32-22:24:53 & Xinglong 60 cm & 15 & 367 & 0.12& Golden\\
      & Feb 05, 2016 & 15:40:51-21:46:28 & Xinglong 60/90 cm Schmidt & 120 & 144 & 0.17& Golden\\
      & Feb 20, 2017 & 14:33:04-18:14:48 & Xinglong 60/90 cm Schmidt & 110 & 111 & 0.22& Mediocre\\
      & Mar 06, 2017 & 15:48:40-19:12:57 & Xinglong 60/90 cm Schmidt & 120 & 73 & 0.15& Golden\\
      & Apr 19, 2017 & 12:26:07-16:36:06 & Xinglong 60/90 cm Schmidt & 160 & 80 & 0.18& Golden\\
      & Dec 10, 2017 & 17:44:45-22:39:58 & Xinglong 60 cm & 30 & 325 & 0.17& Golden\\
      & Feb 13, 2018 & 17:01:06-20:38:11 & Xinglong 60 cm & 25 & 208 & 0.20& Golden\\
      & Mar 15, 2018 & 13:33:15-17:38:18 & Xinglong 60 cm & 139 & 83 & 0.17& Golden\\
\hline
WASP-24 b & Feb 22, 2016 & 17:57:38-22:08:20 & Xinglong 60/90 cm Schmidt & 80 & 124 & 0.26& Mediocre\\
\hline
WASP-28 b & Aug 19, 2014 & 15:16:26-18:06:52 & Xinglong 60 cm & 90 & 102 & 0.24& Mediocre\\
      & Nov 14, 2017 & 11:05:38-15:30:57 & Xinglong 60 cm & 25 & 313 & 0.23& Mediocre\\
\hline
WASP-36 b & Jan 17, 2014 & 14:39:45-19:20:42 & Xinglong 60 cm & 60 & 264 & 0.29& Mediocre\\
      & Jan 20, 2014 & 17:07:56-20:31:16 & Xinglong 60 cm & 60 & 188 & 0.36& Mediocre\\
      & Jan 08, 2016 & 15:45:33-20:07:08 & Xinglong 60 cm & 40 & 337 & 0.22& Mediocre\\
      & Feb 14, 2016 & 13:10:45-16:17:46 & Xinglong 60 cm & 150 & 69 & 0.21& Mediocre\\
      & Feb 14, 2016 & 13:11:30-16:05:21 & Xinglong 60/90 cm Schmidt & 160 & 60 & 0.27& Mediocre\\
      & Mar 02, 2016 & 12:01:44-14:19:24 & Xinglong 60/90 cm Schmidt & 250 & 33 & 0.47& Bad\\
      & Apr 18, 2017 & 11:51:53-13:43:47 & Xinglong 60 cm & 50 & 92 & 0.29& Mediocre\\
\hline
WASP-37 b & Apr 13, 2016 & 15:50:06-20:26:14 & Xinglong 60 cm & 55 & 249 & 0.22& Mediocre\\
      & Mar 26, 2017 & 16:37:22-19:50:35 & Xinglong 60 cm & 40 & 43 & 0.72& Bad\\
      & Apr 20, 2017 & 17:13:06-20:38:04 & Xinglong 60 cm & 40 & 180 & 0.21& Mediocre\\
      & Apr 20, 2018 & 15:02:26-17:08:44 & Xinglong 60 cm & 249 & 26 & 0.71& Bad\\
\hline
WASP-38 b & May 12, 2016 & 17:31:50-17:31:32 & Xinglong 60 cm & 12 & 1056 & 0.11& Golden\\
\hline
WASP-43 b & Jan 09, 2016 & 19:42:49-22:43:56 & Xinglong 60 cm & 45 & 219 & 0.2& Golden\\
      & Jan 18, 2016 & 19:42:52-21:18:44 & Xinglong 60 cm & 150 & 38 & 0.15& Golden\\
      & Feb 14, 2016 & 16:24:49-18:37:08 & Xinglong 60 cm & 150 & 51 & 0.27& Mediocre\\
      & Feb 14, 2016 & 17:07:08-19:24:23 & Xinglong 60/90 cm Schmidt & 160 & 49 & 0.30& Mediocre\\
      & Mar 29, 2016 & 13:55:29-16:27:52 & Xinglong 60/90 cm Schmidt & 60 & 126 & 0.22& Mediocre\\
      & Nov 27, 2016 & 18:46:02-22:01:22 & Xinglong 60/90 cm Schmidt & 100 & 98 & 0.18& Golden\\
      & Mar 13, 2017 & 13:01:32-16:40:18 & Xinglong 60/90 cm Schmidt & 90 & 118 & 0.30& Mediocre\\
      & Mar 26, 2017 & 13:23:27-16:47:56 & Xinglong 60/90 cm Schmidt & 100 & 111 & 0.19& Golden\\
\hline
WASP-52 b & Jan 12, 2016 & 10:17:30-11:42:14 & Xinglong 60 cm & 100 & 49 & 0.32& Mediocre\\
      & Oct 09, 2016 & 14:39:28-18:06:37 & Xinglong 60/90 cm Schmidt & 150 & 77 & 0.21& Mediocre\\
\hline
WASP-56 b & Feb 06, 2016 & 16:37:17-21:21:43 & Xinglong 60 cm & 45 & 333 & 0.13& Golden\\
      & Feb 06, 2016 & 16:29:28-20:34:20 & Xinglong 60/90 cm Schmidt & 90 & 141 & 0.19& Golden\\
      & Feb 14, 2017 & 15:44:07-21:09:08 & Xinglong 60 cm & 15 & 375 & 0.29& Mediocre\\
      & Apr 29, 2017 & 12:31:36-17:37:16 & Xinglong 60/90 cm Schmidt & 140 & 104 & 0.24& Mediocre\\
      & May 08, 2018 & 13:04:20-16:09:08 & Xinglong 60 cm & 249 & 45 & 0.23& Mediocre\\
\hline
WASP-60 b & Nov 11, 2013 & 10:38:18-15:40:59 & Xinglong 60/90 cm Schmidt & 100 & 125 & 0.12& Golden\\
\hline
WASP-65 b & Jan 21, 2014 & 15:45:12-21:41:30 & Xinglong 60 cm & 60 & 302 & 0.17& Golden\\
      & Jan 21, 2014 & 14:29:34-20:47:23 & Xinglong 60/90 cm Schmidt & 80 & 180 & 0.17& Golden\\
      & Jan 10, 2016 & 14:09:05-16:17:48 & Xinglong 60 cm & 30 & 213 & 0.12& Golden\\
      & Nov 03, 2016 & 18:13:11-21:47:02 & Xinglong 60 cm & 20 & 271 & 0.21& Mediocre\\
      & Dec 10, 2016 & 17:09:59-21:47:31 & Xinglong 60 cm & 20 & 373 & 0.16& Golden\\
      & Jan 23, 2017 & 15:41:58-20:10:18 & Xinglong 60 cm & 22.5 & 323 & 0.16& Golden\\
      & Apr 21, 2017 & 11:42:30-15:01:11 & Xinglong 60 cm & 40 & 134 & 0.18& Golden\\
\hline
WASP-80 b & May 12, 2016 & 18:38:09-19:57:34 & Xinglong 60 cm & 25 & 158 & 0.14& Golden\\
      & Sep 12, 2017 & 11:49:35-14:50:21 & Xinglong 60 cm & 7 & 336 & 0.30& Mediocre\\
\hline
XO-2 b & Feb 17, 2017 & 12:36:48-17:02:57 & Xinglong 60/90 cm Schmidt & 40 & 217 & 0.12& Golden\\
\hline
XO-5 b & Jan 12, 2016 & 17:58:54-22:10:58 & Xinglong 60 cm & 100 & 132 & 0.24& Mediocre\\
      & Jan 29, 2016 & 12:20:36-16:06:44 & Xinglong 60/90 cm Schmidt & 150 & 84 & 0.23& Mediocre\\
\hline
\enddata
\end{deluxetable*}

\begin{deluxetable*}{lcccc}
\tablecaption{\label{tab:lcs} Photometry of TEMP Targets}
\tablehead{\colhead{Planet} & \colhead{$ \mathrm{BJD}_\mathrm{TDB}$} & \colhead{Relative Flux} & \colhead{Uncertainty} & \colhead{Filter} }
\startdata
GJ 436 b& 2457456.053648 & 1.0030 & 0.0019 & R   \\
GJ 436 b& 2457456.054238 & 1.0002 & 0.0019 & R   \\
GJ 436 b& 2457456.054840 & 0.9999 & 0.0019 & R   \\
GJ 436 b& 2457456.055430 & 0.9982 & 0.0019 & R   \\
GJ 436 b& 2457456.056032 & 0.9986 & 0.0019 & R   \\
GJ 436 b& 2457456.056622 & 1.0016 & 0.0019 & R   \\
GJ 436 b& 2457456.057224 & 1.0004 & 0.0019 & R   \\
GJ 436 b& 2457456.057815 & 1.0015 & 0.0019 & R   \\
GJ 436 b& 2457456.058416 & 1.0013 & 0.0019 & R   \\
GJ 436 b& 2457456.059018 & 0.9983 & 0.0019 & R   \\
...    & ...            & ...    & ...    & ...
\enddata
\tablenotetext{}{Note: This table is available in its entirety in machine-readable format.
A portion of the table is shown here for guidance regarding its form and content. Table 2, 3 and 4 can be downloaded via this \href{http://vospace.china-vo.org/vospace/sharefile?Ravu36E\%2F2ja\%2B9Sa8a9jQu5wj9oWLvenACi6PEI7KYLWlIRqRRBhUJJN6P1U5331g}
{link} .} 
\end{deluxetable*}

\begin{deluxetable*}{lcccccc}
\tablewidth{0pt}
\tablecaption{Transit Mid-Times for TEMP Targets  \label{tab:midtimes}}
\tablehead{ 
 \colhead{Planet} &\colhead{Epoch Number} & \colhead{T$_{\rm 0}$} & \colhead{$\sigma$} & \colhead{O-C} \\ 
 \colhead{} & \colhead{} & \colhead{(BJD$_{\rm TDB}$)} & \colhead{(s)} & \colhead{(s)}}
\startdata 
GJ 436 b&-42&2454222.616809&65.30&-7.04\\
GJ 436 b&-41&2454225.260917&71.90&11.29\\
GJ 436 b&-41&2454225.261190&66.25&34.86\\
GJ 436 b&-41&2454225.261356&68.20&49.19\\
GJ 436 b&-39&2454230.548755&70.76&15.32\\
GJ 436 b&-36&2454238.480865&202.79&51.88\\
GJ 436 b&-34&2454243.767054&51.72&-86.56\\
GJ 436 b&-33&2454246.410962&148.03&-85.43\\
GJ 436 b&-33&2454246.411611&152.83&-29.37\\
GJ 436 b&-31&2454251.700230&67.85&42.17\\
GJ 436 b&-20&2454280.782620&21.49&2.27\\
GJ 436 b&1181&2457456.101184&103.67&0.52\\
...  & ... & ... & ... &  \\
\enddata
\tablenotetext{}{Note: This table is available in its entirety in machine-readable format.
A portion of the table is shown here for guidance regarding its form and content.} 
\end{deluxetable*}

\begin{longrotatetable}
\begin{deluxetable*}{lcccccccc}
\tablecaption{System Parameters for TEMP Targets obtained in This Study are Compared with Literature Values.}
\tablehead{\colhead{~~~Parameter} & \colhead{Units} & \colhead{Values} &\colhead{Previous values} & Refs  & Agreement ($\sigma$)}
\startdata
\smallskip                      \\\multicolumn{6}{c}{WASP-36}&\smallskip
\smallskip                      \\\multicolumn{2}{l}{Stellar Parameters:}&\smallskip                                                                                           \\
~~~~$M_*$\dotfill              &Mass (\msun)\dotfill                                    &$1.030^{+0.033}_{-0.036}$                &$1.081\pm0.034$           &\cite{Mancini2016}         & 1.08 \\
~~~~$R_*$\dotfill              &Radius (\rsun)\dotfill                                  &$0.966^{+0.013}_{-0.014}$                &$0.985\pm0.014$           &\cite{Mancini2016}         & 0.99 \\
~~~~$L_*$\dotfill              &Luminosity (\lsun)\dotfill                              &$1.202^{+0.089}_{-0.081}$                &...                         &...                         &... \\ 
~~~~$\rho_*$\dotfill           &Density (cgs)\dotfill                                   &$1.609^{+0.042}_{-0.036}$                &$1.595\pm0.045$             &\cite{Mancini2016}         & 0.24 \\
~~~~$\log{g}$\dotfill          &Surface gravity (cgs)\dotfill                           &$4.4807^{+0.0086}_{-0.0085}$             &$4.486\pm0.009$           &\cite{Mancini2016}         & 0.43 \\
~~~~$T_{\rm eff}$\dotfill      &Effective Temperature (K)\dotfill                       &$6150^{+110}_{-100}$                     &$5959\pm134$              &\cite{Mancini2016}         & 1.14 \\
~~~~$[{\rm Fe/H}]$\dotfill     &Metallicity (dex)\dotfill                               &$-0.31^{+0.20}_{-0.27}$                  &$-0.26\pm0.1$             &\cite{Mancini2016}         & 0.17 \\
~~~~$[{\rm Fe/H}]_{0}$\dotfill &Initial Metallicity \dotfill                            &$-0.31^{+0.18}_{-0.23}$                  &...                         &...                         &... \\ 
~~~~$Age$\dotfill              &Age (Gyr)\dotfill                                       &$1.01^{+1.1}_{-0.68}$                    &$1.4^{+0.4}_{-0.3}$         &\cite{Mancini2016}         & 0.34 \\
~~~~$EEP$\dotfill              &Equal Evolutionary Phase \dotfill                       &$315^{+20}_{-34}$                        &...                         &...                         &... \\ 
~~~~$A_V$\dotfill              &V-band extinction (mag)\dotfill                         &$0.190^{+0.078}_{-0.079}$                &...                         & ...                        &... \\ 
~~~~$\sigma_{SED}$\dotfill     &SED photometry error scaling \dotfill                   &$1.75^{+0.69}_{-0.43}$                   &...                         &  ...                       &... \\ 
~~~~$\varpi$\dotfill           &Parallax (mas)\dotfill                                  &$2.640\pm0.033$                          &$2.560\pm0.035$             & \cite{Stassun2019}                & 1.66 \\
~~~~$d$\dotfill                &Distance (pc)\dotfill                                   &$378.8^{+4.7}_{-4.6}$                    &$386.346^{+5.261}_{-5.123}$ &\cite{Stassun2019}         & 1.09 \\
\smallskip                      \\\multicolumn{2}{l}{Planetary Parameters:}&b\smallskip                              & ...                       &                            \\
\bf{~~~~$P$\dotfill                }&\bf{Period (days)\dotfill                                   }&\bf{$1.53736533\pm0.00000014$                }&\bf{$1.53736596\pm0.00000024$   }&\bf{\cite{Mancini2016}         }&\bf{2.27} \\
\bf{~~~~$R_P$\dotfill              }&\bf{Radius (\rj)\dotfill                                    }&\bf{$1.270^{+0.018}_{-0.019}$                }&\bf{$1.327\pm0.021$             }&\bf{\cite{Mancini2016}          }&\bf{2.06} \\
~~~~$M_P$\dotfill              &Mass (\mj)\dotfill                                      &$2.281^{+0.070}_{-0.071}$                &$2.361\pm0.070$             &\cite{Mancini2016}          & 0.81 \\
~~~~$T_C$\dotfill              &Time of conjunction (\bjdtdb)\dotfill                   &$2455569.83795\pm0.00011$                & ...                        &   ...                      &... \\ 
~~~~$T_0$\dotfill              &Optimal conjunction Time (\bjdtdb)\dotfill              &$2456678.278350^{+0.000047}_{-0.000046}$ &  ...                       & ...                         &... \\ 
~~~~$a$\dotfill                &Semi-major axis (AU)\dotfill                            &$0.02635^{+0.00028}_{-0.00031}$          &$0.02677\pm0.00028$         &\cite{Mancini2016}           & 1.06 \\
~~~~$i$\dotfill                &Inclination (Degrees)\dotfill                           &$83.42^{+0.12}_{-0.11}$                  &$83.15\pm0.13$              &\cite{Mancini2016}          & 1.59 \\
~~~~$e$\dotfill                &Eccentricity \dotfill                                   &$0.0087^{+0.0097}_{-0.0061}$             &...                         & ...                       &... \\ 
~~~~$\omega_*$\dotfill         &Argument of Periastron (Degrees)\dotfill                &$-43\pm89$                               & ...                      & ...                        &... \\ 
~~~~$T_{eq}$\dotfill           &Equilibrium temperature (K)\dotfill                     &$1796^{+32}_{-31}$                       &$1733\pm19$                 & \cite{Mancini2016}           & 1.73 \\
~~~~$\tau_{\rm circ}$\dotfill  &Tidal circularization timescale (Gyr)\dotfill           &$0.01638^{+0.00086}_{-0.00079}$          &  ...                      &...                          &... \\ 
~~~~$K$\dotfill                &RV semi-amplitude (m/s)\dotfill                         &$391.0^{+8.0}_{-8.1}$                    &$391.500\pm8.300$            &  0.04                        & 0.04 \\
~~~~$\log{K}$\dotfill          &Log of RV semi-amplitude \dotfill                       &$2.5922^{+0.0088}_{-0.0091}$             & ...                       &  ...                        &... \\ 
\bf{~~~~$R_P/R_*$\dotfill          }&\bf{Radius of planet in stellar radii \dotfill              }&\bf{$0.13515^{+0.00027}_{-0.00028}$          }&\bf{$0.13677\pm0.00056$         }&\bf{\cite{Mancini2016}          }&\bf{2.61} \\
~~~~$a/R_*$\dotfill            &Semi-major axis in stellar radii \dotfill               &$5.862^{+0.050}_{-0.044}$                &$5.8480^{+0.0552}_{-0.0542}$&\cite{Mancini2016}           & 0.2 \\
~~~~$\delta$\dotfill           &Transit depth (fraction)\dotfill                        &$0.018266^{+0.000074}_{-0.000076}$       & ...                     & ...                       &... \\ 
\bf{~~~~$Depth$\dotfill            }&\bf{Flux decrement at mid transit \dotfill                  }&\bf{$0.018266^{+0.000074}_{-0.000076}$       }&\bf{ $0.01916\pm0.0002$        }&\bf{ \cite{Mancini2016}          }&\bf{4.19} \\
~~~~$\tau$\dotfill             &Ingress/egress transit duration (days)\dotfill          &$0.01577^{+0.00023}_{-0.00024}$          &  ...                      &...                         &... \\ 
\bf{~~~~$T_{14}$\dotfill           }&\bf{Total transit duration (days)\dotfill                   }&\bf{$0.07732\pm0.00019$                      }&\bf{ $0.07566\pm0.00042$        }&\bf{ \cite{Smith2012}           }&\bf{3.6} \\
~~~~$T_{FWHM}$\dotfill         &FWHM transit duration (days)\dotfill                    &$0.06155\pm0.00019$                      &  ...                  &   ...                       &... \\ 
~~~~$b$\dotfill                &Transit Impact parameter \dotfill                       &$0.6736^{+0.0055}_{-0.0058}$             & $0.657^{+0.029}_{-0.033}$  & \citet{Maciejewski2016}     & 0.56 \\
~~~~$b_S$\dotfill              &Eclipse impact parameter \dotfill                       &$0.6699^{+0.0094}_{-0.012}$              & ...                       &   ...                       &... \\ 
~~~~$\tau_S$\dotfill           &Ingress/egress eclipse duration (days)\dotfill          &$0.01561^{+0.00041}_{-0.00049}$          & ...                        &  ...                        &... \\ 
~~~~$T_{S,14}$\dotfill         &Total eclipse duration (days)\dotfill                   &$0.07716^{+0.00045}_{-0.00065}$          & ...                        &  ...                         &... \\ 
~~~~$T_{S,FWHM}$\dotfill       &FWHM eclipse duration (days)\dotfill                    &$0.06150\pm0.00023$                      &  ...                        & ...                       &... \\ 
~~~~$\delta_{S,3.6\mu m}$\dotfill &Blackbody eclipse depth at 3.6$\mu$m (ppm)\dotfill   &$1988\pm46$                              & ...                       &  ...                        &... \\ 
~~~~$\delta_{S,4.5\mu m}$\dotfill &Blackbody eclipse depth at 4.5$\mu$m (ppm)\dotfill   &$2515\pm46$                              &  ...                      &   ...                      &... \\ 
~~~~$\rho_P$\dotfill           &Density (cgs)\dotfill                                   &$1.380^{+0.050}_{-0.047}$                & $1.260\pm0.060$           &  \cite{Mancini2016}       & 1.57 \\
~~~~$logg_P$\dotfill           &Surface gravity \dotfill                                &$3.545\pm0.012$                          & ...                        &  ...                        &... \\ 
~~~~$\Theta$\dotfill           &Safronov Number \dotfill                                &$0.0918\pm0.0023$                        & $0.0880\pm0.0023$          &\cite{Mancini2016}       & 1.17 \\
~~~~$\fave$\dotfill            &Incident Flux (\fluxcgs)\dotfill                        &$2.36^{+0.17}_{-0.16}$                   & ...                       &   ...                      &... \\ 
~~~~$T_P$\dotfill              &Time of Periastron (\bjdtdb)\dotfill                    &$2455569.28^{+0.39}_{-0.38}$             &  ...                        & ...                         &... \\ 
~~~~$T_S$\dotfill              &Time of eclipse (\bjdtdb)\dotfill                       &$2455570.6086^{+0.0094}_{-0.0054}$       & ...                        & ...                         &... \\ 
~~~~$T_A$\dotfill              &Time of Ascending Node (\bjdtdb)\dotfill                &$2455569.4537^{+0.0049}_{-0.0047}$       &  ...                       &  ...                       &... \\ 
~~~~$T_D$\dotfill              &Time of Descending Node (\bjdtdb)\dotfill               &$2455570.2242^{+0.0077}_{-0.0039}$       &  ...                      &   ...                     &... \\ 
~~~~$e\cos{\omega_*}$\dotfill  & \dotfill                                               &$0.0020^{+0.0095}_{-0.0055}$             &  ...                       & ...                         &... \\ 
~~~~$e\sin{\omega_*}$\dotfill  & \dotfill                                               &$-0.0018^{+0.0056}_{-0.0097}$            &  ...                       & ...                         &... \\ 
~~~~$M_P\sin i$\dotfill        &Minimum mass (\mj)\dotfill                              &$2.266^{+0.069}_{-0.071}$                &  ...                       & ...                        &... \\ 
~~~~$M_P/M_*$\dotfill          &Mass ratio \dotfill                                     &$0.002116\pm0.000050$                    &  ...                        & ...                         &... \\ 
~~~~$d/R_*$\dotfill            &Separation at mid transit \dotfill                      &$5.873^{+0.10}_{-0.072}$                 &  ...                       &  ...                        &... \\ 
~~~~$P_T$\dotfill              &A priori non-grazing transit prob \dotfill              &$0.1473^{+0.0018}_{-0.0025}$             &  ...                     &  ...                        &... \\ 
~~~~$P_{T,G}$\dotfill          &A priori transit prob \dotfill                          &$0.1933^{+0.0024}_{-0.0032}$             &  ...                      & ...                        &... \\ 
~~~~$P_S$\dotfill              &A priori non-grazing eclipse prob \dotfill              &$0.14798^{+0.00081}_{-0.00082}$          &  ...                      &  ...                        &... \\ 
~~~~$P_{S,G}$\dotfill          &A priori eclipse prob \dotfill                       A   &$0.1942\pm0.0012$                        &  ...                      &  ...                        &... \\ 
~~~~...         &...             &...                        &   ...                      &  ...                        &... \\ 
\enddata
\label{tab:WASP-36.}
\tablenotetext{}{Note: This table is available in its entirety in machine-readable format.
A portion of the table is shown here for guidance regarding its form and content.} 
\end{deluxetable*}
\end{longrotatetable}

\clearpage
\begin{deluxetable*}{lcccccccccccc}
\tablecaption{Stellar Parameters  \label{tab:stellar_parameters}}
\tablehead{ 
 \colhead{Star}  &\colhead{\mstar}   &\colhead{\rstar}    &\colhead{\logg}    &\colhead{\teff}    &\colhead{\feh}  \\
 \colhead{}  &\colhead{(\msun)}   &\colhead{(\rsun)}    &\colhead{(cgs)}    &\colhead{(K)}    &\colhead{}}
\startdata 
GJ 436& $0.426^{+0.027}_{-0.017}$ & $0.415^{+0.014}_{-0.010}$ & $4.833^{+0.013}_{-0.013}$ & $3370^{+94}_{-95}$ & $-0.46^{+0.31}_{-0.24}$ \\
HAT-P-12& $0.719^{+0.016}_{-0.016}$ & $0.7084^{+0.0095}_{-0.0095}$ & $4.594^{+0.013}_{-0.013}$ & $4710^{+49}_{-49}$ & $-0.240^{+0.055}_{-0.062}$ \\
HAT-P-16& $1.187^{+0.058}_{-0.065}$ & $1.217^{+0.022}_{-0.022}$ & $4.342^{+0.024}_{-0.027}$ & $6196^{+59}_{-71}$ & $0.11^{+0.15}_{-0.11}$ \\
HAT-P-18& $0.750^{+0.015}_{-0.014}$ & $0.7202^{+0.0095}_{-0.010}$ & $4.599^{+0.013}_{-0.013}$ & $4835^{+39}_{-35}$ & $0.044^{+0.060}_{-0.051}$ \\
HAT-P-19& $0.863^{+0.029}_{-0.025}$ & $0.851^{+0.013}_{-0.013}$ & $4.514^{+0.019}_{-0.017}$ & $5049^{+42}_{-65}$ & $0.283^{+0.081}_{-0.079}$ \\
HAT-P-20& $0.764^{+0.042}_{-0.051}$ & $0.685^{+0.016}_{-0.021}$ & $4.650^{+0.018}_{-0.018}$ & $4622^{+57}_{-49}$ & $0.23^{+0.25}_{-0.24}$ \\
HAT-P-22& $1.051^{+0.045}_{-0.050}$ & $0.990^{+0.018}_{-0.018}$ & $4.4687^{+0.0096}_{-0.013}$ & $5426^{+55}_{-52}$ & $0.325^{+0.084}_{-0.081}$ \\
HAT-P-24& $1.228^{+0.035}_{-0.036}$ & $1.372^{+0.018}_{-0.018}$ & $4.252^{+0.017}_{-0.017}$ & $6488^{+70}_{-68}$ & $-0.056^{+0.050}_{-0.054}$ \\
HAT-P-27& $0.907^{+0.040}_{-0.036}$ & $0.896^{+0.015}_{-0.015}$ & $4.491^{+0.024}_{-0.023}$ & $5162^{+41}_{-41}$ & $0.210^{+0.097}_{-0.097}$ \\
HAT-P-3& $0.960^{+0.031}_{-0.034}$ & $0.8627^{+0.0090}_{-0.012}$ & $4.550^{+0.015}_{-0.016}$ & $5133^{+36}_{-31}$ & $0.332^{+0.074}_{-0.073}$ \\
HAT-P-30& $1.314^{+0.053}_{-0.056}$ & $1.381^{+0.020}_{-0.022}$ & $4.277^{+0.019}_{-0.021}$ & $6216^{+56}_{-48}$ & $0.155^{+0.077}_{-0.079}$ \\
HAT-P-37& $0.953^{+0.035}_{-0.040}$ & $0.8635^{+0.0072}_{-0.0070}$ & $4.545^{+0.016}_{-0.019}$ & $5479^{+37}_{-52}$ & $0.22^{+0.12}_{-0.12}$ \\
HAT-P-39& $1.414^{+0.039}_{-0.039}$ & $1.723^{+0.028}_{-0.028}$ & $4.116^{+0.017}_{-0.016}$ & $6412^{+56}_{-58}$ & $0.118^{+0.074}_{-0.066}$ \\
HAT-P-4& $1.260^{+0.043}_{-0.042}$ & $1.573^{+0.015}_{-0.015}$ & $4.145^{+0.015}_{-0.015}$ & $5927^{+57}_{-47}$ & $0.258^{+0.070}_{-0.074}$ \\
HAT-P-42& $1.145^{+0.052}_{-0.052}$ & $1.27^{+0.11}_{-0.11}$ & $4.288^{+0.073}_{-0.067}$ & $5773^{+45}_{-40}$ & $0.312^{+0.072}_{-0.074}$ \\
HAT-P-43& $1.042^{+0.036}_{-0.035}$ & $1.076^{+0.025}_{-0.025}$ & $4.392^{+0.026}_{-0.023}$ & $5656^{+37}_{-29}$ & $0.267^{+0.061}_{-0.064}$ \\
HAT-P-5& $1.056^{+0.067}_{-0.067}$ & $1.085^{+0.012}_{-0.012}$ & $4.391^{+0.029}_{-0.031}$ & $5844^{+54}_{-52}$ & $0.14^{+0.18}_{-0.15}$ \\
HAT-P-52& $0.891^{+0.023}_{-0.022}$ & $0.886^{+0.031}_{-0.030}$ & $4.492^{+0.030}_{-0.029}$ & $5144^{+47}_{-48}$ & $0.319^{+0.076}_{-0.078}$ \\
HAT-P-53& $1.078^{+0.036}_{-0.037}$ & $1.127^{+0.039}_{-0.036}$ & $4.367^{+0.030}_{-0.032}$ & $5942^{+36}_{-40}$ & $-0.001^{+0.077}_{-0.064}$ \\
HAT-P-54& $0.706^{+0.027}_{-0.025}$ & $0.6703^{+0.0098}_{-0.0097}$ & $4.635^{+0.019}_{-0.018}$ & $4369^{+62}_{-71}$ & $0.157^{+0.076}_{-0.075}$ \\
HAT-P-56& $1.319^{+0.033}_{-0.033}$ & $1.429^{+0.024}_{-0.023}$ & $4.248^{+0.017}_{-0.018}$ & $6564^{+47}_{-46}$ & $0.11^{+0.14}_{-0.10}$ \\
HAT-P-8& $1.270^{+0.030}_{-0.030}$ & $1.491^{+0.016}_{-0.014}$ & $4.1956^{+0.0095}_{-0.013}$ & $6410^{+140}_{-140}$ & $-0.018^{+0.072}_{-0.056}$ \\
KELT-3& $1.301^{+0.046}_{-0.046}$ & $1.583^{+0.036}_{-0.036}$ & $4.153^{+0.024}_{-0.024}$ & $6306^{+36}_{-35}$ & $0.030^{+0.072}_{-0.066}$ \\
Qatar-4& $0.856^{+0.029}_{-0.030}$ & $0.800^{+0.015}_{-0.014}$ & $4.565^{+0.018}_{-0.020}$ & $5174^{+33}_{-35}$ & $0.095^{+0.076}_{-0.088}$ \\
TrES-1& $0.886^{+0.021}_{-0.022}$ & $0.8200^{+0.0079}_{-0.0078}$ & $4.5577^{+0.0096}_{-0.010}$ & $5316^{+43}_{-42}$ & $0.017^{+0.045}_{-0.041}$ \\
WASP-104& $1.003^{+0.030}_{-0.032}$ & $0.9277^{+0.0089}_{-0.0085}$ & $4.504^{+0.014}_{-0.016}$ & $5392^{+69}_{-73}$ & $0.348^{+0.076}_{-0.076}$ \\
WASP-24& $1.177^{+0.025}_{-0.025}$ & $1.312^{+0.019}_{-0.020}$ & $4.273^{+0.015}_{-0.014}$ & $6489^{+89}_{-74}$ & $-0.37^{+0.12}_{-0.18}$ \\
WASP-28& $1.082^{+0.068}_{-0.071}$ & $1.132^{+0.019}_{-0.018}$ & $4.365^{+0.027}_{-0.033}$ & $5976^{+54}_{-60}$ & $0.06^{+0.17}_{-0.12}$ \\
WASP-36& $1.030^{+0.033}_{-0.036}$ & $0.966^{+0.013}_{-0.014}$ & $4.4807^{+0.0086}_{-0.0085}$ & $6150^{+110}_{-100}$ & $-0.31^{+0.20}_{-0.27}$ \\
WASP-37& $0.926^{+0.039}_{-0.034}$ & $1.071^{+0.019}_{-0.018}$ & $4.346^{+0.023}_{-0.021}$ & $5795^{+69}_{-64}$ & $-0.098^{+0.050}_{-0.060}$ \\
WASP-38& $1.200^{+0.029}_{-0.029}$ & $1.368^{+0.014}_{-0.014}$ & $4.2450^{+0.0081}_{-0.0083}$ & $6168^{+57}_{-58}$ & $-0.001^{+0.065}_{-0.064}$ \\
WASP-43& $0.723^{+0.028}_{-0.031}$ & $0.6747^{+0.0086}_{-0.0097}$ & $4.639^{+0.010}_{-0.010}$ & $4286^{+34}_{-40}$ & $0.28^{+0.19}_{-0.21}$ \\
WASP-52& $0.886^{+0.039}_{-0.043}$ & $0.825^{+0.012}_{-0.012}$ & $4.553^{+0.019}_{-0.024}$ & $5070^{+35}_{-40}$ & $0.37^{+0.13}_{-0.17}$ \\
WASP-56& $1.092^{+0.023}_{-0.023}$ & $1.129^{+0.021}_{-0.020}$ & $4.371^{+0.018}_{-0.018}$ & $5646^{+49}_{-48}$ & $0.160^{+0.057}_{-0.057}$ \\
WASP-60& $1.082^{+0.032}_{-0.031}$ & $1.384^{+0.039}_{-0.037}$ & $4.190^{+0.026}_{-0.027}$ & $5926^{+77}_{-76}$ & $0.023^{+0.067}_{-0.055}$ \\
WASP-65& $1.098^{+0.046}_{-0.052}$ & $1.056^{+0.017}_{-0.018}$ & $4.430^{+0.016}_{-0.015}$ & $5749^{+45}_{-48}$ & $0.447^{+0.086}_{-0.13}$ \\
WASP-80& $0.614^{+0.014}_{-0.012}$ & $0.6023^{+0.0066}_{-0.0052}$ & $4.6672^{+0.0092}_{-0.011}$ & $4158^{+41}_{-49}$ & $-0.011^{+0.034}_{-0.017}$ \\
XO-2& $0.967^{+0.029}_{-0.028}$ & $1.004^{+0.013}_{-0.013}$ & $4.420^{+0.017}_{-0.016}$ & $5360^{+50}_{-50}$ & $0.436^{+0.043}_{-0.046}$ \\
XO-5& $1.053^{+0.064}_{-0.062}$ & $1.065^{+0.026}_{-0.026}$ & $4.406^{+0.029}_{-0.029}$ & $5515^{+69}_{-68}$ & $0.447^{+0.079}_{-0.12}$ \\
\enddata
\end{deluxetable*}

\begin{longrotatetable}
\begin{deluxetable}{lcccccccccccc}
\tablecaption{Planetary Physical and Orbital Parameters for TEMP Targets Obtained in This Study  \label{tab:planetary_parameters}}
\tablehead{ 
 \colhead{Planet}  &\colhead{$P$} &\colhead{$T_0$}   &\colhead{$e$}     &\colhead{$R_{P}/R_{*}$} &\colhead{$a/R_{*}$} & \colhead{$T_{14}$}  &\colhead{$\rp$}    &\colhead{$\mplanet$}  \\
  \colhead{} &\colhead{(BJD$_{\rm TDB}$)}  &\colhead{} &\colhead{}       &\colhead{} &\colhead{} & \colhead{(days)}  &\colhead{($\rj$)}    &\colhead{($\mj$)}
 }
\startdata 
GJ 436 b & $2.64389565^{+0.00000055}_{-0.00000054}$ & $2454346.879828^{+0.000099}_{-0.000100}$ & $0.1684^{+0.0099}_{-0.0096}$ & $0.08536^{+0.00076}_{-0.00064}$ & $14.60^{+0.25}_{-0.25}$ & $0.04236^{+0.00040}_{-0.00038}$ & $0.3443^{+0.014}_{-0.0098}$ & $0.0659^{+0.0029}_{-0.0020}$ \\
HAT-P-12 b & $3.21305856^{+0.00000017}_{-0.00000017}$ & $2456003.233217^{+0.000059}_{-0.000059}$ & $0.018^{+0.018}_{-0.012}$ & $0.14005^{+0.00057}_{-0.00059}$ & $11.59^{+0.16}_{-0.15}$ & $0.09771^{+0.00032}_{-0.00032}$ & $0.965^{+0.015}_{-0.015}$ & $0.2070^{+0.0077}_{-0.0076}$ \\
HAT-P-16 b & $2.77596752^{+0.00000091}_{-0.00000094}$ & $2455238.56602^{+0.00026}_{-0.00027}$ & $0.0400^{+0.0030}_{-0.0031}$ & $0.10615^{+0.00081}_{-0.00082}$ & $7.24^{+0.16}_{-0.17}$ & $0.12804^{+0.00092}_{-0.00086}$ & $1.257^{+0.028}_{-0.028}$ & $4.13^{+0.13}_{-0.15}$ \\
HAT-P-18 b & $5.50802919^{+0.00000068}_{-0.00000069}$ & $2455144.64858^{+0.00013}_{-0.00014}$ & $0.0064^{+0.0082}_{-0.0046}$ & $0.13826^{+0.00092}_{-0.00096}$ & $16.56^{+0.24}_{-0.22}$ & $0.11347^{+0.00059}_{-0.00059}$ & $0.969^{+0.017}_{-0.018}$ & $0.196^{+0.012}_{-0.013}$ \\
HAT-P-19 b & $4.00878735^{+0.00000072}_{-0.00000073}$ & $2455528.49236^{+0.00019}_{-0.00019}$ & $0.021^{+0.024}_{-0.015}$ & $0.1386^{+0.0013}_{-0.0013}$ & $11.88^{+0.22}_{-0.19}$ & $0.11858^{+0.00097}_{-0.00092}$ & $1.148^{+0.023}_{-0.023}$ & $0.300^{+0.016}_{-0.016}$ \\
HAT-P-20 b & $2.87531801^{+0.00000021}_{-0.00000021}$ & $2456374.820509^{+0.000083}_{-0.000084}$ & $0.0157^{+0.0027}_{-0.0025}$ & $0.1351^{+0.0010}_{-0.0010}$ & $11.40^{+0.21}_{-0.19}$ & $0.07780^{+0.00052}_{-0.00054}$ & $0.900^{+0.025}_{-0.031}$ & $7.37^{+0.27}_{-0.33}$ \\
HAT-P-22 b & $3.21223369^{+0.00000043}_{-0.00000043}$ & $2455559.81843^{+0.00015}_{-0.00015}$ & $0.0049^{+0.0034}_{-0.0032}$ & $0.09883^{+0.00064}_{-0.00061}$ & $9.426^{+0.075}_{-0.12}$ & $0.11837^{+0.00061}_{-0.00057}$ & $0.952^{+0.020}_{-0.018}$ & $2.360^{+0.067}_{-0.076}$ \\
HAT-P-24 b & $3.35524419^{+0.00000076}_{-0.00000075}$ & $2455626.31703^{+0.00021}_{-0.00021}$ & $0.00078^{+0.0012}_{-0.00058}$ & $0.09743^{+0.00056}_{-0.00056}$ & $7.36^{+0.13}_{-0.12}$ & $0.15513^{+0.00072}_{-0.00070}$ & $1.301^{+0.021}_{-0.021}$ & $0.742^{+0.024}_{-0.025}$ \\
HAT-P-27 b & $3.03957984^{+0.00000089}_{-0.00000090}$ & $2455483.89831^{+0.00017}_{-0.00017}$ & $0.057^{+0.048}_{-0.038}$ & $0.1194^{+0.0023}_{-0.0022}$ & $9.54^{+0.21}_{-0.20}$ & $0.0718^{+0.0015}_{-0.0017}$ & $1.041^{+0.027}_{-0.026}$ & $0.645^{+0.036}_{-0.036}$ \\
HAT-P-3 b & $2.89973755^{+0.00000018}_{-0.00000018}$ & $2455874.509791^{+0.000082}_{-0.000082}$ & $0.019^{+0.014}_{-0.010}$ & $0.10995^{+0.00056}_{-0.00057}$ & $9.80^{+0.14}_{-0.14}$ & $0.08729^{+0.00041}_{-0.00041}$ & $0.923^{+0.012}_{-0.015}$ & $0.618^{+0.017}_{-0.018}$ \\
HAT-P-30 b & $2.81060217^{+0.00000059}_{-0.00000066}$ & $2456131.01029^{+0.00026}_{-0.00026}$ & $0.020^{+0.019}_{-0.014}$ & $0.1151^{+0.0011}_{-0.0010}$ & $6.65^{+0.12}_{-0.12}$ & $0.0935^{+0.0010}_{-0.0010}$ & $1.546^{+0.030}_{-0.030}$ & $0.766^{+0.027}_{-0.028}$ \\
HAT-P-37 b & $2.79744006^{+0.00000062}_{-0.00000062}$ & $2455983.43142^{+0.00016}_{-0.00016}$ & $0.024^{+0.017}_{-0.016}$ & $0.1350^{+0.0011}_{-0.0011}$ & $9.52^{+0.13}_{-0.15}$ & $0.09615^{+0.00073}_{-0.00071}$ & $1.135^{+0.014}_{-0.014}$ & $1.18^{+0.12}_{-0.12}$ \\
HAT-P-39 b & $3.5438762^{+0.0000011}_{-0.0000011}$ & $2455587.94585^{+0.00029}_{-0.00029}$ & $0.0052^{+0.0091}_{-0.0040}$ & $0.10092^{+0.00069}_{-0.00071}$ & $6.37^{+0.11}_{-0.11}$ & $0.1782^{+0.0010}_{-0.0010}$ & $1.692^{+0.034}_{-0.034}$ & $0.578^{+0.078}_{-0.076}$ \\
HAT-P-4 b & $3.0565240^{+0.0000011}_{-0.0000011}$ & $2454499.50645^{+0.00032}_{-0.00031}$ & $0.0140^{+0.014}_{-0.0095}$ & $0.08362^{+0.00061}_{-0.00059}$ & $6.085^{+0.077}_{-0.078}$ & $0.17423^{+0.00090}_{-0.00091}$ & $1.280^{+0.016}_{-0.015}$ & $0.676^{+0.025}_{-0.025}$ \\
HAT-P-42 b & $4.6418374^{+0.0000050}_{-0.0000050}$ & $2456110.34894^{+0.00056}_{-0.00057}$ & $0.111^{+0.12}_{-0.078}$ & $0.0842^{+0.0020}_{-0.0016}$ & $9.64^{+0.85}_{-0.72}$ & $0.1654^{+0.0038}_{-0.0029}$ & $1.040^{+0.10}_{-0.089}$ & $0.83^{+0.17}_{-0.20}$ \\
HAT-P-43 b & $3.33268034^{+0.00000097}_{-0.00000096}$ & $2456247.32242^{+0.00019}_{-0.00021}$ & $0.082^{+0.19}_{-0.060}$ & $0.11900^{+0.0011}_{-0.00083}$ & $8.84^{+0.24}_{-0.21}$ & $0.13551^{+0.0011}_{-0.00084}$ & $1.246^{+0.031}_{-0.030}$ & $0.34^{+0.36}_{-0.34}$ \\
HAT-P-5 b & $2.78847362^{+0.00000037}_{-0.00000037}$ & $2455348.80101^{+0.00010}_{-0.00010}$ & $0.027^{+0.025}_{-0.019}$ & $0.11399^{+0.00081}_{-0.00086}$ & $7.82^{+0.18}_{-0.19}$ & $0.12273^{+0.00065}_{-0.00066}$ & $1.204^{+0.017}_{-0.017}$ & $0.989^{+0.053}_{-0.053}$ \\
HAT-P-52 b & $2.7535988^{+0.0000014}_{-0.0000014}$ & $2455995.29102^{+0.00031}_{-0.00031}$ & $0.027^{+0.040}_{-0.019}$ & $0.1149^{+0.0016}_{-0.0017}$ & $8.98^{+0.31}_{-0.29}$ & $0.1006^{+0.0012}_{-0.0012}$ & $0.991^{+0.043}_{-0.042}$ & $0.812^{+0.057}_{-0.069}$ \\
HAT-P-53 b & $1.96162393^{+0.00000070}_{-0.00000072}$ & $2456270.81371^{+0.00032}_{-0.00033}$ & $0.063^{+0.052}_{-0.039}$ & $0.1117^{+0.0013}_{-0.0011}$ & $6.00^{+0.20}_{-0.21}$ & $0.1161^{+0.0013}_{-0.0011}$ & $1.224^{+0.049}_{-0.043}$ & $1.467^{+0.076}_{-0.083}$ \\
HAT-P-54 b & $3.79985429^{+0.00000082}_{-0.00000083}$ & $2456656.49059^{+0.00015}_{-0.00014}$ & $0.078^{+0.027}_{-0.024}$ & $0.15675^{+0.00078}_{-0.00080}$ & $13.62^{+0.24}_{-0.23}$ & $0.07493^{+0.00057}_{-0.00056}$ & $1.022^{+0.016}_{-0.016}$ & $0.802^{+0.098}_{-0.098}$ \\
HAT-P-56 b & $2.79083132^{+0.00000010}_{-0.00000011}$ & $2447653.65348^{+0.00055}_{-0.00058}$ & $0.062^{+0.063}_{-0.042}$ & $0.1009^{+0.0014}_{-0.0014}$ & $6.41^{+0.12}_{-0.12}$ & $0.0928^{+0.0014}_{-0.0014}$ & $1.403^{+0.031}_{-0.030}$ & $1.81^{+0.24}_{-0.24}$ \\
HAT-P-8 b & $3.07634338^{+0.00000054}_{-0.00000054}$ & $2455769.73201^{+0.00013}_{-0.00013}$ & $0.0092^{+0.011}_{-0.0066}$ & $0.09192^{+0.00035}_{-0.00034}$ & $6.473^{+0.057}_{-0.083}$ & $0.16467^{+0.00044}_{-0.00043}$ & $1.334^{+0.015}_{-0.013}$ & $1.342^{+0.032}_{-0.032}$ \\
KELT-3 b & $2.7033927^{+0.0000014}_{-0.0000013}$ & $2456426.28703^{+0.00045}_{-0.00044}$ & $0.041^{+0.041}_{-0.029}$ & $0.0947^{+0.0012}_{-0.0012}$ & $5.64^{+0.14}_{-0.14}$ & $0.1320^{+0.0017}_{-0.0017}$ & $1.458^{+0.042}_{-0.041}$ & $1.56^{+0.12}_{-0.11}$ \\
Qatar-4 b & $1.80536494^{+0.00000092}_{-0.00000093}$ & $2457899.55166^{+0.00020}_{-0.00020}$ & $0.046^{+0.064}_{-0.034}$ & $0.13911^{+0.0010}_{-0.00096}$ & $7.42^{+0.13}_{-0.14}$ & $0.08753^{+0.00069}_{-0.00064}$ & $1.083^{+0.022}_{-0.021}$ & $5.26^{+0.22}_{-0.21}$ \\
TrES-1 b & $3.03006967^{+0.00000011}_{-0.00000011}$ & $2455083.630963^{+0.000067}_{-0.000068}$ & $0.0103^{+0.011}_{-0.0072}$ & $0.13542^{+0.00052}_{-0.00049}$ & $10.322^{+0.094}_{-0.095}$ & $0.10466^{+0.00032}_{-0.00030}$ & $1.081^{+0.012}_{-0.012}$ & $0.698^{+0.026}_{-0.025}$ \\
WASP-104 b & $1.75540646^{+0.00000028}_{-0.00000028}$ & $2456788.79015^{+0.00011}_{-0.00011}$ & $0.014^{+0.019}_{-0.010}$ & $0.12117^{+0.00070}_{-0.00068}$ & $6.607^{+0.086}_{-0.090}$ & $0.07515^{+0.00054}_{-0.00049}$ & $1.094^{+0.013}_{-0.013}$ & $1.205^{+0.049}_{-0.044}$ \\
WASP-24 b & $2.34122303^{+0.00000060}_{-0.00000059}$ & $2455781.40485^{+0.00013}_{-0.00013}$ & $0.0123^{+0.013}_{-0.0086}$ & $0.10158^{+0.00042}_{-0.00044}$ & $5.971^{+0.093}_{-0.088}$ & $0.11418^{+0.00057}_{-0.00057}$ & $1.297^{+0.022}_{-0.022}$ & $1.082^{+0.027}_{-0.028}$ \\
WASP-28 b & $3.40883577^{+0.00000043}_{-0.00000021}$ & $2455754.00767^{+0.00031}_{-0.00032}$ & $0.020^{+0.023}_{-0.014}$ & $0.1198^{+0.0012}_{-0.0011}$ & $8.65^{+0.21}_{-0.24}$ & $0.1379^{+0.0013}_{-0.0011}$ & $1.319^{+0.028}_{-0.026}$ & $0.948^{+0.051}_{-0.052}$ \\
WASP-36 b & $1.53736533^{+0.00000014}_{-0.00000014}$ & $2456678.27835^{+0.000047}_{-0.000046}$ & $0.0087^{+0.0097}_{-0.0061}$ & $0.13515^{+0.00027}_{-0.00028}$ & $5.862^{+0.050}_{-0.044}$ & $0.07732^{+0.00019}_{-0.00019}$ & $1.270^{+0.018}_{-0.019}$ & $2.281^{+0.070}_{-0.071}$ \\
WASP-37 b & $3.5774729^{+0.0000025}_{-0.0000025}$ & $2458096.85691^{+0.00019}_{-0.00019}$ & $0.069^{+0.025}_{-0.028}$ & $0.11713^{+0.0010}_{-0.00095}$ & $8.97^{+0.19}_{-0.18}$ & $0.13107^{+0.00100}_{-0.00086}$ & $1.220^{+0.025}_{-0.024}$ & $1.77^{+0.12}_{-0.12}$ \\
WASP-38 b & $6.8718942^{+0.0000037}_{-0.0000036}$ & $2456332.34464^{+0.00051}_{-0.00052}$ & $0.0272^{+0.0042}_{-0.0040}$ & $0.08518^{+0.00062}_{-0.00062}$ & $11.824^{+0.094}_{-0.096}$ & $0.2028^{+0.0013}_{-0.0013}$ & $1.134^{+0.015}_{-0.015}$ & $2.597^{+0.045}_{-0.045}$ \\
WASP-43 b & $0.813474229^{+0.000000047}_{-0.000000047}$ & $2455793.247658^{+0.000031}_{-0.000031}$ & $0.0059^{+0.0053}_{-0.0039}$ & $0.15978^{+0.00044}_{-0.00045}$ & $4.883^{+0.041}_{-0.038}$ & $0.05166^{+0.00016}_{-0.00016}$ & $1.049^{+0.014}_{-0.016}$ & $2.054^{+0.055}_{-0.060}$ \\
WASP-52 b & $1.74978344^{+0.00000032}_{-0.00000032}$ & $2456960.785928^{+0.000061}_{-0.000060}$ & $0.040^{+0.022}_{-0.023}$ & $0.16319^{+0.00059}_{-0.00061}$ & $7.12^{+0.12}_{-0.15}$ & $0.07783^{+0.00023}_{-0.00023}$ & $1.310^{+0.020}_{-0.019}$ & $0.458^{+0.027}_{-0.026}$ \\
WASP-56 b & $4.6170605^{+0.0000028}_{-0.0000028}$ & $2457148.23869^{+0.00046}_{-0.00046}$ & $0.138^{+0.062}_{-0.060}$ & $0.1039^{+0.0021}_{-0.0022}$ & $10.65^{+0.21}_{-0.21}$ & $0.1509^{+0.0029}_{-0.0027}$ & $1.141^{+0.032}_{-0.033}$ & $0.588^{+0.040}_{-0.041}$ \\
WASP-60 b & $4.3050032^{+0.0000055}_{-0.0000055}$  & $2456853.41727^{+0.00029}_{-0.00029}$ & $0.038^{+0.042}_{-0.026}$ & $0.08982^{+0.00081}_{-0.00093}$ & $8.26^{+0.24}_{-0.24}$ & $0.1497^{+0.0013}_{-0.0014}$ & $1.209^{+0.039}_{-0.038}$ & $0.506^{+0.052}_{-0.051}$ \\
WASP-65 b & $2.31142098^{+0.00000039}_{-0.00000039}$ & $2456545.23453^{+0.00011}_{-0.00011}$ & $0.0126^{+0.014}_{-0.0089}$ & $0.11334^{+0.00047}_{-0.00045}$ & $7.174^{+0.11}_{-0.080}$ & $0.11507^{+0.00038}_{-0.00037}$ & $1.165^{+0.020}_{-0.020}$ & $1.719^{+0.062}_{-0.064}$ \\
WASP-80 b & $3.06785314^{+0.00000044}_{-0.00000043}$ & $2456416.864196^{+0.000025}_{-0.000025}$ & $0.0107^{+0.011}_{-0.0075}$ & $0.17212^{+0.00042}_{-0.00043}$ & $12.55^{+0.11}_{-0.13}$ & $0.08901^{+0.00015}_{-0.00015}$ & $1.0091^{+0.011}_{-0.0095}$ & $0.571^{+0.020}_{-0.020}$ \\
XO-2 b & $2.61585881^{+0.00000029}_{-0.00000029}$ & $2454762.4765^{+0.00012}_{-0.00012}$ & $0.0095^{+0.0092}_{-0.0065}$ & $0.10372^{+0.00062}_{-0.00064}$ & $7.87^{+0.13}_{-0.12}$ & $0.11294^{+0.00048}_{-0.00047}$ & $1.013^{+0.018}_{-0.017}$ & $0.589^{+0.015}_{-0.014}$ \\
XO-5 b & $4.18775571^{+0.00000049}_{-0.00000049}$  & $2454510.79378^{+0.000024}_{-0.000024}$ & $0.0101^{+0.012}_{-0.0072}$ & $0.1022^{+0.0010}_{-0.0010}$ & $10.45^{+0.29}_{-0.28}$ & $0.12765^{+0.00083}_{-0.00073}$ & $1.058^{+0.033}_{-0.032}$ & $1.188^{+0.053}_{-0.052}$ \\
\enddata
\end{deluxetable}
\end{longrotatetable}

\clearpage
\section{Acknowledgments}
We thank Jason Eastman for his useful discussions. We also would like to thank Metrics for advice on figure designs that has led to a considerable improvement of the manuscript. S.W. acknowledges support from Indiana University, Yale University, and the Heising-Simons Foundation. M.R. is supported by the National Science Foundation Graduate Research Fellowship Program under Grant Number DGE-1752134. TCH acknowledges financial support from the National Research Foundation (NRF; No. 2019R1I1A1A01059609). M.B. and Y.W. thank the support of the National Natural Science Foundation of China (Grant No. 12073092). Y.W. thanks the fellowship of the China Postdoctoral Science Foundation (Grant No. 2020M672936). H.Z. thanks the support of the National Natural Science Foundation of China (Grant No. 12073010). This work is supported by the National Natural Science Foundation of China (NSFC) (11803055), the Joint Research Fund in Astronomy (U1731125, U1731243, U1931132) under cooperative agreement between the NSFC and Chinese Academy of Sciences (CAS), the 13th Five-year Informatization Plan of Chinese Academy of Sciences (No. XXH-13514, XXH13503-03-107). This work is supported by Astronomical Big Data Joint Research Center, co-founded by National Astronomical Observatories, Chinese Academy of Sciences and Alibaba Cloud. We acknowledge the support of the staff of the Xinglong 60 cm telescope. This work was partially supported by the Open Project Program of the Key Laboratory of Optical Astronomy, National Astronomical Observatories, Chinese Academy of Sciences.

\end{document}